\documentclass[11pt,oneside,letterpaper]{article}
\pdfoutput=1
\usepackage{amssymb}
\usepackage{amsmath}
\usepackage[dvips]{graphicx}
\usepackage{setspace}
\usepackage{mathtools} 
\usepackage{amsfonts}
\usepackage{fancyhdr} 
\usepackage{xcolor} 
\usepackage{graphicx} 
\usepackage{rotating}
\usepackage{comment}
\usepackage{color}
\usepackage{caption}
\usepackage{subcaption}
\usepackage[percent]{overpic}
\usepackage{cite}
\usepackage{braket}
\definecolor{darkgreen}{rgb}{0,0.5,0}
\definecolor{darkblue}{rgb}{0,0,0.6}
\definecolor{purple}{rgb}{0.4,.2,0.7}
\newcommand{\p}{\partial}

\newcommand{\f}{\frac}

\newcommand{\be}{\begin{equation}}
\newcommand{\ee}{\end{equation}}

\usepackage[colorlinks=true,citecolor=darkgreen,linkcolor=black,urlcolor=purple]{hyperref}

\usepackage{pdfsync}

\makeatletter
\newcommand*{\defeq}{\mathrel{\rlap{%
                     \raisebox{0.3ex}{$\m@th\cdot$}}%
                     \raisebox{-0.3ex}{$\m@th\cdot$}}%
                     =} 
\makeatother

\def\be{\begin{eqnarray}}
\def\ee{\end{eqnarray}}

\newcommand{\ra}{\rangle}
\newcommand{\tr}{\textrm{Tr}\,}

\newcommand{\bea}{\begin{eqnarray}}
\newcommand{\eea}{\end{eqnarray}}
\def\ben{\begin{equation}}
\def\een{\end{equation}}

   \let\d=\delta 
\let\z=\zeta  \let\q=\theta 
\let\l=\lambda \let\m=\mu \let\n=\nu  \let\p=\phi \let\r=v
\let\sg=\sigma   

\let\y=\psi
\let\w=\omega

\def\no{\nonumber \\}
\let\f=\frac

\let\pa=\partial
\def\be{\begin{equation}}
\def\ee{\end{equation}}
\def\ba{\begin{eqnarray}}
\def\ea{\end{eqnarray}}
\def\del{\partial}

\def\bal#1\eal{\begin{align}#1\end{align}}
\def\bs#1\es{\begin{split}#1\end{split}}

\newcommand{\AAl}[1]{{\textbf{\textcolor{red}{#1}}}}

\renewcommand{\p}{\partial}

\allowdisplaybreaks  

\numberwithin{equation}{section}

\def\m{{\mu}}
\def\w{{\omega}}

\def\n{{\nu}}

\def\ep{{\epsilon}}
\def\d{{\delta}}

\def\p{{\phi}}

\def\sg{{\sigma}}

\def\s{\sqrt}

\def\CO{{\cal O}}

\newcommand{\bz}{\bar{z}}

\def\pp{\partial}

\def\be{\begin{equation}}
\def\ee{\end{equation}}
\def\ba{\begin{eqnarray}}
\def\ea{\end{eqnarray}}
\def\bal#1\eal{\begin{align}#1\end{align}}

\def\r{\rightarrow}

\def\LR{\Leftrightarrow}
\def\f {\frac}

\def\no{\nonumber \\}

\def\l{\left}
\def\r{\right}
\def\ep{\epsilon}
\def\ra{\rightarrow}

\def\q{\quad}

\def\qqq{\quad\quad\quad}

\def\z{\bar{z}}
\def \bA {\bar{A}}
\def\y{\bar{y}}
\def\w{\bar{w}}

\def \be {\begin{equation}}
\def \ee {\end{equation}}
	
\def \JM#1 {{\color{blue}  JM: #1 }}
\def \AAl#1 {{\color{red}  AA: #1 }}

\newcommand{\ext}{\mbox{ext}}
\newcommand{\Ssemi}{S_{\rm QFT}}

\renewcommand{\p}{\partial}
\renewcommand{\min}{{\rm min}}
\newcommand{\bh}{\bar{h}}
\newcommand{\by}{\bar{y}}

\newcommand{\Smatter}{S_{\rm QFT}}

\usepackage{framed}

\begin{document}
\begin{spacing}{1.1}
\begin{center}

~
\vskip1cm

{\LARGE {
Replica wormholes for an evaporating 2D black hole
\ \\
}}

\vskip10mm

Kanato Goto,$^{1,2}$\ \ Thomas Hartman,$^{1}$  and Amirhossein Tajdini$^{3}$

\vskip5mm

{\it $^1$ Department of Physics, Cornell University, Ithaca, New York, USA
}  \\ \bigskip
{\it $^2$ RIKEN Interdisciplinary Theoretical and Mathematical Sciences (iTHEMS),\\ 
Wako, Saitama 351-0198, Japan } \\  \bigskip
{\it $^3$ Department of Physics, University of California, Santa Barbara, CA 93106, USA
}
\vskip5mm

{\tt  kanato.goto@riken.jp, hartman@cornell.edu, ahtajdini@ucsb.edu}

\vskip5mm

\end{center}

\vspace{4mm}
\begin{abstract}
\noindent
 Quantum extremal islands reproduce the unitary Page curve of an evaporating black hole. 
This has been derived by including replica wormholes in the gravitational path integral, but for the transient, evaporating black holes most relevant to Hawking's paradox, these wormholes have not been analyzed in any detail. In this paper we study replica wormholes for black holes formed by gravitational collapse in Jackiw-Teitelboim gravity, and confirm that they lead to the island rule for the entropy. The main technical challenge is that replica wormholes rely on a Euclidean path integral, while the quantum extremal islands of an evaporating black hole exist only in Lorentzian signature. Furthermore, the Euclidean equations for the Schwarzian mode are non-local, so it is unclear how to connect to the local, Lorentzian dynamics of an evaporating black hole.
We address these issues with Schwinger-Keldysh techniques and show how the non-local equations reduce to the local `boundary particle' description in special cases.

 \end{abstract}

\pagebreak
\pagestyle{plain}

\setcounter{tocdepth}{2}
{}
\vfill

\end{spacing}
\tableofcontents

\newpage
\begin{spacing}{1.1}

\section{Introduction}

The entropy of Hawking radiation is a diagnostic of information loss. It was long believed that this entropy could only be computed in the ultraviolet theory. Recently, however, the entropy was calculated in the low-energy theory by Almheiri, Engelhardt, Marolf and Maxfield \cite{Almheiri:2019psf} and simultaneously by Penington \cite{Penington:2019npb} using an extension of the gravitational entropy formula \cite{Penington:2019npb,Almheiri:2019psf, Ryu:2006bv, Hubeny:2007xt, Lewkowycz:2013nqa, Barrella:2013wja, Faulkner:2013ana, Engelhardt:2014gca}. They discovered that after the Page time, a quantum extremal island appears in the black hole interior. The gravitational entropy formula in this context, called the `island formula' \cite{Almheiri:2019hni}, is
\be\label{islandRT}
S(R) = \min \; \ext_I \left[ \f{\mbox{Area}(\pp{I})}{4G_N}+\Ssemi(I\cup R)  \right]
\ee
where $R$ is the radiation, $I$ is the island, and $\Ssemi$ is the entropy of quantum fields calculated by the traditional methods of quantum field theory in curved spacetime. In two-dimensional gravity, `Area' means the value of the dilaton. 

Almheiri et. al. \cite{Almheiri:2019psf} did the calculation in a two-dimensional model of Jackiw-Teitelboim (JT) gravity \cite{JACKIW1985343,TEITELBOIM198341} glued to non-dynamical flat spacetime. Penington \cite{Penington:2019npb} also gave some general arguments that the quantum extremal surface should exist for higher-dimensional black holes. In further work, the entropy formula has  been applied to various other setups and it consistently produces a unitary Page curve \cite{Almheiri:2019hni, Almheiri:2019yqk, Almheiri:2019qdq, Penington:2019kki, Hartman:2020swn, Hartman:2020khs,Chen:2020tes,Chen:2020uac,Chen:2020jvn,Chen:2020hmv ,Balasubramanian:2020xqf,Hollowood:2020cou,Hollowood:2020kvk,Geng:2020qvw,Alishahiha:2020qza,Hashimoto:2020cas,Anegawa:2020ezn,Gautason:2020tmk } (see \cite{Almheiri:2020cfm} for an introductory review).

The island formula \eqref{islandRT} was originally proposed based on holographic reasoning, and therefore assuming unitarity. It was then derived without holography from the gravitational path integral in \cite{Almheiri:2019qdq,Penington:2019kki}, using the gravitational replica method  developed earlier in \cite{Lewkowycz:2013nqa, Dong:2016hjy}. This entails calculating $\tr (\rho_R)^n$ by a Euclidean path integral, and analytically continuing $n\to 1$. At finite $n$, there are `replica wormholes' connecting the different copies of the black hole through their interiors. In the limit $n \to 1$, the mouth of the wormhole degenerates to the island region, $I$.

In this paper we undertake a detailed analysis of replica wormholes for a black hole in two-dimensional JT gravity that begins near the vacuum state, forms by gravitational collapse, and then evaporates. We focus on $n \sim 1$. In this example we can be much more explicit about the global, real-time solutions in Lorentzian signature, as compared to \cite{Lewkowycz:2013nqa,Dong:2016hjy} and the general arguments in \cite{Almheiri:2019qdq,Penington:2019kki}. Along the way, we will gain a better understanding of how to apply gravitational path integral techniques to collapsing black holes. This is important because the replica wormhole derivation of the island formula requires a state prepared by a Euclidean path integral --- but this seems to exclude ordinary black holes formed by collapse, because they have no time-reflection symmetry and therefore no real Euclidean continuation. We will overcome this by viewing the collapsing black hole as a limit of a Euclidean solution that scales away the time-reversed process.

We will also repeat the derivation of the island formula in this explicit example. That is, we will fully define the global replica equations in this Lorentzian setup, analyze them as $n \to 1$, and use the Schwarzian equations to show that the gravitational path integral in the low energy theory reproduces \eqref{islandRT}. We assume a large-$N$ matter sector. As with all derivations of gravitational entropy from the replica method, this is on a similar conceptual footing to the Gibbons-Hawking derivation of the area law ---  it computes the entropy without telling us what microstates are responsible, or whether such microstates really exist. 

Finally, this example illustrates some conceptual points that are hidden in the derivation of the island rule based on the action \cite{Almheiri:2019qdq, Penington:2019kki}, such as the role of conformal welding, the relation between the local analysis \cite{Lewkowycz:2013nqa} and the Schwarzian theory, and other subtleties that arise in Lorentzian signature. None of these subtleties lead to any surprises in the end, but there could be other contexts, like cosmology \cite{Dong:2020uxp,Krishnan:2020fer,Chen:2020tes,Hartman:2020khs,VanRaamsdonk:2020tlr,Balasubramanian:2020coy} or applications to other observables \cite{Marolf:2020xie,Engelhardt:2020qpv,Stanford:2020wkf,Akers:2020pmf,Marolf:2020rpm,Goel:2020yxl}, where the answers are less clear \textit{a priori} and these subtleties come into play.

Our setup is similar to Almheiri, Engelhardt, Marolf, and Maxfield (AEMM) \cite{Almheiri:2019psf}: an evaporating black hole in AdS$_2$ glued to non-dynamical flat spacetime. (One important difference is that we create the black hole with an operator insertion, instead of a joining quench.) The Lorentzian theory is described by the Schwarzian action coupled to an external system, which adds a source term in the Schwarzian equation of motion \cite{Maldacena:2016upp,Engelsoy:2016xyb,Jensen:2016pah}. Our main new results are as follows:
\begin{itemize}
\item We construct an evaporating black hole from the Euclidean path integral. This is nontrivial because the Euclidean equations in the Schwarzian theory glued to flat space involve the non-local (and generally intractable) conformal welding problem. We start with a shockwave created by operators inserted in the Euclidean path integral. Then we take a scaling limit where the nonlinear conformal welding problem becomes exactly solvable. This turns out to reproduce the local Lorentzian equations for the Schwarzian `boundary particle' derived in \cite{Maldacena:2016upp,Engelsoy:2016xyb} and studied in AEMM \cite{Almheiri:2019psf}. (See section \ref{sec:3}).
\item We derive the replica equations for the Schwarzian theory defined on a Schwinger-Keldysh contour. This is necessary to study replica wormholes for evaporating black holes because the nontrivial island we seek does not exist in Euclidean signature.  (See section \ref{sec:repworm}.)
\item We derive the extremality conditions for the island from the replica equations of motion in the Schwarzian theory as $n \to 1$. (See section \ref{sec:QESSch}.) The derivation applies to any state in this theory created by a Euclidean path integral with local operator insertions, generalizing the eternal black hole analysis in \cite{Almheiri:2019qdq}. In the above-mentioned scaling limit, this includes black holes that form by collapse, then evaporate.
\item Using the Ward identity for CFT coupled to gravity, we show that the extremality conditions can be integrated to yield the entropy formula \eqref{islandRT}. This argument is equivalent to the derivation of the entropy in \cite{Almheiri:2019qdq,Penington:2019kki} from the defect action, but easier in practice. (See section \ref{sec:QESward}.)
\item We also analyze the two-interval replica wormhole for an eternal black hole at late times, elaborating on a calculation in \cite{Almheiri:2019qdq}. We show that at late times the wormhole factorizes into two copies of the one-interval solution. This result, which is logically independent from the rest of the paper, confirms a physical argument made in \cite{Almheiri:2019qdq} that wormholes factorize in the OPE limit of the twist operators. This assumption was necessary in order to apply replica wormholes to the information paradox of an eternal black hole. (See section \ref{sec:fac}.)
\end{itemize}
Our results give an explicit realization of the Schwinger-Keldysh approach to the gravitational entropy formula developed in \cite{Dong:2016hjy}, incorporating quantum effects. 
In the rest of this introduction, we describe our setup in more detail, highlight some technical challenges, and summarize how things unfold.

\subsection{Summary}
\begin{figure}
      \begin{center}
     \includegraphics[height=7cm]{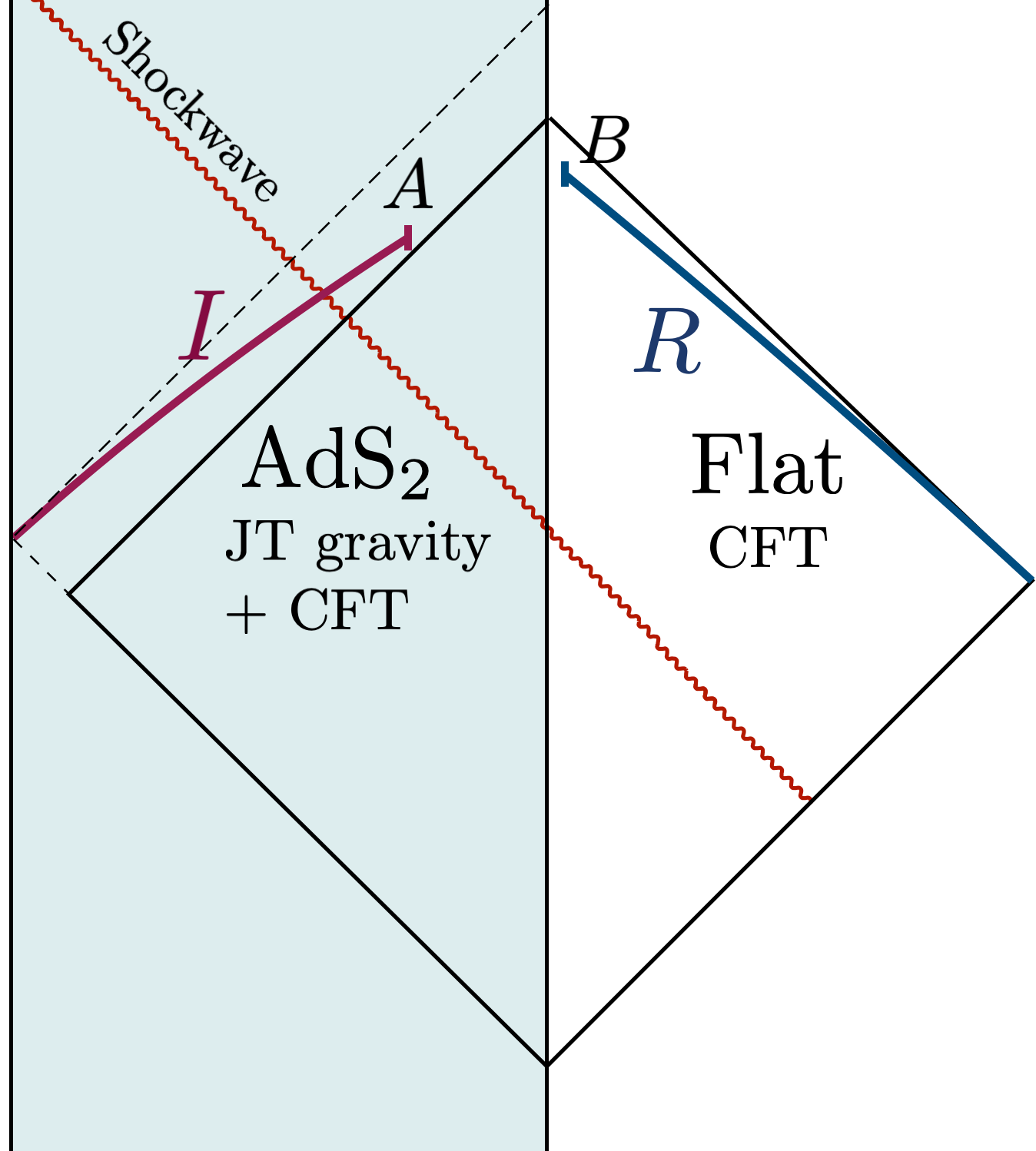}
  \caption{\small A black hole in AdS$_2$ that is created by a shockwave, then evaporates. The AdS$_2$ region is initially in vacuum. We will calculate the von Neumann entropy of region $R$.}
  \label{fig:setup-onesided}
 \end{center}
 \end{figure}
 
We are primarily interested in the Page curve for the black hole in figure \ref{fig:setup-onesided}. The theory is JT gravity in AdS$_2$ coupled to a large-$N$ CFT, glued to a flat spacetime as in \cite{Engelsoy:2016xyb, Almheiri:2019psf}. The CFT lives everywhere while gravity lives only in AdS$_2$, so the flat spacetime is non-dynamical. A delta-function shockwave is sent from ${\cal I}^-$ to form a black hole. The black hole evaporates and Hawking radiation escapes toward ${\cal I}^+$. We will calculate the entropy of region $R$. The final answer \cite{Almheiri:2019psf} is given by the gravitational entropy formula \eqref{islandRT}, with an island that appears after the Page time.

For $t>0$, this solution is nearly identical to the evaporating black hole studied in \cite{Almheiri:2019psf} (AEMM). In the AEMM setup, the AdS$_2$ and flat regions are initially disconnected, and the shockwave is created when these two regions are suddenly joined at $t=0$. We will not take this route because we found it difficult to study this setup with Euclidean path integrals.  Fortunately our $t>0$ solution for the Schwarzian mode in Lorentzian signature is identical, after taking a limit described below, so for the island analysis we can borrow the results of AEMM (and its extension to finite $\beta$ in \cite{Chen:2020jvn,Hollowood:2020kvk,Hollowood:2020cou  }).
\begin{figure}
      \begin{center}
 \includegraphics[width=0.7\textwidth]{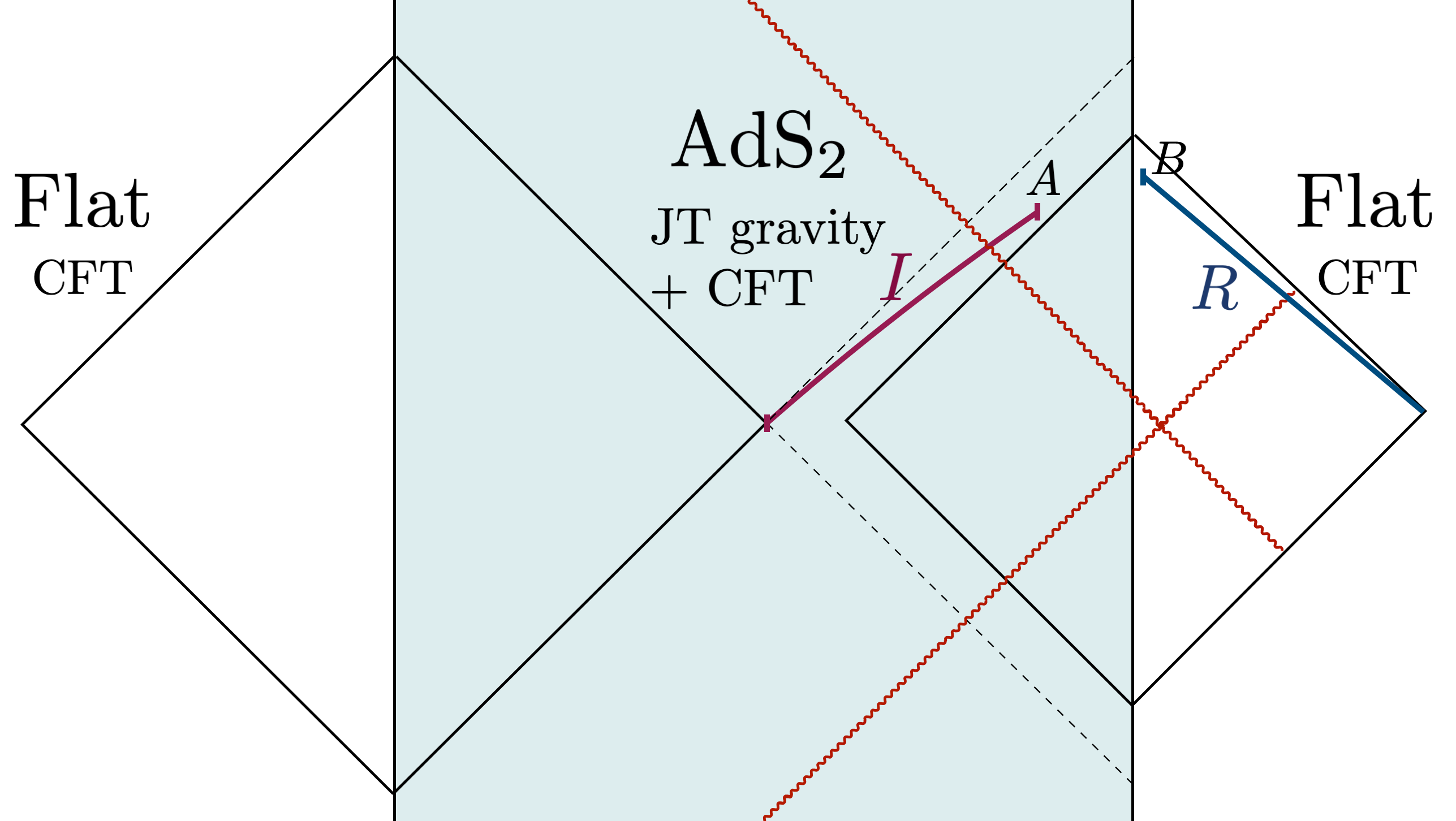}
  \caption{\small
  A shockwave thrown into an AdS$_2$ black hole. In this case, the AdS$_2$ region starts at finite temperature. Here we show a time-symmetric shockwave, with both ingoing and outgoing shocks, which has a straightforward Euclidean continuation.
  }
  \label{fig:setup-twosided2}
 \end{center}
 \end{figure}

Instead of figure \ref{fig:setup-onesided}, we will start with the more general setup in figure \ref{fig:setup-twosided2}. At $t=0$ the gravity region is a black hole at temperature $\beta$. The solution is time-symmetric, so there are ingoing and outgoing shockwaves. The shockwaves are not delta functions because we want the observables, and in particular the entropy, to be analytic functions of position (up to the usual singularities associated to coincident points in Euclidean and lightcones in Lorentzian which are handled by the $i\epsilon$ prescription). So the shockwaves are smeared over a width $\delta$. This  state is defined by inserting the CFT operators
\be
\psi(y_1) \psi(y_2)
\ee
into the Hartle-Hawking path integral, with
\be
y_1 =L+i\d,\q  y_2 = L-i\d\, .
\ee
Here $y$ is a complex coordinate defined below. $L$ is the distance from the AdS boundary, and the shift by $i\delta$ is an offset in imaginary time that smears out the shockwave.
The shockwave operator $\psi$ is a scalar primary with conformal weights $h_\psi = \bh_\psi$, related to the energy of the shockwave by
\be
E_\psi =  \frac{h_\psi}{\delta} \ .
\ee
The CFT state created this way has been studied extensively in \cite{Nozaki:2014hna,Asplund:2014coa,Roberts:2014ifa,Hartman:2015lfa,Caputa:2015waa,Anous:2016kss,Afkhami-Jeddi:2017rmx}, mostly for applications to the AdS$_3$/CFT$_2$ correspondence, with the shockwave operators inserted in the boundary CFT. Here we will use similar techniques but there is no holography; the operators are inserted in the matter CFT, which is directly coupled to gravity in the AdS$_2$ region.

The operators are inserted in the background of the Euclidean eternal black hole. But even at $t=0$, the geometry backreacts, so the gravitational solution is not exactly an eternal black hole. The reason is simply that the CFT stress tensor produced by the operator insertions is everywhere non-vanishing. This backreacts on the geometry, which feeds back into the matter stress tensor and thus leads to a complicated set of non-local equations coupling the geometry to the matter stress tensor. The equations involve an implicit solution to the conformal welding problem, and cannot be solved analytically at finite $\delta$. We will not solve these equations, but we will write them down, and solve them in the limit $\delta \to 0$.

Even before getting to the replica problem, this addresses a puzzle raised by comparing the two papers \cite{Almheiri:2019psf} and \cite{Almheiri:2019qdq}. In \cite{Almheiri:2019psf}, the Lorentzian geometry associated to a shockwave was found by solving a relatively simple local equation for the Schwarzian mode \cite{Engelsoy:2016xyb} at the boundary of AdS$_2$. By contrast the Euclidean equations of \cite{Almheiri:2019qdq} are non-local due to conformal welding whenever there are nontrivial operators inserted, including the operators that produce the shockwave. How does conformal welding in Euclidean signature lead to nice, local equations in Lorentzian signature? The answer we will find is that as $\delta \to 0$, the conformal welding equations have an exact nonlinear solution in Lorentzian signature that agrees with \cite{Almheiri:2019psf}. This is somewhat surprising, since it is impossible to find exact nonlinear solutions to conformal welding in Euclidean signature, but it had to be the case for consistency between \cite{Almheiri:2019psf} and \cite{Almheiri:2019qdq}. We will postpone the more technical discussion until later, but the basic observation is that nonlinear conformal welding is tractable when the Schwarzian mode is purely positive or purely negative frequency, a situation that is possible only in Lorentzian signature. In the $\delta \to 0$ limit the outgoing shockwave decouples from the ingoing shockwave --- this simplifies the conformal welding problem, and then the Euclidean equations reduce exactly to the `boundary particle' equations for the Schwarzian theory studied in \cite{Maldacena:2016upp,Engelsoy:2016xyb,Almheiri:2019psf}.\footnote{Taking $\delta \to 0$ has two related effects. First, it makes the shockwaves delta-function localized. Second, it prevents the outgoing shock from overlapping with the ingoing shock. Only the second effect is necessary for this simplification. There is also a local boundary particle description of a general incoming matter distribution.} The boundary particle equation does not apply in the more general case with simultaneous ingoing and outgoing matter.

At finite $\delta$, the black holes in figs. \ref{fig:setup-onesided}-\ref{fig:setup-twosided2} are described as solutions of the Schwarzian equation on a Schwinger-Keldysh contour. The equations can be solved as $\delta \to 0$. The one-sided shockwave in figure \ref{fig:setup-onesided} is obtained by sending the operator insertions to ${\cal I}^-$. (Alternatively we could smear the insertion against a wavepacket, but we will not do this in detail.)

Another puzzle in figure \ref{fig:setup-onesided} is the role of the left endpoint of the island. It appears to sit on the left boundary of AdS$_2$. However this is potentially problematic because in calculating the entropy of the radiation region $R$, we should \textit{not} include the boundary point on the left side of the Penrose diagram. We could therefore question whether the replica manifolds associated to figure \ref{fig:setup-onesided}  obey the correct boundary conditions on the left boundary. This puzzle is eliminated by going to finite temperature in figure \ref{fig:setup-twosided2}. The left endpoint of the island is now a second quantum extremal surface that sits near the bifurcation point of the original black hole. Now it is manifest that the correct boundary conditions are obeyed.

So far we have described the background solution, or the $n=1$ replica manifold. We now turn to finite $n$, which is relatively straightforward now that we have framed the problem in a convenient way. Our strategy (see section \ref{sec:repworm}) will be to write the gravitational equation of motion (i.e., the Schwarzian equation) in Lorentzian signature and solve it in the limit $\delta \to 0$, $n \sim 1$. This gives the extremality condition for the quantum extremal surface. This derivation (see section \ref{sec:qesreplica}) is very general for JT gravity, not limited to the shockwave state. It combines elements of \cite{Almheiri:2019qdq} and \cite{Penington:2019kki}. Conformal welding enters this calculation in the intermediate steps, but it ultimately drops out of the extremality condition. We will also discuss how this derivation relates to the local analysis of the dilaton equations of motion around the defect that was used by Dong and Lewkowycz \cite{Dong:2017xht} to derive the quantum extremality condition. 

Finally, we must calculate the entropy from the gravitational action plus the 1-loop effective action of the matter fields. There is a shortcut: we demonstrate using the Ward identity that once the extremality conditions have been derived, the entropy automatically agrees with \eqref{islandRT} (see section \ref{sec:QESward}). This is similar in spirit to Cardy and Calabrese's derivation of entanglement entropy in CFT from the conformal Ward identity \cite{Calabrese:2004eu}, but the argument is modified in the presence of gravity.

\section{Evaporating Black Holes in JT  gravity plus a CFT}\label{sec2}

In this section we will review the formulation of the information paradox in JT gravity along the lines of \cite{Almheiri:2019psf}, in Lorentzian signature. Readers familiar with \cite{Almheiri:2019psf} can skip to section \ref{sec:3}.

There are two differences in our setup compared to \cite{Almheiri:2019psf}. The first is that we retain a finite temperature parameter $\beta$ for the initial black hole; in \cite{Almheiri:2019psf} the temperature is taken to zero, but the case of finite $\beta$ has been studied in \cite{Hollowood:2020cou,Chen:2020jvn}. The second difference is that our shockwave is produced by local operator insertions, rather than a joining quench. The Lorentzian solution is identical for $t>0$, so this will not make any difference until the next section.

\subsection{Jackiw-Teitelboim gravity theory plus a CFT}

We begin with the Lorentzian theory in AdS$_2$ coupled to a CFT on just one side of AdS. We will generalize to the 2-sided gluing below.
The action of Jackiw-Teitelboim gravity in AdS$_2$ coupled to a CFT is 
\bal
I_{{\rm JT}+{\rm CFT}}[g_{\m\n},\phi,\, \chi]=I_{{\rm JT}}[g_{\m\n},\phi]+I_{{\rm CFT}}[g_{\m\n},\chi]
\eal
where (in Lorentzian signature)
\bal
I_{\rm JT}[g_{\mu\nu}, \phi] &=\f{\phi_0}{16\pi G_N}\int_{\Sigma_2}R + \f{\phi_0}{8\pi G_N}\int _{\del \Sigma_2}K\no
&\qquad +\f{1}{16\pi G_N}\int_{\Sigma_2}\phi\l(R+
2\r)+\f{1}{8\pi G_N}\int _{\del \Sigma_2}\phi_b\l(K-
1\r)\, ,
\eal
where we set $\ell_{\rm AdS}=1$.
The first line is topological. We take the matter action to be independent of the dilaton. We couple this system to the same CFT living in a portion of rigid Minkowski spacetime and impose transparent boundary conditions for the CFT at the interface \cite{Almheiri:2019psf,Almheiri:2019qdq}. 
Variation with respect $\phi$ yields the equation $R+2=0$ which requires the metric to be locally AdS$_2$.
We take Poincare coordinates $x^\pm$ for the AdS$_2$ geometry in the interior region and $y^{\pm}=t\pm \sg$ for the flat space in the exterior region,
\ba\label{coordinate}
ds_{\rm int }^2=-\f{4
}{(x^+-x^-)^2}dx^+dx^-\, ,\q ds_{\rm ext }^2=-\f{1}{\ep^2}dy^+dy^-\, .
\ea
The two spacetimes are glued together near the AdS$_2$ boundary according to a map $x(t)$ that will be determined dynamically. Given $x(t)$, the gluing identifies the curve $y^+ = y^- = t$ in the center of Minkowski spacetime to the curve
\be
x^+ = x(t) - \epsilon 
x'(t) , \quad x^- = x(t) + \epsilon 
x'(t)
\ee
in AdS. In the limit $\epsilon \to 0$ the gluing is simply along $x^\pm = x(t)$. At the interface, the boundary conditions for the metric and the dilaton are
\ba
g_{tt}|_{\rm bdy}=-\f{1}{\ep^2}\,,\q \phi|_{\rm bdy}=\phi_b=\f{\phi_r}{\ep}\, .
\ea

The variation with respect to the metric yields the dilaton equations
\begin{align}
\partial_{x^{+}} \partial_{x^{-}} \phi+\frac{2}{\left(x^{+}-x^{-}\right)^{2}} \phi &=8\pi G_{N} T_{x^{+} x^{-}} \\-\frac{1}{\left(x^{+}-x^{-}\right)^{2}} \partial_{x^{\pm}}\left(\left(x^{+}-x^{-}\right)^{2} \partial_{x^{\pm}} \phi\right) &=8 \pi G_{N} T_{x^{\pm} x^{\pm}}\, 
\end{align}
where  $T_{\mu\nu}=-\f{2}{i\s{-g}}\f{\d }{\d g^{\mu\nu}} \log Z_{\rm CFT}$. The trace $T_{x^+x^-}$ is set by the anomaly and can be absorbed into the definition of $\phi_0$, so we will adopt this convention and drop it from the first equation. Then the general solution of these equations is \cite{Almheiri:2014cka}
\bal\label{dilaton}
\phi(x^+,x^-)=-\f{2\pi
\phi_r}{\beta}\f{x^++x^-}{x^+-x^-}-\f{8\pi G_N}{x^+-x^-}\int^{x^-}_0dx (x^+-x)(x^--x)T_{x^-x^-}(x)\no +\f{8\pi G_N}{x^+-x^-}\int^{x^+}_0dx (x^+-x)(x^--x)T_{x^+x^+}(x)
\eal
up to $SL(2)$ transformations. 

As argued in \cite{Maldacena:2016upp}, it is convenient to express the dynamics of JT gravity in terms of the shape of the boundary curve, or gluing map, $x(t)$. The dynamics of $x(t)$ are governed by the Schwarzian action. After setting $R = -2$, the JT action is a pure boundary term that evaluates to
\be
I_{\rm JT} = -\f{\phi_r
}{8\pi G_N}\int _{\del \Sigma_2} dt \{x(t),t\}\,  + \mbox{topological} \ .
\ee
The ADM energy of the gravity region is given by
\ba
M(t)=-\f{\phi_r
}{8\pi G_N}\{x(t),t\}\, .
\ea
Conservation of energy relates $dM/dt$ to the net flux of energy across the interface,
\ba\label{Schwarzianeq}
\f{dM}{dt}=-\f{d}{dt}\l(\f{\phi_r 
}{8\pi G_N}\{x(t),t\}\r)=T_{y^+y^+}-T_{y^-y^-}\, .
\ea
This is the equation of motion for the gluing map $x(t)$.

We will be particularly interested in situations where $T_{x^- x^-} = 0$ at the interface. In this case we can use the equation of motion to re-express the dilaton \eqref{dilaton} in terms of the gluing map as 
\ba\label{dilaton2}
\phi(x^+,x^-)=-\phi_r\l[\f{2x'(y^+)}{x^+-x^-}-\f{x''(y^+)}{x'(y^+)}\r]\, , \quad
\mbox{where} \quad y^+ = x^{-1}(x^+)  \ .
\ea
If AdS$_2$ is glued to Minkowski spacetime at both of the AdS$_2$ boundaries, then this expression for the dilaton holds in a region on the right side of AdS$_2$ that is spacelike separated from the left interface.

\subsection{Eternal black hole coupled to a bath}
\begin{figure}
     \begin{center}
 \includegraphics[height=7cm]{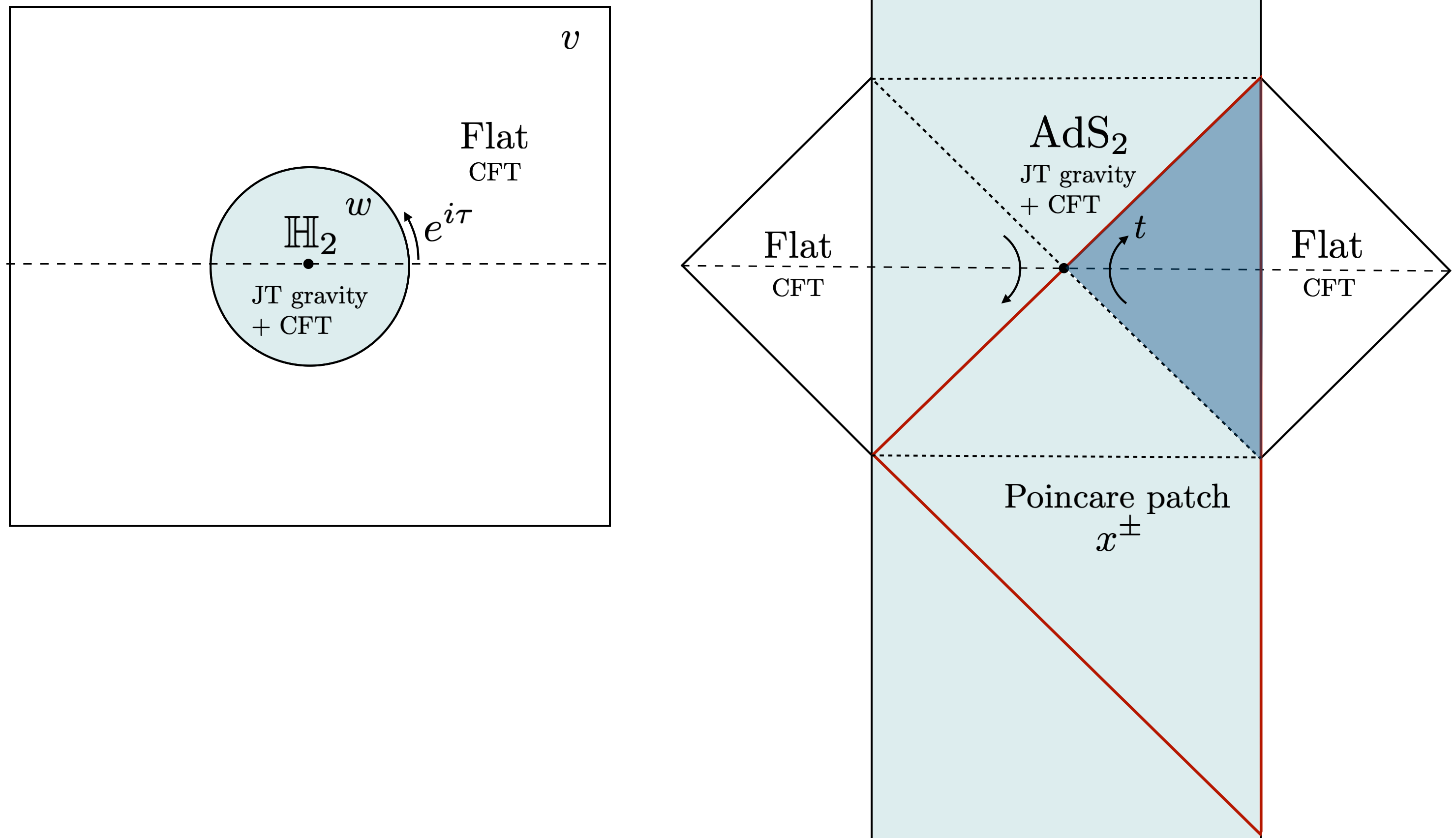} 
  \caption{ Eternal black hole glued to flat space in Euclidean (left) and Lorentzian (right). The region within the red triangle is covered by the Poincare coordinates $x^\pm$. The coordinates $y^\pm = t \pm \sigma$  cover the blue wedge and the right Minkowski half-space. \label{systemEL}}
 \end{center}
 \end{figure}

We now review the eternal black hole coupled to an external bath system which is analyzed in \cite{Almheiri:2019psf}. We couple an eternal black hole at inverse temperature $\beta$ to Minkowski spacetime at the same temperature. In the Lorentzian picture, the geometry  is an eternal black hole in AdS$_2$ connected to two non-gravitating half-spaces, one on each side as drawn in figure \ref{systemEL}. The gravitational system is in equilibrium with the bath. Therefore the net flux is zero, and the Schwarzian equation of motion is
\ba\label{eq:eternal}
\p_t \{x(t),t\} =0\, .
\ea
The solution corresponding to inverse temperature $\beta$ is
\be\label{solueternal}
x(t) = e^{2\pi t / \beta} \ .
\ee
We can also act on this with an SL$(2)$ transformation, which would correspond to a different coordinate system in AdS$_2$.

\textit{A priori}, the gluing map $x(t)$ is defined only at the interface, so it is a real function of a real variable, and the $y^\pm$ coordinates are defined only in the Minkowski region. But we can extend the $y^{\pm}$ coordinate system  into the gravity region by the coordinate change
\be\label{xyrel}
x^+ = x(y^+) , \quad x^- = x(y^-) \ .
\ee
This can be done for any gluing map, but it is only useful in certain cases (in particular we will find this does not suffice to describe the replica wormholes or finite-$\delta$ shockwaves). In the eternal black hole the reason this is useful is that the gluing map can be analytically continued to a holomorphic function on the Euclidean disk (as we will see in more detail below), and this provides a natural Hartle-Hawking state for the CFT in which $T_{x^\pm x^\pm}$ is given entirely by the conformal anomaly. In this state, 
\be\label{stress_eternal}
T_{y^\pm y^\pm} = - \frac{c}{24\pi}\{ x(y^\pm), y^\pm \} = \frac{\pi c}{12\beta^2} \ .
\ee
This has zero flux, consistent with the Schwarzian equation of motion  \eqref{eq:eternal}.

 Putting the solution (\ref{solueternal}) into (\ref{coordinate}) and (\ref{dilaton2}), we obtain the following physical metric for the inside region and the dilaton profile
\ba
ds_{\text{in}}^2=-\l(\f{2\pi
}{\beta}\r)^2\f{dy^+dy^-}{\sinh^2\f{\pi}{\beta}(y^--y^+)}\, ,\q \phi=\phi_r\f{2\pi 
}{\beta}\f{1}{\tanh\f{\pi}{\beta}(y^--y^+)}\, .
\ea
An $SL(2)$ transformation acting on $x^\pm$ relates different choices for how to embed the $(y^+,y^-)$ coordinates into the Poincare patch. 
Our choice (\ref{xyrel}) is depicted in figure \ref{systemEL}.\footnote{Another popular choice in the literature is $\tilde{x}^{\pm} =  \f{\beta}{\pi}\tanh \frac{\pi y^{\pm}}{\beta} =  \f{\beta}{\pi}\frac{x^{\pm} - 1}{x^\pm + 1}$.
}
The coordinates $(y^+,y^-)$ cover the right wedge of the eternal black hole. The left wedge corresponds to the region $x^\pm<0$ which can be obtained by the analytic continuation $y^\pm\ra y^\pm+i\beta/2$. 
The Euclidean continuation is
\be\label{eucon}
w = 1/x^- , \quad \bar{w} = x^+ \ , 
\ee
where $w$ is a complex coordinate. The gravity region is the interior of the unit disk, with the hyperbolic metric
\be
ds^2_{\rm int} = \frac{4}{(1-|w|^2)^2} dw d\bar{w} \ .
\ee
The hyperbolic disk is glued at the boundary to a flat spacetime with the metric
\be
ds^2_{\rm ext} = \frac{\beta^2}{4\pi^2 \epsilon^2} \frac{dw d\bar{w}}{w \bar{w}} 
\ee
for $|w| >1$.
See figure \ref{systemEL}. The Euclidean continuation of the solution (\ref{solueternal}) is given by
\ba
x(t=-i\tau)=e^{-2\pi i \tau/\beta}\, ,
\ea
which is the map between the Euclidean cylinder and the plane coordinate.
The black hole mass and the Bekenstein-Hawking entropy are given by
\ba
M =-\f{\phi_r 
}{8\pi G_N}\{x(t),t\}=\f{\phi_r
}{4G_N} \frac{\pi}{\beta^2}\, 
\ea
and
\ba\label{S_eternal}
S^{\beta}_{\rm Bek}=S_0+\f{\phi_h}{4G_N}\, , \q {\rm with} \ \phi_h\equiv \phi_r\f{2\pi 
}{\beta}\, ,
\ea
respectively. Here $S_0$ is the extremal entropy given by $S_0=\f{\phi_0}{4G_N}$, and $\phi_h$ represents the dilaton value at the horizon.

\subsection{Evaporating black hole}\label{sec:evap}
We will now add a shockwave to the eternal black hole, leading to an evaporating solution. At times $|t| < L$, we have the eternal black hole at temperature $\beta$ in equilibrium with the bath. The evaporating black hole is created by injecting a shockwave from the flat spacetime region at time $t=L$ along the AdS boundary. To make the solution time-symmetric, we also have a shockwave exiting the AdS region at $t = -L$. For now, these shockwaves will be treated as delta function sources, but below we will describe how they can be obtained by the singular limit of a smooth state with operator insertions in Euclidean signature. Since the solution is time-symmetric, we will focus on $t>0$. 

Injection of the ingoing shockwave increases the temperature of the black hole and also changes the gluing function $x(t)$ as compared to the eternal black hole. While the ingoing stress tensor remains thermal with the original temperature $\beta$ away from the shockwave, the outgoing stress tensor is changed according to the new gluing function:
\bal\label{shockT}
T_{y^+y^+}= \f{\pi c }{12\beta^2}+E_\psi\d (y^+-L)\ ,\q T_{y^-y^-}=-\f{c}{24\pi}\{x(y^-),y^-\}\, ,\q (t>0)
\eal
where $E_\psi$ is the energy of the shockwave.

The Schwarzian equation of motion for $t>0$ is
\ba\label{Schwarzeq_evap}
-\pp_{t}\l(\f{\phi_r
}{8\pi G_N}\{x(t),t\}\r) =E_\psi \d(t-L)+\f{c}{24\pi}\{x(t),t\}+\f{c\pi }{12\beta^2}\, .
\ea
Integrating once gives
\be\label{Sch_exp}
\{x(t), t\} =  - \frac{2\pi^2}{\beta^2} - \frac{24\pi \kappa E_\psi}{c}\Theta(t-L)e^{-\kappa(t-L)} \ .
\ee
We have set the integration constant to agree with the eternal black hole Schwarzian at early times, $\{ e^{2\pi t/\beta}, t\}  = - \frac{2\pi^2}{\beta^2}$, and defined a parameter
\ba
\kappa=\f{c}{24 \pi}\f{8\pi G_N}{\phi_r
}\, ,
\ea
which controls the evaporation rate. This represents a black hole that increases in mass when the shockwave enters, then decays back toward the original mass,
\bal\label{massfunct}
M(t)=-\f{\phi_r
}{8\pi G_N}\{x(t),t\}=\f{\phi_r
}{4G_N} \frac{\pi}{\beta^2}+\Theta(t-L)E_\psi e^{-\kappa (t-L)}\q (t>0)\, .\eal
From the above expression, we can define the temperature of the evaporating black hole as
\ba
T(t)=T\s{1+\Theta(t-L) u_0^2  e^{-\kappa (t-L)}}\, ,
\ea
 where we introduced a dimensionless parameter 
 \ba 
 u_0=\beta \s{\f{12\kappa E_\psi}{c\pi}}\, ,
 \ea
 which controls the increase in temperature of the black hole due to the shockwave with energy $E_\psi$. 
In this paper, we will consider the weak gravity limit 
\ba
\kappa\ll1 \q {\rm and}\q  c\gg 1 \q {\rm with} \q
\kappa E_\psi/c {\rm \  fixed}\, ,
\ea
such that the change in temperature is finite.

The solution $x(t)$ to the equation of motion (\ref{Schwarzeq_evap}) is explicitly given by 
\bal\label{sol_evap}
x(t)=\begin{dcases*} e^{\f{2\pi}{\beta}t}\vphantom{\frac{0}{0}} \qqq\qqq\qqq\qqq\qqq\qqq\q\ \ \q 0<t<L\\
e^{\f{2\pi}{\beta}L}\l[1+\f{2}{u_0}\f{-K_{\nu}(\nu u_0)I_{\nu}( \nu u)+I_{\nu}( \nu u_0)K_{\nu}(\nu u)}{K_{\nu+1}(\nu u_0)I_{\nu}( \nu u)+I_{\nu+1}(\nu u_0) K_{\nu}(\nu u)}\r]\q \q t >L 
\end{dcases*}
\eal
where $u=u_0e^{-\f{\kappa}{2}(t-L)}$ and $\nu=\f{2\pi}{\beta \kappa}$. $I_\n$ and $K_\nu$ are the modified Bessel functions of the first and the second kind respectively. To obtain the solution for $t>L$, we imposed the matching conditions $x(L)=e^{\f{2\pi}{\beta}L}, x'(L)=\f{2\pi}{\beta}e^{\f{2\pi}{\beta}L}$ and $x''(L)=\l(\f{2\pi}{\beta}\r)^2e^{\f{2\pi}{\beta}L}$ at $t=L$. The boundary curve terminates at a finite position
 \bal
  x_\infty&\equiv x(t=\infty) \no&= e^{\f{2\pi}{\beta}L}\l[1+\f{2}{ u_0}\f{I_{\nu}(\nu u_0)}{I_{\nu+1}(\nu u_0) }\r]\approx e^{\f{2\pi}{\beta} L}\l[1+\f{2}{\s{1+u_0^2}-1}\r]\, ,
  \eal
  which indicates that the new horizon sits outside the old horizon of the original black hole before the shockwave as depicted in figure \ref{fig:setup-twosided2}.
For the arguments in the following subsections, we need to know the late-time behavior of the solution $x(t)$. At late times $t\sim \CO(1/\kappa)$, the solution can be approximated as
 \bal\label{Asymptx}
{\rm log}\l(\f{ x_\infty-x(t)}{2(x_\infty-x(L))}\r) 
 &\sim -2\nu(\eta(u_0)-\eta(u))+\CO(e^{-2\nu(\eta(u_0)-\eta(u))})\, ,
 \eal
where we used the the asymptotic behavior of the modified Bessel functions
 \ba
\f{K_{\nu}(\nu u)}{\pi I_{\nu}(\nu u)} \sim e^{-2\nu \eta(u)}\l(1-\f{1}{12\nu}\f{1}{(1+u^2)^{3/2}}(3u^2-2)+\CO(\nu^{-2})\r).
 \ea
 by  fixing $\kappa (t-L)$ and $\kappa E_\psi$ in the $\kappa\ra 0$ limit. Here we introduced a function $\eta(u)$ 
  \ba
  \eta(u)=\s{1+u^2}+{\rm log}\f{u}{1+\s{1+u^2}}\, .
  \ea
The low-temperature limit $\beta \to \infty$ corresponds to $\eta(u) \to u$ and $\nu \to 0$, and in this limit we recover the solution studied in AEMM \cite{Almheiri:2019psf}.
Taking derivatives with respect to $t$, we also have the following expressions
\ba\label{Asymptx2}
\f{x'(t)}{x_\infty-x(t)}\sim 
\f{2\pi}{\beta}\s{1+u^2}=2\pi T(t)\, ,\q \f{x''(t)}{x'(t)}\sim -
\f{2\pi}{\beta}\s{1+u^2}=-2\pi T(t)\, ,
\ea
where we used $\eta'(u)=\f{\s{1+u^2}}{u}$. 
From these expressions, we can check the solution satisfies (\ref{Sch_exp}) at the level of this late-time approximation.
 
 \newcommand{\epsilonuv}{\epsilon_{\rm uv}}
\subsection{Hawking calculation of the entropy}
In the metric $-\Omega^{-2}dz^+ dz^-$, the vacuum entropy of a CFT on an interval $[z_1, z_2]$ is
\be\label{cftvac}
S_{\rm CFT} =  \frac{c}{6}\log \left( \frac{(z_1^+ - z_2^+)(z_1^- - z_2^-) }{\epsilonuv^2 \Omega(z_1)\Omega(z_2)} \right) \ .
\ee
This is the entropy in the vacuum state with respect to the $z$ coordinate. 
To measure the entropy of the Hawking radiation of the evaporating black hole, we will apply this to an interval $R: [\sigma_B, \sigma_{B'}]$ placed in the right flat region at time $t$. The endpoint $\sigma_{B'}$ is an IR regulator which is taken to be large.

In the vicinity of point $B$, as we can see from the stress tensor \eqref{shockT}, the CFT is in the vacuum state with respect to the coordinate
\be
z_B^- = x(y^-_B) , \quad z_B^+ = e^{2\pi y^+/\beta} \ .
\ee
That is, left-movers are in a thermal state at inverse temperature $\beta$, while right-movers have a thermal contribution at the same temperature plus a contribution from Hawking radiation. Near the point $B'$, both left- and right-movers are thermal at inverse temperature $\beta$, so
\be
z_{B'}^- = e^{2\pi y_{B'}^-/\beta} \ , \quad z_{B'}^+ = e^{2\pi y_{B'}^+/\beta} \ .
\ee
The conformal factors in the metric $-dy^+ dy^-$ are $\Omega^2 = z'
(y^+)z'
(y^-)$.
Therefore, applying \eqref{cftvac}, the von Neumann entropy of the radiation in the standard semi-classical approximation is
\be\label{shawkingT}
\Smatter(R) = S_{\rm Hawking}+ \frac{\pi c}{3\beta}(\sigma_{B'} - \sigma_B) + S_{\rm shock} 
\ee
where
\be\label{shawkingB}
S_{\rm Hawking} = \frac{c}{6} \log \frac{e^{-\pi y_B^-/\beta}x(y^-_B)}{\sqrt{x'(y^-_B)}}  + \mbox{const.}
\ee
is the entropy of the Hawking radiation, with $x(t)$ the gluing map obtained in \eqref{sol_evap}. The second term in \eqref{shawkingT} $\frac{\pi c}{3\beta}(\sigma_{B'} - \sigma_B)$ is the thermal entropy of the initial equilibrium state. $S_{\rm shock}$ is here because we are not in the vacuum state -- it is the entanglement of the matter in the ingoing shockwave with the matter in the outgoing shockwave. In a coherent state, this would vanish, $S_{\rm shock}=0$ \cite{Fiola:1994ir}. In a state created by local operator insertions, $S_{\rm shock}$ is non-zero, but assuming a large-$c$ limit (with $h_\psi/c$ held fixed but small), it is a constant that grows only logarithmically with the energy \cite{Asplund:2014coa,Caputa:2015waa}. This is subleading compared to other contributions to the entropy and we will therefore neglect it.

At late  times $t\sim\CO(1/\kappa)$, the entropy of the Hawking radiation increases as
  \ba
    \f{dS_{\rm Hawking}}{dt}\sim \f{\pi c}{6} (T(t)-T)\, ,
    \ea
    and finally asymptotes to a finite value\footnote{Plus terms from `const.' in \eqref{shawkingB}, including UV divergences from the endpoints. These terms cancel in a more careful comparison to the generalized entropy of $R^C$.}
    \ba
      S_{\rm Hawking}(t=\infty)=
      2\f{\phi_h}{4G_N}\l(\s{1+u^2_0}+\log\f{2}{1+\s{1+u_0^2}}-1\r)\, .
    \ea
As a diagnostic for information loss we will compare this to the entropy of the black hole. Actually, this setup is slightly more complicated than an ordinary evaporating black hole, because the Hawking radiation is entangled with both the black hole and the left Minkowski wedge. To allow for this additional entanglement we should compare to the generalized entropy of the complementary region $R^C$, which extends from spatial infinity in the left Minkowski region to the point $B$. That is, we consider it a `paradox' if 
\be
\Smatter(R) > S_{\rm gen}(R^C) \ .
\ee
The generalized entropy of $R^C$ is calculated in a quasi-static approximation, where we sum the gravitational entropy at the two event horizons and the entropy of the quantum fields in the union of the two exterior regions (one extending from left infinity to the left event horizon, and the other extending from the right event horizon to $B$). The important piece of this entropy is the gravitational contributions at the left and right event horizons, yielding
\ba\label{sbhii}
S_{\rm BH} \sim S_{\rm gen}(R^C) \sim 2S_0+  \f{\phi_h}{4G_N}+\f{\phi_h}{4G_N}\s{1+u_0^2e^{-\kappa (t-L)}}\, 
\ea
at late times $t\sim\CO(1/\kappa)$. This asymptotes to twice the entropy of the original black hole entropy before the shockwave.\footnote{We have dropped subleading matter contributions, which are suppressed by $\kappa$, and UV and IR divergences from the matter entropy, which cancel against identical divergences in $\Smatter(R)$. We have done this calculation by the old-fashioned, pre-Ryu-Takayanagi method, using the quasi-static approximation in order to avoid using the island rule in the statement of the paradox. The more accurate semiclassical calculation of the entropy of $R^C$ would actually use the island rule, \textit{i.e.,} the gravitational contributions would be evaluated at the quantum extremal surfaces. The difference between these two calculations is subleading at small $\kappa$, so it is sufficient to use the old-fashioned calculation of $S_{\rm gen}(R^C)$ to state the paradox.}

For large $u_0$, we find $\Smatter(R) > S_{\rm gen}(R^C)$, so the black hole does not have enough entropy to purify the Hawking radiation. This is the information paradox.

\subsection{Island calculation of the entropy}
The island formula \cite{Penington:2019npb,Almheiri:2019psf,Almheiri:2019hni,Almheiri:2020cfm,Penington:2019kki} states that the true von Neumann entropy of region $R$ is given not by Hawking's calculation, but by an extremum of the generalized entropy,
  \ba\label{islandRTdiv}
 S(R)={\rm min}\, {\rm ext}_{I}\l[
 \f{\mbox{Area}(\pp{I})}{4G_N}+S_{\rm QFT}(I\cup R) - S_{\rm div}(\p I) 
 \r]\, .
 \ea
$I$ is called the `island', and the endpoints $ \p I$ are quantum extremal surfaces (QES's) \cite{Engelhardt:2014gca}.  The term $S_{\rm div}$ is a counterterm subtracting the UV divergences in $\Smatter$ coming from the boundary of the island.\footnote{In the introduction we did not write the subtraction explicitly, as it is often left implicit in this formula. It can be viewed as coming from the renormalization of Newton's constant. Thus the $G_N$ in \eqref{islandRT} is bare, while the $G_N$ in \eqref{islandRTdiv} is renormalized.}  In two dimensions, `Area' means the value of the dilaton. At early times $I$ is trivial and this formula leads to the usual Hawking entropy. After the Page time, a non-trivial island appears inside the black hole and the entropy starts to decrease, leading eventually to the same entropy as the original black hole, as required by unitarity.

\begin{figure}
     \begin{center}
 \includegraphics[height=6.5cm]{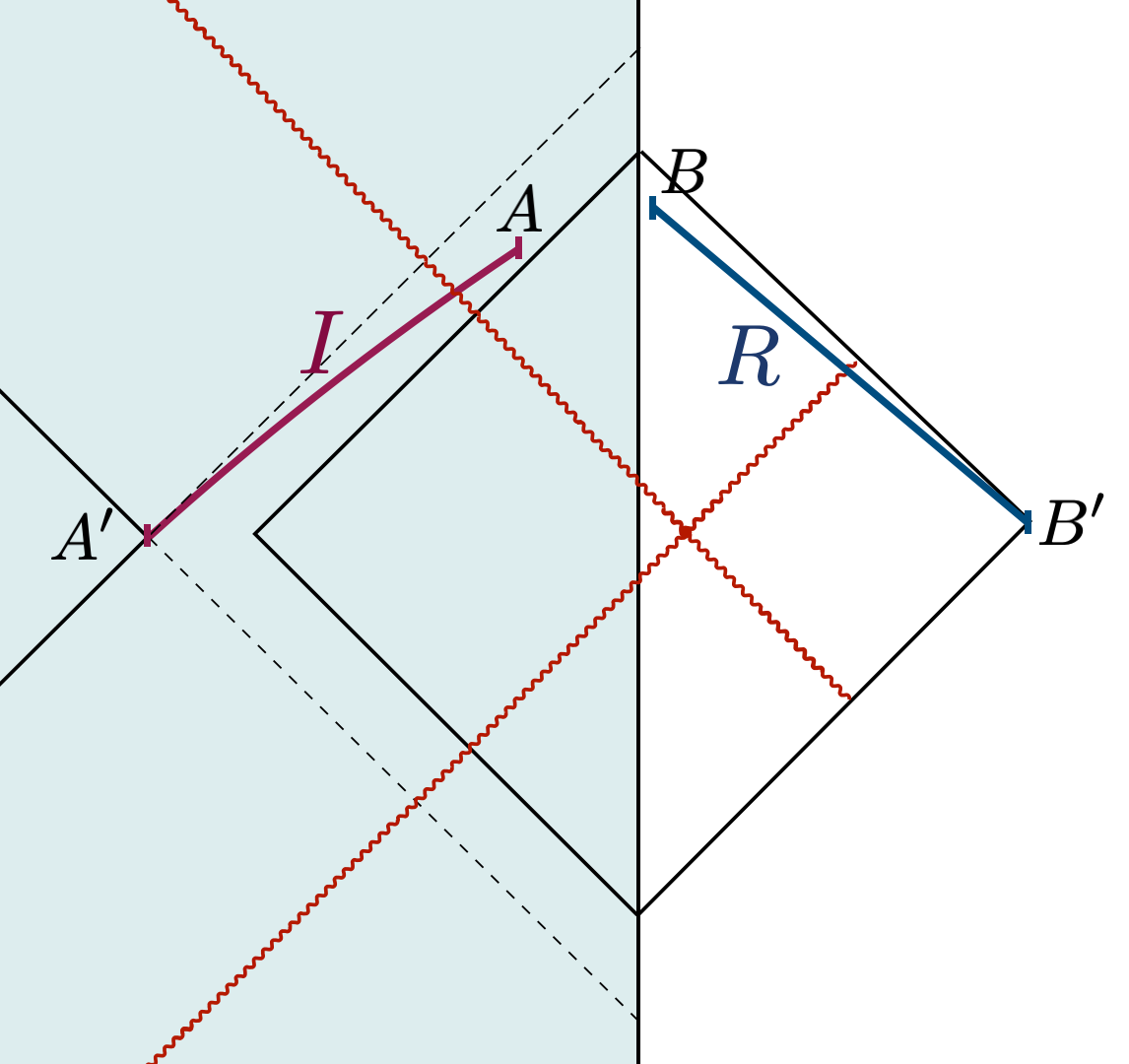}
  \caption{The island $I$ for a single interval placed in the right flat region: $R=[B,B']$ in an eternal black hole. The island has two ends $A$ and $A'$ determined by the extremal conditions. As we will see, the island is placed outside the original horizon of the right black hole, which is depicted as the dashed null lines.}
  \label{conf_island}
 \end{center}
 \end{figure}
  Now let us briefly explain how to compute the entropy of the Hawking radiation using the island formula. We assume the energy of the shockwave is large $u_0\gg 1$.  To find a non-trivial QES, we also assume late enough times $t \sim O(1/\kappa)$ when the the black hole has lost an order one fraction of its mass (see \cite{Almheiri:2019psf}) , but no too lates times  $u_0 e^{-\kappa(t-L)/2} \gg 1$ so that the mass is still much bigger than the original one. We take a single interval $R$ which lies in the right flat region. In this case, the island has two ends; we call the right end $A$ and the left end $A'$ as depicted in figure \ref{conf_island}.    The area terms associated to $A$ and $A'$ can be easily computed from the formula (\ref{dilaton2}) by using the asymptotic expansion of the solution $x(t)$ (\ref{Asymptx}), (\ref{Asymptx2}). For the variations of the matter entanglement entropy $S_{\rm QFT}(I\cup R)$, we assume the factorization 
     \ba\label{factorization}
\pp_{A} S_{\rm QFT}(I\cup R)\simeq \pp_{A}  S_{\rm QFT}([A,B])\, ,\
\pp_{A'} S_{\rm QFT}(I\cup R)\simeq \pp_{A'}  S_{\rm QFT}([A',B'])\, ,
\ea
up to small corrections from the non-factorizable part of the entropy which we can neglect when evaluated at the QES. We will check that \textit{a posteriori} once we have solved for the QES. The extremization for each endpoint of the island is then independent of the other.
We also assume the points $A'$ and $B'$ are both placed before the shockwave, where the background solution is locally equivalent to the black hole at inverse temperature $\beta$.  In the limit $\sg_{B'}\ra\infty$, the entropy $S_{\rm QFT}(A',B')$ is extremized at $\sg_{A'}=-\infty$, i.e., the bifurcation point of the original black hole.\footnote{With $R$ in the right flat region, the island should be outside the original eternal black hole, since $R$ cannot be used to reconstruct information on the left. This follows from the explicit calculation, and is also guaranteed by the quantum normal conditions derived in \cite{Almheiri:2019psf,Hartman:2020khs}. It also follows from the quantum focusing conjecture \cite{Almheiri:2019yqk}: If the left end of the island $A'$ is placed behind the original horizon of the right black hole, one can send a light ray from $A'$ to the left physical boundary where $\phi\sim\phi_r/\ep$. The quantum focusing conjecture, taking into account of $\pp S_{\rm gen}(I \cup R)=0$ at the QES $A'$,  implies that $S_{\rm gen}$ would monotonically decrease toward the left boundary along the light ray. This contradicts with the fact that the dilaton, and so $S_{\rm gen}$ as well, diverge at the left boundary.} Therefore the generalized entropy for the region $[A',B']$ is
 \ba
S_{\rm gen}([A',B'])|_{\rm QES}\simeq S_0+\f{\phi_h}{4G_N}
 + \frac{\pi c}{3\beta}\sigma_{B'} + \frac{c}{6}\log \frac{\beta}{\pi\epsilonuv\ep}\, .
\ea
The term $\f{c\pi}{3\beta}\sg_{B'}$ is the IR divergent thermal contribution, far from the black hole.

Next, we will extremize the right QES, $A$. The calculation of the matter entropy for $[A,B]$ is similar to the calculation for $[B,B']$ around \eqref{shawkingT}, except that now both endpoints are after the shockwave, so in \eqref{cftvac} we use the relations
\be \label{zzbar}
z^- = x^- , \quad
z^+ = v^+ = v(y^+)
\ee
for both endpoints, with $v(y) = e^{2\pi y/\beta}$. The other difference is that because point $A$ is inside AdS, it has the conformal factor $\Omega = \frac{1}{2}(x^- - x^+)\sqrt{ |\frac{dz^+}{dx^+}  \frac{dz^-}{dx^-} |}$. This leads to the generalized entropy 
 \bal
&S_{\rm gen}([A,B])=S_0+\f{\phi(A)}{4G_N}+S_{\rm QFT}([A,B]) \no
&=S_0-\f{\phi_r}{4G_N} \l[\f{2x'(y^+_A)}{x^+_A-x^-_A}-\f{x''(y^+_A)}{x'(y^+_A)}\r]+
\frac{c}{6}{\rm log}\f{2(v^+_B-v^+_A)(x^-_A-x^-_B)}{\epsilonuv\ep(x^-_A-x^+_A)\s{v'(y^+_B)x'(y_B^-)}}\s{\f{x'(y_A^+)}{v'(y^+_A)}} \, .
 \eal
 The derivative of the generalized entropy with respect $A$ can be expressed as
   \bal
\pp_{x^+_A}S_{\rm gen}
 &\sim\f{c}{24}\f{1}{x_\infty-x^+_A}-\f{c}{6}\f{\nu u_A}{(x_\infty-x^+_A)^2}(x_A^--x_\infty)=0,\no
  \pp_{x^-_A}S_{\rm gen}&\sim - \f{c}{12}\f{2\nu u_A}{x_\infty-x_A^+}+\f{c}{6}\f{1}{x_A^--x_B^-}=0\, ,
 \eal
 at leading order for large $u_0$. Here we defined $u_A=u_0 e^{-\f{\kappa}{2}(y_A^+ - L)}$.
 \begin{figure}
   \begin{center}
  \includegraphics[height=6cm]{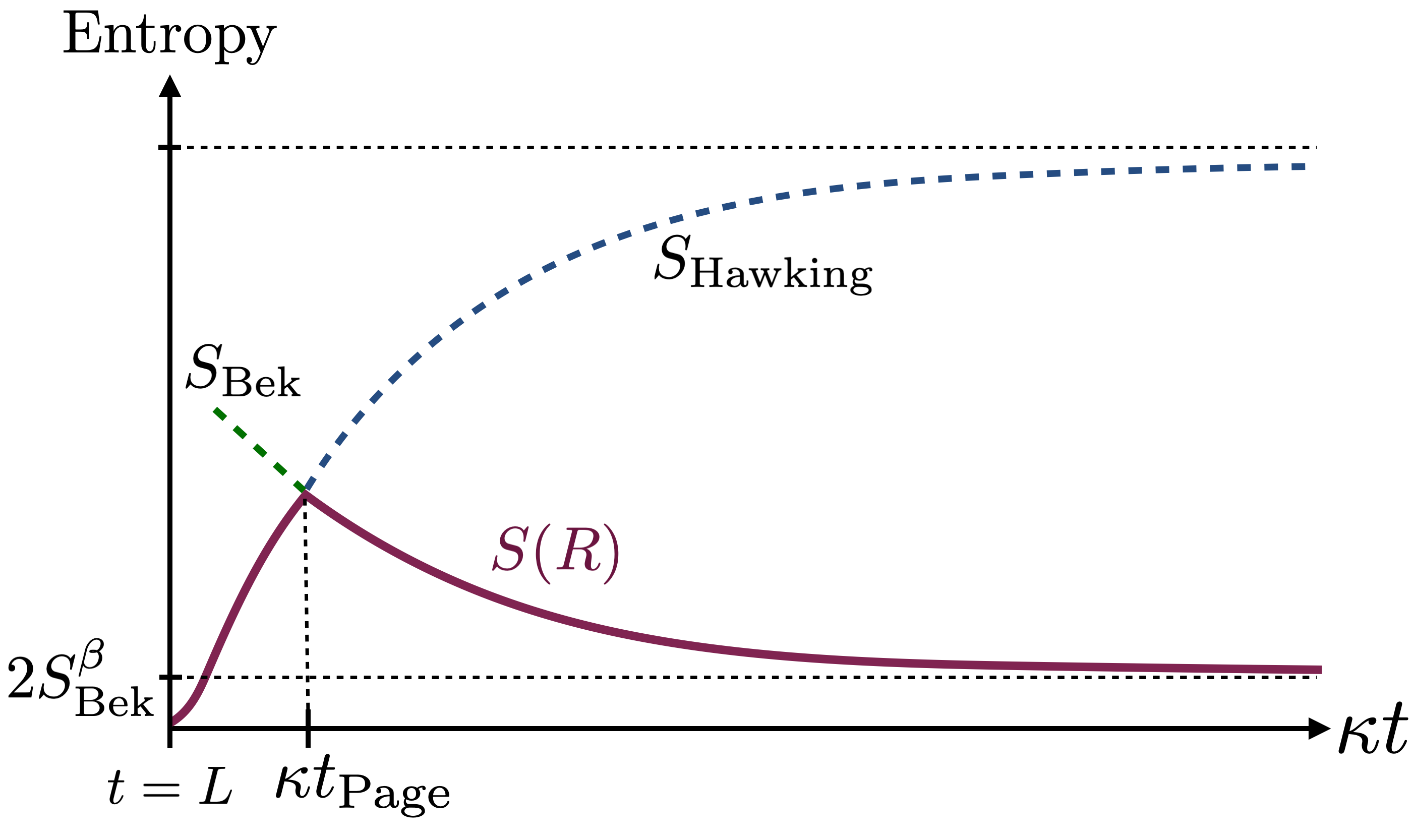}
  \caption{The Page curve is reproduced by the island formula for the entropy of the Hawking radiation.}
  \end{center}
\end{figure}
Thus the location of the right QES satisfies
 \bal \label{locQES2}
x_\infty- x^+_A=\f{4\nu u_A}{3}(x_\infty-x^-_B)\, ,\q
 x^-_A-x_\infty=\f{1}{3}(x_\infty-x^-_B)\, .
 \eal
 As observed in AEMM, by sending the point $B$ to the AdS boundary $y^-_B=t$, the outgoing coordinate of the quantum extremal surface $y^+_A$ is associated to the boundary time $t$ as
 \bal\label{etaABrela}
y^+_A\sim t-\f{1}{2\pi T(t)}{\rm log}\l(\f{16 (S_{\rm Bek}(t)-S_0)}{c}\r)\, ,
 \eal
up to small corrections $\CO(1/u_0)$ and $\CO(\kappa)$. Here $S_{\rm Bek}(t)$ is the Bekenstein-Hawking entropy of the right evaporating black hole $S_{\rm Bek}(t)=S_0+\f{\phi_r}{4G_N}2\pi T(t)\approx S_0+\f{c}{12} \f{ u_0}{\kappa}e^{-\f{\kappa}{2}(t-L)}$.  This time delay can be interpreted as the scrambling time in the thermal system with temperature $T(t)$ and gives a nice geometric realization of the Hayden-Preskill protocol \cite{Hayden:2007cs, Penington:2019npb, Almheiri:2019psf}.

Plugging the solution for the quantum extremal surface back into the entropy, after the Page time we have $S(R)\approx S_{\rm BH}$ where the black hole entropy $S_{\rm BH}$ was given in \eqref{sbhii}. So with the island prescription, the entropy of the radiation obeys the unitary Page curve.

We assumed factorization of the entanglement entropy to determine the QES. Now, we must go back and check that this assumption is self-consistent, given the resulting QES. This analysis is done in appendix \ref{app:matterentropy}. Actually we find that the entropy does \textit{not} quite factorize at the position of the QES, but the additional terms do not affect the extrema at leading order in $\kappa$. That is, any large non-factorized contributions to the entropy drop out when we take the derivatives with respect $A$ or $A'$.  

\section{Shockwave in Euclidean signature}\label{sec:3}

Our goal in this section is to obtain the evaporating black hole, joined to an external bath, from the Euclidean path integral.  As usual, once the quantum state and equations of motion for geometry and matter are constructed in Euclidean signature, real time observables are given by analytic continuation.  We will focus on the state created by a local operator insertion, but it should be straightforward to generalize the methods to arbitrary states created by a Euclidean path integral with operators or sources inserted in the non-gravitating region.

\subsection{Euclidean Setup}

\begin{figure}
\centering 
\includegraphics[scale=0.5]{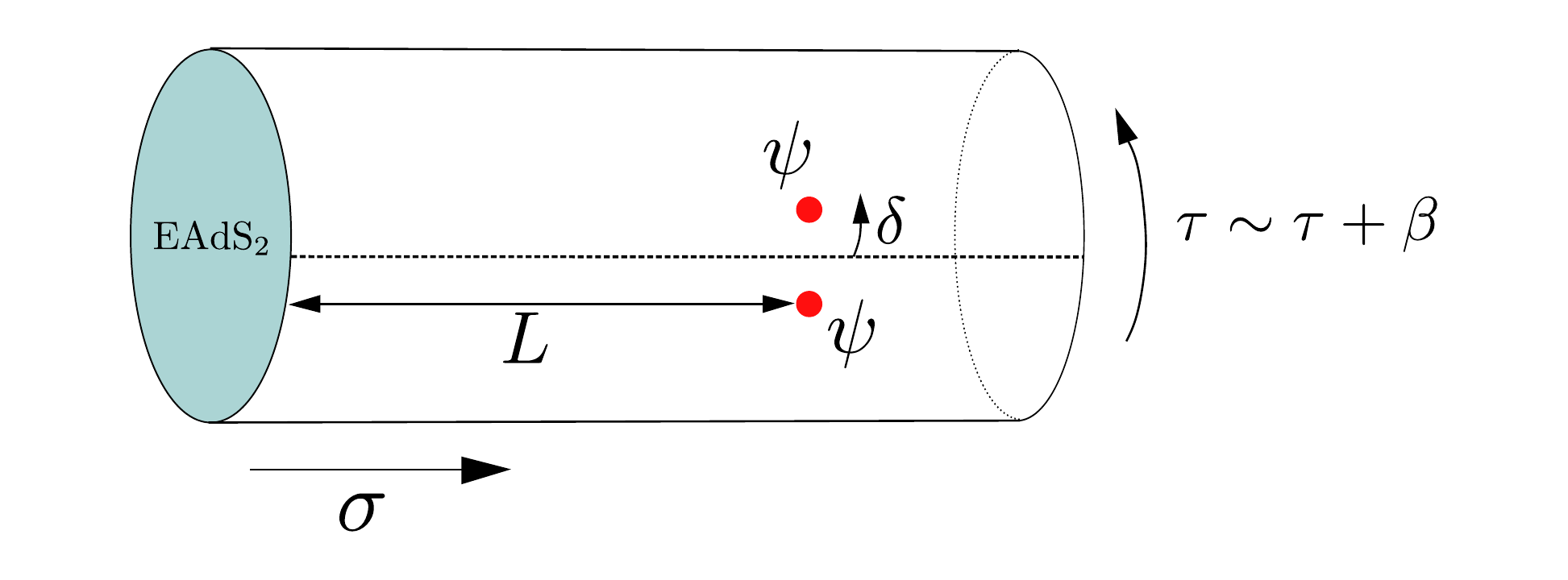}
\caption{The semi-classical realization of the Euclidean path integral which corresponds to the shockwave geometry after analytical continuation to Lorentzian time.   }\label{fig:cyl}
\end{figure}

At finite temperature we prepare the Euclidean path integral by imposing periodic boundary conditions in Euclidean time for quantum fields and gravity. A semi-classical picture of the path integral is shown in figure \ref{fig:cyl}.  We require that matter fields and the metric be continuous at the interface between the two regions.

 In the flat region, we use coordinates $ y= \sigma + i \tau , \bar{y} = \sigma - i \tau$, 
 \ba
ds_{\text{ext}}^2 = \frac{d y d \bar{y}}{\epsilon^2}.
\ea
We set the inverse temperature to $\beta = 2\pi$. 
The shockwave is created by a scalar primary operator $\psi$ inserted in the flat region. In the path integral we have two operator insertions (for the bra and ket),
\bal
\psi (y_1) \psi(y_2),
\eal
at 
\bal
  y_1 = \bar{y}_2= L + i \delta \ .
\eal
The shockwave operator $\psi$  has conformal weights $(h_\psi, h_\psi)$ and scaling dimension  $2 h_\psi$.
 This is similar to the local operator quench \cite{Asplund:2014coa, Caputa:2014eta} studied in the literature of 2d CFT. However, in the present setup gravity is dynamical.

 In the interior of the disk corresponding to the gravity region, the coordinate is denoted by $w$. In JT gravity, the metric is hyperbolic, so for the disk topology we have
 \ba\label{metinwplane}
ds_{\text{int}}^2= \frac{4 dw d\bar{w}}{ (1-|w|^2)^2}.
\ea
 The boundary of the interior region in the Schwarzian limit $\epsilon \to 0$ is denoted by $
w|_{|w|=1} = e^{ i\theta}$.  The dynamics is encoded via the Schwarzian action in a non-trivial diffeomorphism between the boundary of the disk $\theta$ and the time $ \tau$ along the boundary of the flat space region. The equation of motion for the gluing function $\theta(\tau)$ involves the energy flux across the interface, which in turns depends on the manifold and therefore on $\theta(\tau)$.

So far two different coordinates $w$ and $y$ corresponding to the interior and the exterior of the disk were introduced.  Gluing the two regions amounts to finding a coordinate system which covers both regions. The coordinate system is not unique and could be explicitly written for non-holomorphic extensions of the form $(w(y, \bar{y}) ,\bar{w}(y, \bar{y}))$.  However, a non-holomorphic coordinate extension would not allow us to compute the CFT stress tensor. Instead, we look for a holomorphic coordinate which covers both regions.  In practice, this means given $e^{i \theta (\tau)}$, we implicitly define a coordinate $z$ that covers both gravity and flat regions where the metric can be written globally on a plane as $ds^2 = \Omega^{-2} (z,\bar{z}) dz d\bar{z}$. The maps $y \to z$ and $w \to z$ must be holomorphic on their respective domains. Once the metric is known in this form, the stress tensor in the original coordinate system is calculated by the Weyl anomaly.

  Finding the coordinate $z$ given $\theta(\tau)$ is known as the conformal welding problem \cite{Almheiri:2019qdq,Mumford}.  We must find two analytic maps $G ,F$ from the interior and exterior of unit disk respectively, to the $z$ plane such that
\bal
z = &\left\{
\begin{array}{ll}
		G(w)  & \mbox{if } |w| \le 1  \\
		F(v) & \mbox{if } |v| \ge 1 
	\end{array}\label{welding analyticity}
\right. 
\eal
and
\bal
\label{matching cond}&G(e^{ i \theta (\tau)}) = F (e^{ i \tau}),
\eal
where $v= e^{y}$.

\begin{figure}
\centering 
\includegraphics[scale=0.5]{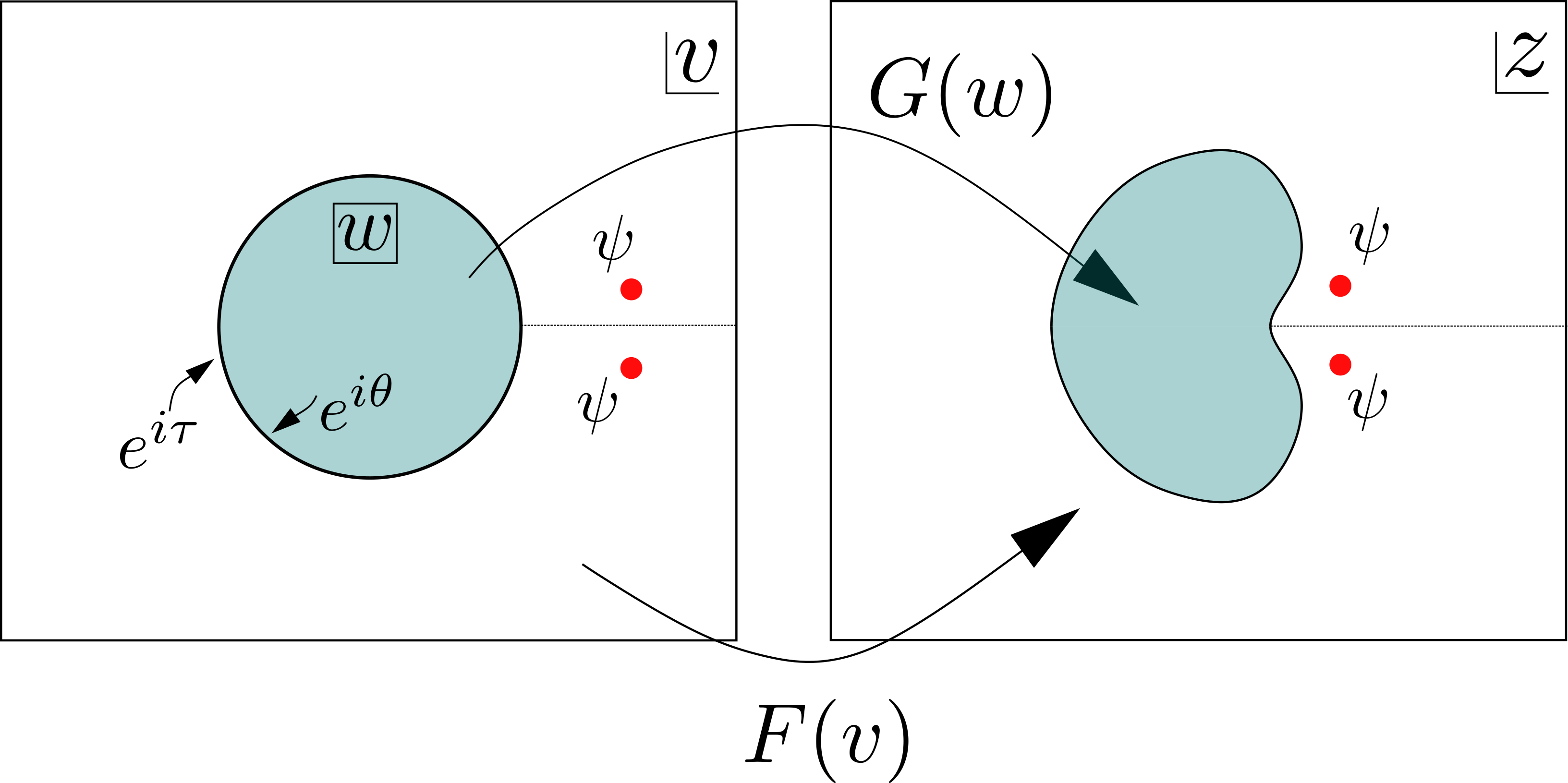}
\label{cyleuc}\caption{The conformal welding problem. For a general diffeomorphism of the unit circle characterized by a real function $\theta(\tau)$, there are holomorphic maps $G, F$ from inside and outside the unit disk, respectively, to the $z$ plane where the two regions join together.  
}
\end{figure}

Here $\theta(\tau)$ is a bijection and in particular  $\theta( \tau+2\pi) = \theta(\tau)+ 2\pi$. Due to the Riemann mapping theorem, the solution for bijective maps $G, F$ always exists and each map is unique up to a PSL$(2,\mathbb{R})$ transformation. However, it is generally not possible to write a solution in closed form.

From the stress tensor on the $z$-plane, the original stress tensor  is determined by transforming back to the $y$-plane. Assuming the functions $F,G$ are known, and using $v=e^{y}$, we have
\bal
T_{yy} (y) &=   \l(\f{dF(e^{y})}{dy}\r)^2   T_{zz} - \frac{c}{24 \pi } \{ F(e^{  y }),y \},
\eal
where $T_{zz}$ is the stress tensor in the $z$-plane with the Weyl-rescaled metric $ds^2=  dz d\bar{z}$.

For the insertion of two operators, the stress tensor on the $z$-plane is determined by conformal invariance and given by 
\bal \label{shockst}
T_{zz}^{\rm shock}(z) =   -\frac{h_\psi}{2\pi}  \frac{(z_1 -z_2)^2}{(z-z_1)^2(z-z_2)^2} \ ,
\eal
 with a similar expression for the anti-holomorphic stress tensor. Here $z_1 =F (e^{y_1}), z_2 = F(e^{y_2})$. 

Finally, using the flux of energy \eqref{shockst} evaluated at the interface, we can write the Schwarzian equation of motion 
for $e^{i \theta (\tau)}$, 
\bal\label{general background}
&\frac{\phi_r}{8\pi G} \pa_{\tau}\{ e^{ i \theta}, \tau \} = i \left. \left( T_{yy} - T_{\bar{y}\bar{y}}  \right) \right|_{y= i \tau } \nonumber \\
& = -\left. i  \left[  \frac{h_\psi}{2\pi}  \frac{ \l(\f{dF(e^{y})}{dy}\r)^2     (F(e^{y_1}) -F(e^{y_2}))^2}{(F(e^{ i \tau})-F(e^{y_1}))^2(F(e^{ i \tau})-F(e^{y_1}))^2}  + \frac{c}{24 \pi } \{ F(e^{  y }),y \} \right]\right|_{y= i \tau} +c.c
\eal
 Note that \eqref{general background} is not simply a differential equation for $\theta(\tau)$, because the welding function $F$  depends non-locally on $\theta(\tau)$. 
  
However, we will see that in the limit of a delta-function localized shockwave, the equations become local and \eqref{general background} can be solved. This is the limit $\delta \to 0$ with 
 \be
 E_\psi = h_\psi / \delta, 
 \ee
 held fixed. In this limit, the stress tensor \eqref{shockst} 
 vanishes away from the insertion points. Therefore the Euclidean gluing map becomes trivial, $\theta (\tau) = \tau$ for real $\tau \in (0, 2\pi)$, and the operators are inserted on the $z$-plane at 
 \be
 z_1 \approx e^L(1 + i\delta) , \quad z_2 \approx e^L(1-i\delta) \ .
 \ee
The stress tensor has delta function contributions supported on the ingoing and outgoing lightcones emanating from the point $y = L$, 
\be
\lim_{\delta \to 0}T_{zz}^{\rm shock} =   E_\psi e^{-L} \delta(z - e^L) \ , \quad
\lim_{\delta \to 0}T_{\bz \bz}^{\rm shock} = E_\psi e^{-L} \delta(\bz - e^L) \ .
 \label{shockdelta}
\ee
 Our strategy will be to setup all of our calculations at finite $\delta$, analytically continue to Lorentzian signature, then take $\delta \to 0$.  Observables in the localized shockwave state are not analytic, but they are analytic at any finite $\delta$, so it is important to do things in this order.

\subsection{The welding solution for small $E_\psi$}\label{subsec:smallE}

As a warmup, let us first consider the case that  $E_\psi$ is small and fixed as $\delta \to 0$. This was treated in appendix B and C of reference \cite{Almheiri:2019qdq} in detail, and here we review that argument.

 We need to solve the welding problem and equation \eqref{general background} to first order in  $E_\psi$.  The holomorphicity conditions on $F, G$ in their respective domains requires them to have the series expansions
\bal
G(w) = \sum_{n=0}^{\infty} g_n w^n, \qquad F(v) = \sum_{m=-\infty}^{2} f_m v^m \ .
\eal 
The $\text{PSL}(2,\mathbb{R})\times \text{PSL}(2,\mathbb{R})$ ambiguity in the welding problem is fixed by setting $f_1=1, g_0= f_2=0$.
  
  The zeroth order solution in $E_\psi$ to the welding problem and \eqref{general background} is given by the trivial solution, with $F$ and $G$ the identity maps,
\bal  
  \theta= \tau, \quad F(v) = v , \quad G(w) = w\ ,\quad z=e^{i \tau}.
\eal
Expanding the matching condition \eqref{matching cond} to first order around the trivial solution $\theta(\tau)  = \tau+ \delta \theta(\tau)$, we find
  \bal\label{weld forder}
  \delta G( e^{i \tau} ) + i e^{i \tau}  \delta \theta(\tau) = \delta F(e^{i \tau}).
  \eal
Note that as a function of $z=e^{i \tau}$, $\delta G$ has a series expansion with only positive powers of $z$, whereas $\delta F$ contains $z^m, m\le 0$. Therefore  $\delta G$ and $\delta F$ are given by
\bal\label{sol wel forder}
\delta F=  i ( e^{ i \tau} \delta \theta)_-, \qquad \delta G = - i ( e^{ i \tau} \delta \theta)_+,
\eal
where $+,-$ indicate positive and negative frequency projections defined with respect to the background welding coordinate, $z = e^{i\tau}$. That is, for any meromorphic function $K = \sum_{n=-\infty}^\infty a_n z^n$, 
\be
K(z) = K(z)_+ + K(z)_- \ , \qquad K(z)_+ = \sum_{n > 0} a_n z^n , \qquad K(z)_- = \sum_{n \leq 0} a_n z^n \ .
\ee
A convenient way to calculate these projections is by the contour integrals
\be
K(z)_{\pm} = \pm \frac{z}{2\pi i} \oint_{|z'|=1} dz' \frac{K(z')}{z'(z'-z)}  \ .
\ee
This expression is valid for $K(z)_+$ when $|z| < 1$ and it is valid for $K(z)_-$ when $|z| > 1$.

Using \eqref{weld forder}, the Schwarzian equation of motion to first order becomes
\bal\label{eom forder}
\pa_\tau  (\delta S_+ + \delta S_-)+ i \kappa (\delta S_+ -  \delta S_-) =  i \f{24 \pi \kappa}{c} \mathcal{F}^{\rm shock},
\eal
where $\delta S= \delta \{ e^{ i\theta}, \tau \}$ and $\mathcal{F}^{\rm shock} = T_{yy}^{\rm shock}(i\tau) - T_{\by \by}^{\rm shock}(-i\tau)$ with 
\bal
T_{yy}^{\rm shock} = \f{h_\psi}{2\pi} \f{\sin^2\delta}{\l(\cos(\delta) - \cosh(L- i \tau)\r)^2}, \qquad  T_{\by \by}^{\rm shock} = \f{h_\psi}{2\pi} \f{\sin^2\delta}{\l(\cos(\delta) - \cosh(L+ i \tau)\r)^2},
\eal
Decomposing the flux into positive and negative modes, \eqref{eom forder} decouples into two separate equations
\bal\label{small delta corr}
\pa_\tau \delta S_{\pm}  \pm i \kappa \delta S_{\pm} =  i \f{ 24\pi \kappa}{c} \mathcal{F}^{\rm shock}_{\pm},
\eal
where the flux from \eqref{shockst} has positive and negative projections
\bal\label{flux Euc}
\mathcal{F}_+^{\rm shock} = T_{yy}^{\rm shock}, \qquad \mathcal{F}_-^{\rm shock} = -T_{\by \by}^{\rm shock}.
\eal
In the limit $\delta \to 0$, the flux vanishes in Euclidean signature, but continuing to Lorentzian with $\tau = i t$ we find
\bal \label{fluxlor1}
&\pa_t \delta S_+ - \kappa \delta S_+ =- \frac{2 4 \pi \kappa}{c}  E_\psi \delta (t+L),  \\
& \label{fluxlor2}\pa_t \delta S_- + \kappa \delta S_- = \f{24\pi \kappa}{c} E_\psi \delta (t- L), 
\eal
where $E_\psi = h_\psi/\delta$. The solution is 
\bal
\delta S_+ = \Theta(-t - L) \f{24\pi \kappa E_\psi}{c} e^{\kappa (t+L)}, \qquad \delta S_- = \Theta(t - L) \f{24 \pi \kappa E_\psi}{c} e^{-\kappa (t-L)},
\eal
where we chose decaying boundary conditions $\delta S \to 0$ as $|t| \to \infty$. 

\begin{figure}[h!]
\centering
\includegraphics[scale=0.3]{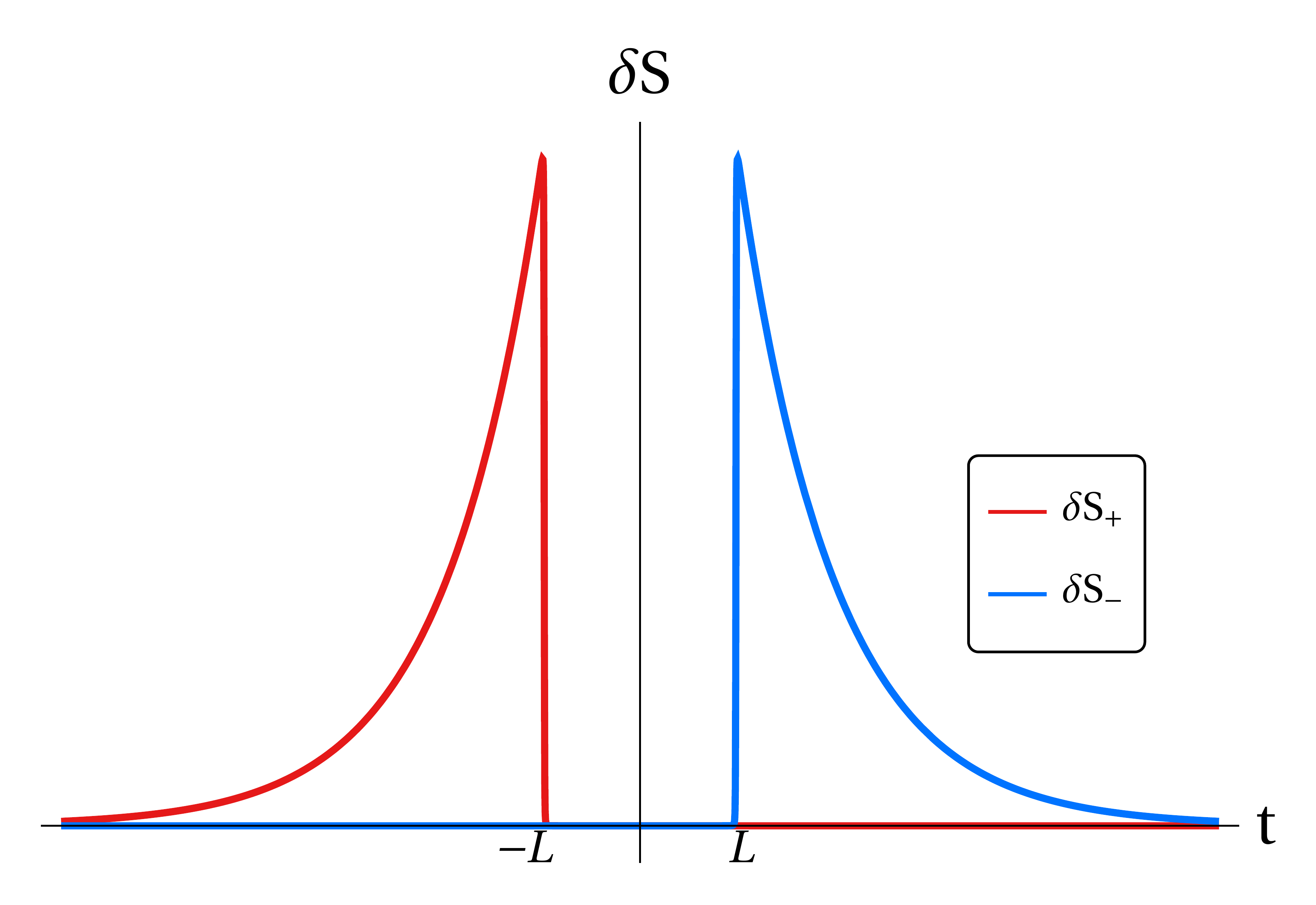}
\caption{Plot of positive and negative Schwarzian $\delta S_+, \delta S_-$ for the shockwave geometry. When $\delta \to0$, there is no overlap between positive and negative modes. The no-mixing condition is crucial to find the full non-linear solution to the welding problem in Lorentzian signature. \label{fig:delSchw}}
\end{figure}

These solutions are plotted in figure \ref{fig:delSchw}. Recall from section \ref{sec2} that $S_+ + S_-$ is proportional to the time-dependent black hole mass, $M(t)$. From the figure we see that $\delta S_-$ describes the formation and decay of a black hole from the infalling shockwave, and $\delta S_+$ is the time-reversed process.

There is no information paradox for a single interval when the geometry is near the eternal black hole solution ($\delta \theta$ small) -- the sharp paradox of AEMM discussed in section \ref{sec2} required the entropy of the black hole created by the shockwave to be large. So we will now turn our attention to non-linear geometries corresponding to fixed $E_\psi$ which is not assumed to be small.

\subsection{Nonlinear solution of welding }
The discussion of section \ref{subsec:smallE} showed that to first  order in $E_\psi$, at small $\delta$, the solution for $\delta S= \delta S_+ +\delta S_-$ consists of only positive or only negative modes at any given $t$. We call this the no-mixing condition: 
\be
[\theta - \tau]_+ [\theta-\tau]_- = 0 \ .
\ee 
At the linearized level of the time-symmetric shockwave, this was a consequence of taking $\delta \to 0$. For finite $\delta$, the positive and negative solutions have overlapping support (see appendix \ref{app:secorderdelta} for a numerical analysis). 

In Euclidean signature, the reality condition for $\theta(\tau)$ makes it impossible to satisfy the no-mixing condition except in the trivial case $\theta = \tau$. However, as evidenced by section \ref{subsec:smallE}, the no-mixing condition has non-trivial solutions in Lorentzian signature. 

For the shockwave configuration, the no-mixing condition is the key to solving the nonlinear welding problem at finite $E_\psi$. For example, if we expand  $\theta(\tau) \approx \tau + \delta^{(1)} \theta+ \delta^{(2)} \theta$, where $\delta^{(1)} \theta ,\delta^{(2)} \theta$ are first order and second order corrections in $E_\psi$, and solve the matching condition \eqref{matching cond} to the second order,  we find
\bal\label{sec weld}
\delta^{(2)} F(e^{ i\tau} ) =& \delta^{(2)}G (e^{ i\tau}) + \delta^{(2)} e^{ i\theta(\tau)} + \delta^{(1)}G'(e^{ i\tau}) \delta^{(1)}e^{ i\theta(\tau)}= \nonumber \\
&\delta^{(2)}G (e^{ i\tau}) + \delta^{(2)}e^{ i\theta(\tau)} - {\left[ \delta^{(1)} e^{ i\theta(\tau)}\right]^\prime}_+ \delta^{(1)}e^{ i\theta(\tau)},
\eal
where equation \eqref{sol wel forder} is used in the first line, and primes denote derivatives with respect to the background coordinate $z=e^{i\tau}$. In order to solve for $\delta^{(2)} F$, we project onto negative modes
\bal\label{sec wel negative}
\delta^{(2)} F= \left[ \delta^{(2)}e^{ i\theta(\tau)} \right]_- - \left [{\left[ \delta^{(1)} e^{ i\theta(\tau)}\right]'}_+ \delta^{(1)}e^{ i\theta(\tau)}\right]_-,
\eal
and similarly for $\delta^{(2)} G$.   If the support of positive and negative solutions have an empty intersection, the second term in equation \eqref{sec wel negative} is zero. 
In the same way, the no-mixing condition can be generalized  to all orders in perturbation theory. For an expansion to  higher orders $\theta(\tau) \approx \tau +\sum_{i=1}^{n} \delta^{(i)} \theta$, terms in \eqref{matching cond} involving   $\left[\delta^{(j)}G^{(k)} \delta^{(i)}\theta\right]_-$, where $\delta^{(j)} G^{(k)}$ is the $k-$th derivative of $\delta G$  at $j$-th order, all vanish.
As a result, the full non-linear solution of welding under the no-mixing condition is
\be\label{nlsol}
\delta F = [\delta e^{i\theta}]_- , \qquad \delta G = -[\delta e^{i\theta}]_+ \ .
\ee
These are identical to the linearized equations \eqref{sol wel forder} but now the variations are nonlinear, $\delta e^{i\theta} \equiv e^{i\theta(\tau)} - e^{i \tau}$.

Here is another perspective. Let us assume the ansatz $[\delta \theta]_+ = 0$ for $t>0$, and $[\delta \theta]_-  = 0$ for $t<0$.  We need to solve the matching condition $G(e^{i\theta}) = F(e^{i\tau})$. The nonlinear solution we have just constructed is simply
\begin{align}
t>0: & \qquad F(e^{i\tau}) = e^{i\theta} , \quad G(e^{i\theta}) = e^{i\theta} \ ,\\
t<0: & \qquad F(e^{i\tau}) = e^{i\tau} , \quad G(e^{i\theta}) = e^{i\tau} \notag \ .
\end{align}
This obviously solves the matching condition, and because of our assumptions on $\delta \theta$ it also obeys the analyticity conditions in \eqref{welding analyticity}. Of course this would not be possible in Euclidean signature because the conditions on $\delta \theta$ would set $\delta \theta = 0$. It is only possible to find a nontrivial solution of this form in Lorentzian signature.

In the Euclidean welding problem, $\theta$ and $\tau$ are real, so the welding functions obey
\be\label{FGreality}
\bar{G}(e^{-i\theta}) = G(e^{i\theta})^* , \qquad
\bar{F}(e^{-i\tau}) = F(e^{i\tau})^* \ .
\ee
In Lorentzian signature, $\tau =i t$, $t \in \mathbb{R}$, and $\theta(it)$ is purely imaginary.  Therefore the solutions \eqref{nlsol} are real for real $t$, and the reality conditions become
\be
\bar{G}(e^{-i\theta}) = G(e^{i\theta}) , \quad \bar{F}(e^t) = F(e^{-t}) \ .
\ee
We can now restate the welding solution for $t>0$ as 
\bal\label{weldingsol}
F^-(e^{t}) = x(t), \qquad  G^{-}(x(t))= x(t),\qquad F^+(e^t)=e^t, \qquad G^+(x(t)) = e^{t}\, ,
\eal
where we introduced the gluing function 
\be
x(t)=e^{-i\theta(it)}
\ee 
which is real for $t\in \mathbb{R}$, and functions $F^+(e^{t})=\bar{F}(e^{t})\, ,\ F^-(e^{t})=1/F(e^{-t})\, ,\ G^+(e^{-i\theta(it)})=\bar{G}(e^{-i\theta(it)})\, ,\ G^-(e^{i\theta(it)})=1/G(e^{i\theta(it)})$ in Lorentzian signature.\footnote{The inversion appearing in these definitions, for example $F^-(e^t) = 1/F(e^{-t})$, is for comparison to section \ref{sec2}. If we continued to Lorentzian signature without this inversion we would find coordinates on the Poincare patch which has $t=0$ at the point of time reflection symmetry. In section \ref{sec2}, we instead used the Poincare patch pictured in figure \ref{systemEL}, and these two patches differ by the null inversion $x^- \to -1/x^-$ (see equation \eqref{eucon}).}
Here $\theta(it)$ has only negative modes when $t>0$ and in particular it does not have to be a small deformation of the eternal black hole solution. Given this ansatz for the welding solution,  the Schwarzian equation of motion in \eqref{general background} becomes
\bal \label{flux-Lor1}
&\pa_t \{ x(t),t \} + \kappa\l( \{ x(t),t \}+\f{1}{2}\r) = - \f{24 \pi  \kappa E_\psi}{c} \delta (t-L), \qquad t>0\\
& \label{flux-Lor2}\pa_t \{ x(t),t \} - \kappa \l( \{ x(t),t \}+\f{1}{2}\r)  =\q \frac{24\pi  \kappa E_\psi}{c} \delta (t+L) , \qquad t<0
\eal
where on the right-hand side we have plugged in the shockwave stress tensor obtained in \eqref{shockdelta}.
These are identical to the equations of motion \eqref{Schwarzeq_evap} obtained by Lorentzian methods.

At this point we have resolved one of the puzzles described in the introduction: The Lorentzian methods of \cite{Maldacena:2016upp,Engelsoy:2016xyb,Almheiri:2019psf} led to relatively simple, local dynamics, whereas the Euclidean approach of \cite{Almheiri:2019qdq} requires non-local solutions of the conformal welding problem. We have shown that these two methods are in perfect agreement. The boundary particle equations of  motion used in \cite{Maldacena:2016upp,Engelsoy:2016xyb,Almheiri:2019psf} correspond to a limit where the no-mixing condition is satisfied, so that the welding problem can be solved exactly and used to write simple Lorentzian equations. In other words, we recover the Lorentzian equation of motion from \cite{Maldacena:2016upp,Engelsoy:2016xyb} when the only outgoing matter is that coming from Hawking radiation, as opposed to the explicit outgoing shockwave. The equations at finite $\delta$ are more complicated because the outgoing shock does not decouple and the equations remain non-local.\footnote{Note that there is nothing inconsistent about having non-local equations of motion for the boundary mode. The theory is still local, since it comes from a local two-dimensional action. 
}

The no-mixing condition is what allowed us to find an exact solution of conformal welding. This condition has a nice interpretation in the $x$-coordinate: It an be satisfied if and only if the stress tensor is chiral, $T_{x^- x^-} = 0$. In the $y$-coordinate, this is equivalent to saying that the only contributions to $T_{y^- y^-}$ come from the Schwarzian, i.e. from Hawking radiation.

For the purposes of studying the black hole background solution, there is no obvious advantage to our Euclidean setup. The advantage comes when we want to apply Euclidean path integral methods, including replica wormholes, to the evaporating black hole, which we will come to in the next section.

\subsection{Schwinger-Keldysh and asymmetric shockwaves}

\begin{figure}[th]
\begin{center}
\includegraphics[scale=0.65]{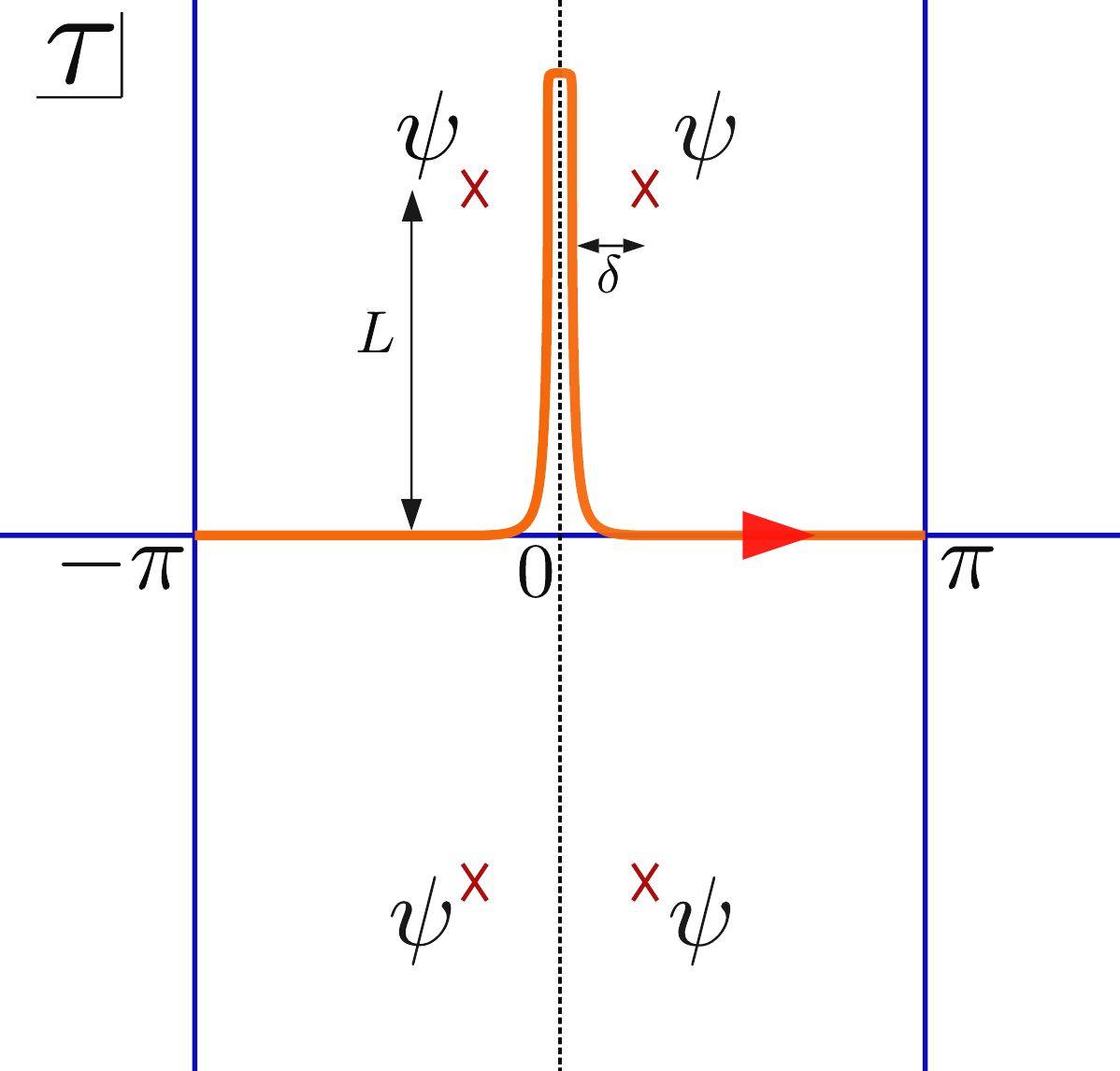}
\end{center}
\caption{Schwinger-Keldysh contour for the shockwave geometry. The gluing function $\theta(\tau)$ is defined along the orange contour. \label{fig:SKbackground}}
\end{figure}

We can view the shockwave geometry as a solution defined on a Schwinger-Keldysh contour\cite{Jana:2020vyx}. The contour in the complex $\tau$ plane is illustrated in figure \ref{fig:SKbackground}. This is the contour on which the gluing function $\theta(\tau)$ is defined. In the Schwarzian equation of motion, there are singularities in the complex $\tau$ plane from the insertions of the $\psi$ operators at $i\tau = y_1, y_2$ and $-i\tau = \by_1,\by_2$. At finite $\delta$, the solution is smooth along the Schwinger-Keldysh contour. As $\delta \to 0$, there is a pinch singularity where the two $\psi$ insertions move into the real-$t$ line, which leads to non-analytic behavior in $\theta(\tau)$ in this limit. 

The solution is obviously symmetric under $t \to -t$. However, by moving the shockwave insertion to ${\cal I}^-$, we can move the outgoing shockwave to past infinity, and the solution at finite $t$ becomes identical to an asymmetric solution with only an ingoing shock.  This simply moves the singularities in the lower-half $\tau$ plane to $\tau = -i\infty$; as $\delta \to 0$, the solution for $t>0$ is unchanged. Therefore we can expect to reproduce all of the physics of the asymmetric shockwave, including the Page curve, by taking this limit of the Euclidean solution.\footnote{The ingoing and outgoing shocks are entangled, so the purely-ingoing shockwave produced in this way is in a mixed state. We are assuming the entanglement entropy of ingoing and outgoing shocks is subleading compared to the entropy of the black hole that forms, as discussed below \eqref{shawkingB}.}

\section{Replica wormhole equations}\label{sec:repworm}

\newcommand{\M}{\mathcal{M}}

In section \ref{sec:3}, we discussed the construction of the shockwave background from the Euclidean path integral. In this section, we write the equations for replica geometries at $n\neq 1$. The equations are implicit due to the non-local welding contributions, but can be simplified somewhat near $n=1$. We set $\beta = 2\pi$. We will not use explicit features of the shockwave solution, so this analysis applies to any state created by a similar Euclidean path integral.

\subsection{Replica geometry setup}
\begin{figure}
   \begin{center}
  \includegraphics[height=6cm]{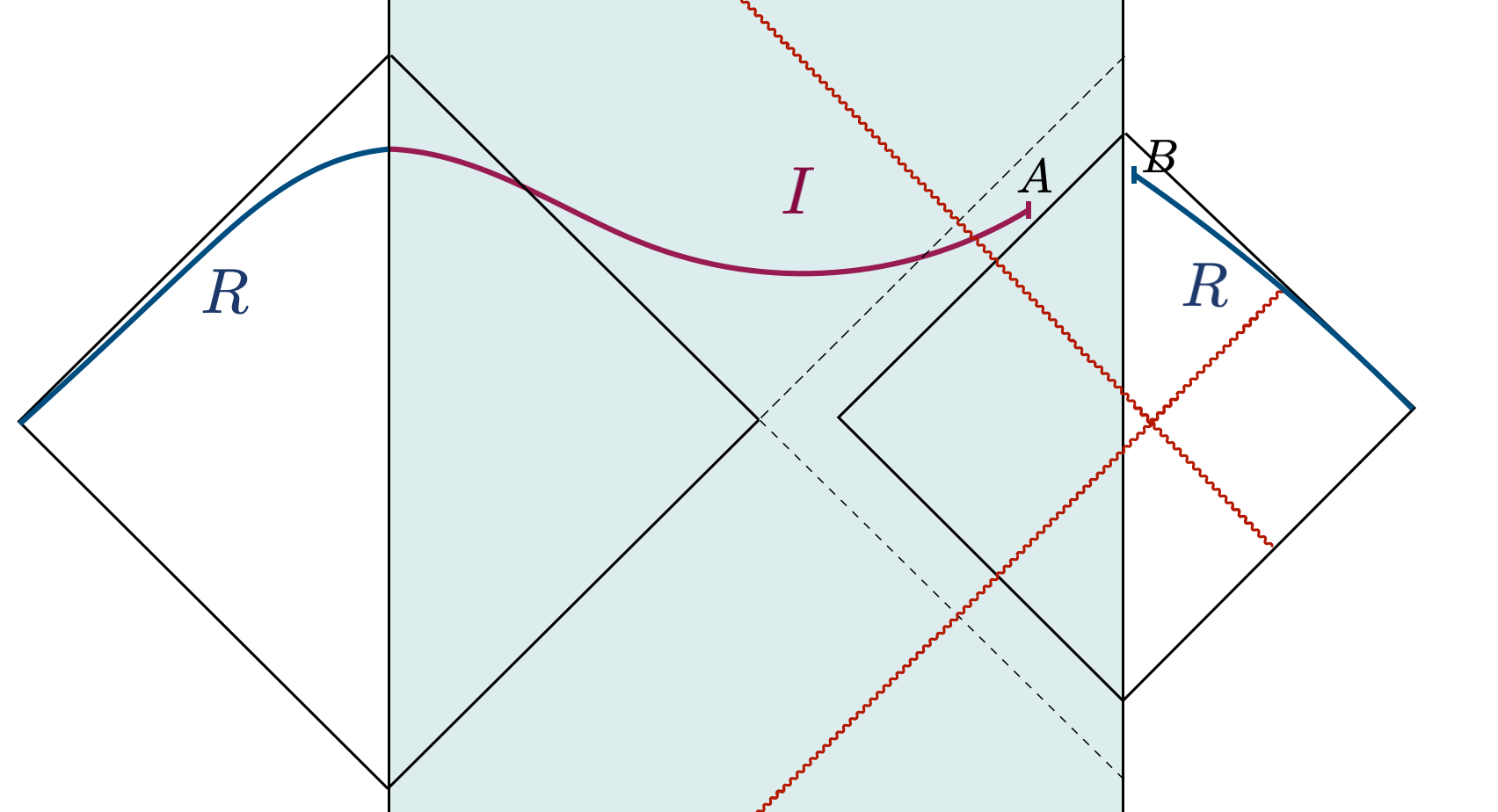}
  \caption{Lorentzian setup discussed in this section. We take $R=[-\infty,0 ]_L\cup [b,\infty]_R$ and the island has a single endpoint in the gravity region.
  \label{fig:Lorentzian_singleinterval}
  }
  \end{center}
\end{figure}
For now, we will focus on a region $R$ which has a single endpoint in the non-gravitating region. That is, on the Euclidean $y$-plane, 
\be
R = [0,b] \ ,
\ee
including the right boundary point at $y=0$. Equivalently we can take the complement region,
\be
R=[-\infty,0 ]_L\cup [b,\infty]_R \ .
\ee
Here $b$ is complex so at the end we can continue to Lorentzian time. 

To compute entropy by the replica method, one needs to first compute the replica partition function $Z_n = \tr (\rho_R)^n$ for $n$ copies of the system. The entanglement entropy is then computed by analytic continuation in $n$,
\be
S = (1- n \p_n) \log \left. Z_n \right|_{n=1} \ .
\ee
In a QFT without gravity, $Z_n$ is given by a path integral on an $n$-sheeted Riemann surface $\widetilde{\M}_n$. The original manifold is the quotient $\M = \widetilde{\M}_n/\mathbb{Z}_n$. The replica partition function can be viewed as a correlation function of twist operators $\langle \Psi| \mathcal{T}_n(y_1) \mathcal{T}_{-n}(y_2)   |\Psi \rangle$, where $|\Psi\rangle$ is a replicated version of the original state, created by inserting the same operators in each copy.  The dimension of the twist field in 2d CFT is $\Delta_{\mathcal{T}} = \frac{c}{12}(n - 1/n)$.

In a theory with dynamical gravity, there are two crucial differences in the replica method. First, the geometry backreacts so that the replica manifold $\widetilde{\M}_n$ solves the gravitational equations of motion. This means that the quotient geometry $\M_n = \widetilde{\M}_n/\mathbb{Z}_n$ now depends on $n$, and it has conical defects at the twist insertions with angular identification $2\pi/n$ (so that $\widetilde{\M}_n$ is smooth). For $n \sim 1$ this leads to the area term in the entropy. The second crucial difference is that higher topologies can contribute to the gravitational path integral. The topologies responsible for the island rule are $\mathbb{Z}_n$-symmetric replica wormholes \cite{Almheiri:2019qdq,Penington:2019kki}. On the quotient manifold, these solutions are realized by the insertion of extra, dynamical twist fields. Any number of dynamical twist fields can appear in the gravity region, so long as the total twist correlator (including the non-dynamical twist fields in the exterior region) is neutral. 

For $n\sim 1$, the dynamical twist fields obey an equation of motion that extremizes the generalized entropy. Thus the twist fields become the endpoints of the island. A general argument for this based on the effective action for twist defects appears in \cite{Lewkowycz:2013nqa, Dong:2017xht,Almheiri:2019qdq, Penington:2019kki}, but in the island context, only the eternal black hole has been treated in detail \cite{Almheiri:2019qdq, Penington:2019kki}. Here we will consider the effects of additional operator insertions in the exterior region in order to see in detail how the replica equations lead to the island rule for the evaporating black hole.

\begin{figure}
   \begin{center}
  \includegraphics[height=6cm]{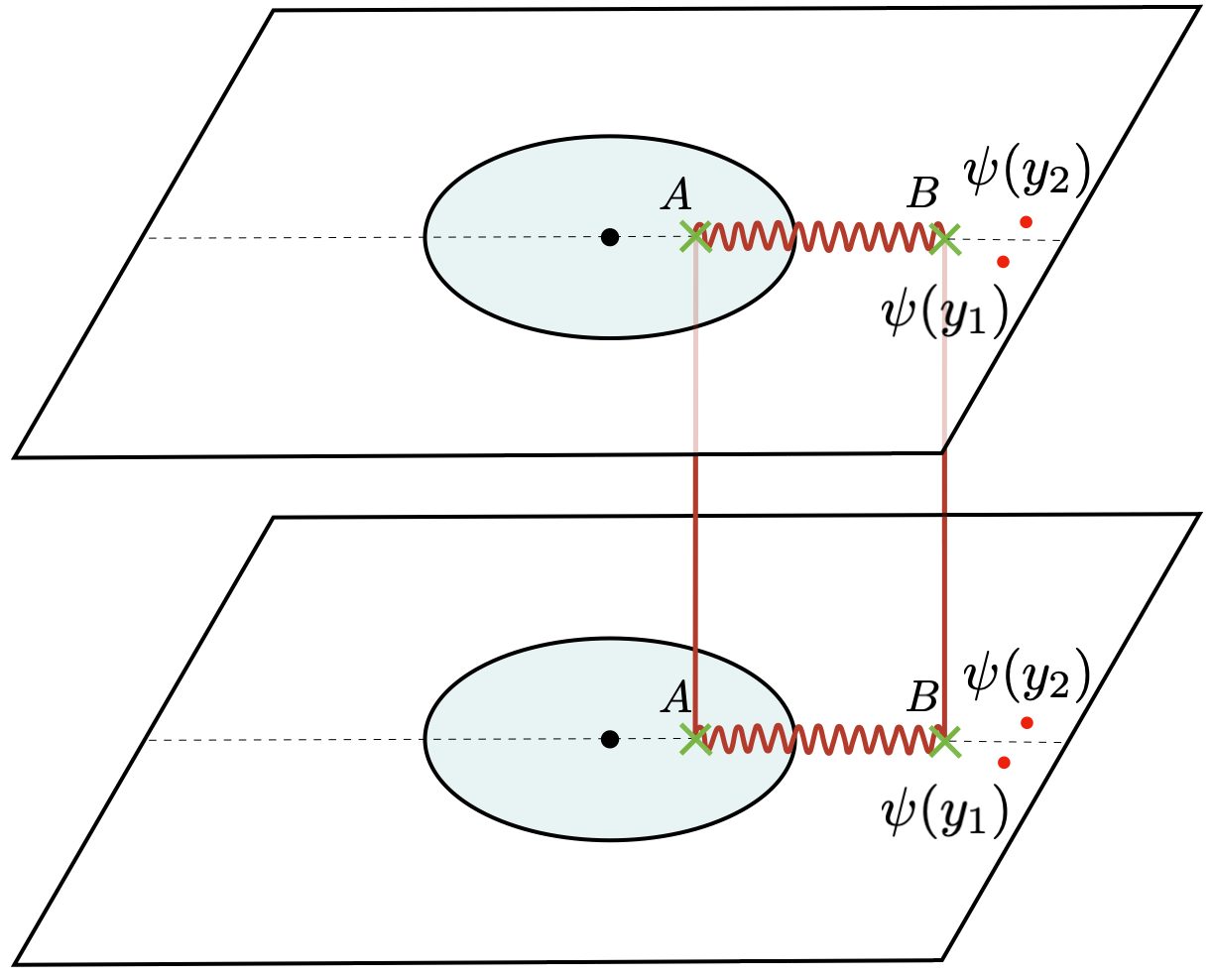}
  \caption{
Replica manifold for $n=2$. The sheets are glued along the cut $[A,B]$.
  \label{fig:replicasetup}
  }
  \end{center}
\end{figure}

For an interval with one endpoint, the important contribution to the path integral comes from a saddlepoint with one dynamical twist field at $y=-a$. See figure \ref{fig:replicasetup} for the replica manifold with $n=2$. The goal is to write the equations of motion for an $n$-fold replica, then take $n \to 1$ to determine the quantum extremal surface, $a$. As in section \ref{sec:3}, we use a coordinate $w$ in the gravity region and $v$ in the flat region. The twist points are at
\be
w = A , \quad v = B = e^{b} \ ,
\ee
and the quantum extremal surface is $A = e^{-a}$ as $n \to 1$.

\subsection{Finite $n$ equation of motion}

 \textbf{Notation:} The background solution is the $n=1$ replica geometry, so in this section the notation $F_1, G_1, \theta_1$, etc., denotes the solutions obtained in section \ref{sec2}. The Euclidean coordinate $\tilde{w}$ is a global coordinate on the interior (gravitating) part of the replica manifold $\widetilde{{\cal M}}_n$, and $w$ is a coordinate on the interior part of the quotient manifold ${\cal M}_n =\widetilde{{\cal M}}_n/ \mathbb{Z}_n$. The function $w_n(\tau)$ is the gluing function, so the gluing in Euclidean signature is $w = w_n(\tau)$. The Euclidean coordinates used here are related to the Lorentzian coordinates of section \ref{sec2} by $w = 1/x^-$, $\bar{w} = x^+$, so the background gluing function is $w_1(\tau) = 1/x_1(\tau)$ with $x_1(\tau)$ the solution found in section \ref{sec2}. The stress tensor is denoted by $T$ for general $n$, and $T^{(1)}$ for the $n=1$ background.

For a single interval, there is a uniformization map from the $n$-fold cover of the unit disk, with a twist field at $w=A= e^{-a}$, to the unit disk itself. The map is an $SU(1,1)$  isometry taking $A$ to the origin, followed by $z \mapsto z^{1/n}$. Denoting the uniformized coordinate by $\tilde{w}$, it is given by
\ba\label{wvstildew}
\tilde{w} = \l(\f{w-A}{1-w \bar{A}} \r)^{1/n} \ .
\ea
In order to solve Einstein's equation, the curvature in the gravity region should satisfy $R=-2$. Therefore, the metric in the unit disk in the covering manifold is  the standard hyperbolic metric
\ba\label{metinfiniten}
ds_{\rm int}^2=\frac{4  d\tilde{w} d\bar{\tilde{w}} }{(1- \tilde{w} \bar{\tilde{w}})^2},
\ea
and the metric in the quotient unit disk with coordinate $w$ is easily found from the coordinate transformation \eqref{wvstildew}.  It is in the $\tilde{w}$ coordinate that the equation of motion is given by \eqref{Schwarzianeq}. That is, if we denote the boundary curve in the $\tilde{w}$ plane by $\tilde{w}_n (\tau)$ and the boundary curve in the $w$ plane by $w_n(\tau)$, the extrinsic curvature in the Schwarzian limit is
\ba
(K-1)/\ep^2= \{ \tilde{w}_n (\tau) , \tau \} = \{ w_n (\tau) ,\tau \} + \frac{1}{2} \l(1-\f{1}{n^2} \r) R(\tau),
\ea
where 
\ba\label{Rvsx}
R_n(\tau)= \f{(1-|A|^2)^2 (w'_n(\tau))^2}{(w_n(\tau) -A)^2 (1-w_n(\tau) \bar{A})^2}.
\ea
The equation of motion that follows from varying the Schwarzian action is therefore
\bal\label{eqforfiniten}
\f{\phi_r}{8\pi G_N}\left[\pa_{\tau} \{w_n, \tau\} + \f{1}{2} \l(1-\f{1}{n^2} \r) \pa_\tau R_n(\tau) \right]= i \l( T_{yy} - T_{\bar{y} \bar{y}} \r)
\eal

In order to have a complete set of equations, we also need to determine the stress tensor. This is done in two steps. First we map to the $z$ plane, using the conformal welding map
\be\label{eomzy1}
z = F_n(e^y) .
\ee
$F_n$ depends implicitly on the gluing function $w_n(\tau)$.
The stress tensor transforms as
\be\label{eomzy2}
T_{yy}(y) = z'(y)^2 T_{zz}(z) - \frac{c}{24\pi}\{z,y\} \ .
\ee
Next we need to find the stress tensor on the $z$-plane. Since $z$ covers the whole plane with a single coordinate system, this can be calculated in principle by the usual CFT methods. There are contributions to $T_{zz}$ from the $\psi$ insertions and from the twist operators. In general these do not decouple, so we cannot calculate $T_{zz}$ exactly in an interacting theory. 
Let us write 
\be\label{tzzsum}
T_{zz}(z) = T_{zz}^{\rm shock}(z) + T_{zz}^{\rm twist}(z)
\ee
where the first terms is the universal contribution given by equation \eqref{shockst}, with the insertion points at 
\be
z_1 = F_n(v=e^{y_1}) , \quad z_2 = F_n(v=e^{y_2}) \ .
\ee
The second term in \eqref{tzzsum} is non-universal --- it encodes the Renyi entropy of the shockwave and therefore depends on the details of the CFT.
Also, the metric in $z$-plane is not flat and the Weyl factor gives an additional contribution to the stress tensor.  The metric in the gravitational region in terms of $z$ coordinate is given by \eqref{metinfiniten},
\bal \label{omegan}
ds_{\text{in}}^2 = \Omega_n^{-2} d|z|^2 =  \f{4}{(1- |\tilde{w}(w)|^2)^2} \f{|\tilde{w}'(w)|^2}{|G_n'(w(z))|^2} |dz|^2,
\eal
  where $\tilde{w}$ is given by \eqref{wvstildew}. Therefore, in this background the stress tensor is related to stress tensor in the flat metric $dz d\bar{z}$ as
\bal  \label{stbackreaction}
  &-2\pi T_{zz}(z) =  -2\pi  T_{zz}^{\rm flat} + \f{c}{6} \f{\pa_z^2 \Omega_n}{\Omega_n}.
  \eal
   In the vicinity of the twist operator at $z_A = G_n(w=A)$ in the flat metric, the stress tensor takes the form
\be\label{stweylvsflat}
-2\pi T_{zz}^{\rm twist,flat}(z) =  \f{c (1-1/n^2)} {24(z-z_A)^2} + \f{(1-n)}{n} \frac{\p_{z_A} S_n^{\rm flat}}{z-z_A} + \cdots
\ee
where  $S_n^{\rm flat}$ is the matter Renyi entropy of region $[A,B]$ in the metric $dz d\bar{z}$. The entanglement entropy for a single interval $[A,B]$ in the flat metric $S^\text{flat}$ is given by $\lim_{n\to 1} S_n^{\text{\rm flat}}= S_{\rm CFT}^{\text{\rm flat}}$.\footnote{
In a large-$c$ CFT, the complete $T_{zz}(z)$ for any $\delta$ has been calculated to leading order in the $1/c$ expansion in \cite{Asplund:2014coa} using the methods of \cite{Hartman:2013mia, Fitzpatrick:2014vua}. Other analytic results are available in rational CFTs \cite{He:2014mwa} and in special kinematic limits \cite{Kusuki:2018wpa}. } More explicitly, by combining \eqref{stbackreaction}, \eqref{stweylvsflat}  for $n\sim 1$, the full stress tensor close to the point $z_A$ is 
\bal \label{wardz}
-2 \pi T_{zz} = -(n-1)  \f{\pa_{z_A} S_{\rm CFT}}{z-z_A} +\cdots,
\eal 
where
\bal
S_{\rm CFT} = S_{\rm CFT}^{\rm flat} - \f{c}{6} \log (\Omega(z_A)), \qquad \Omega(z_A) = \left(\f{1- |A|^2}{2} \right)  \sqrt{G_1'(A) \bar{G}_1'(\bar{A})},
\eal
and $A= G_1^{-1}(z_A)$. Note that the double pole term in \eqref{stweylvsflat} vanishes in \eqref{wardz} due to the non-trivial Weyl factor in \eqref{omegan}. Although we have focused on the shockwave state created by a local $\psi$ insertion, this discussion is general. The only difference in an arbitrary state is that $T_{zz}^{\psi}$ is given by the expectation value in the background state, $T^{\psi}_{zz} = \langle \psi| T_{zz}|\psi\rangle$.

To summarize, the gravitational equation of motion at finite $n$ is determined as follows: Find $T_{zz}$, transform to the $y$-variable using \eqref{eomzy1}-\eqref{eomzy2}, and plug into \eqref{eqforfiniten}. Needless to say, even in situations where $T_{zz}$ is known, this equation is not easy to solve because the welding map depends on $w_{n}(\tau)$. It could perhaps be solved numerically, as was done for the eternal black hole in \cite{Mirbabayi:2020fyk}.

Assuming that $T_{zz}$ can be expressed as an analytic function of $n$, the final equations make sense at non-integer $n$. This is because we assumed replica symmetry and formulated the equations on the quotient manifold $\widetilde{\mathcal{M}}_n/\mathbb{Z}_n$. These are believed to be the dominant contributions in the shockwave state away from the Page transition time, but there are other cases where non-replica-symmetric contributions to the path integral are important \cite{Penington:2019kki, Dong:2020iod, Marolf:2020vsi,Akers:2020pmf}.

\begin{figure}
\begin{center}
\includegraphics[scale=0.6]{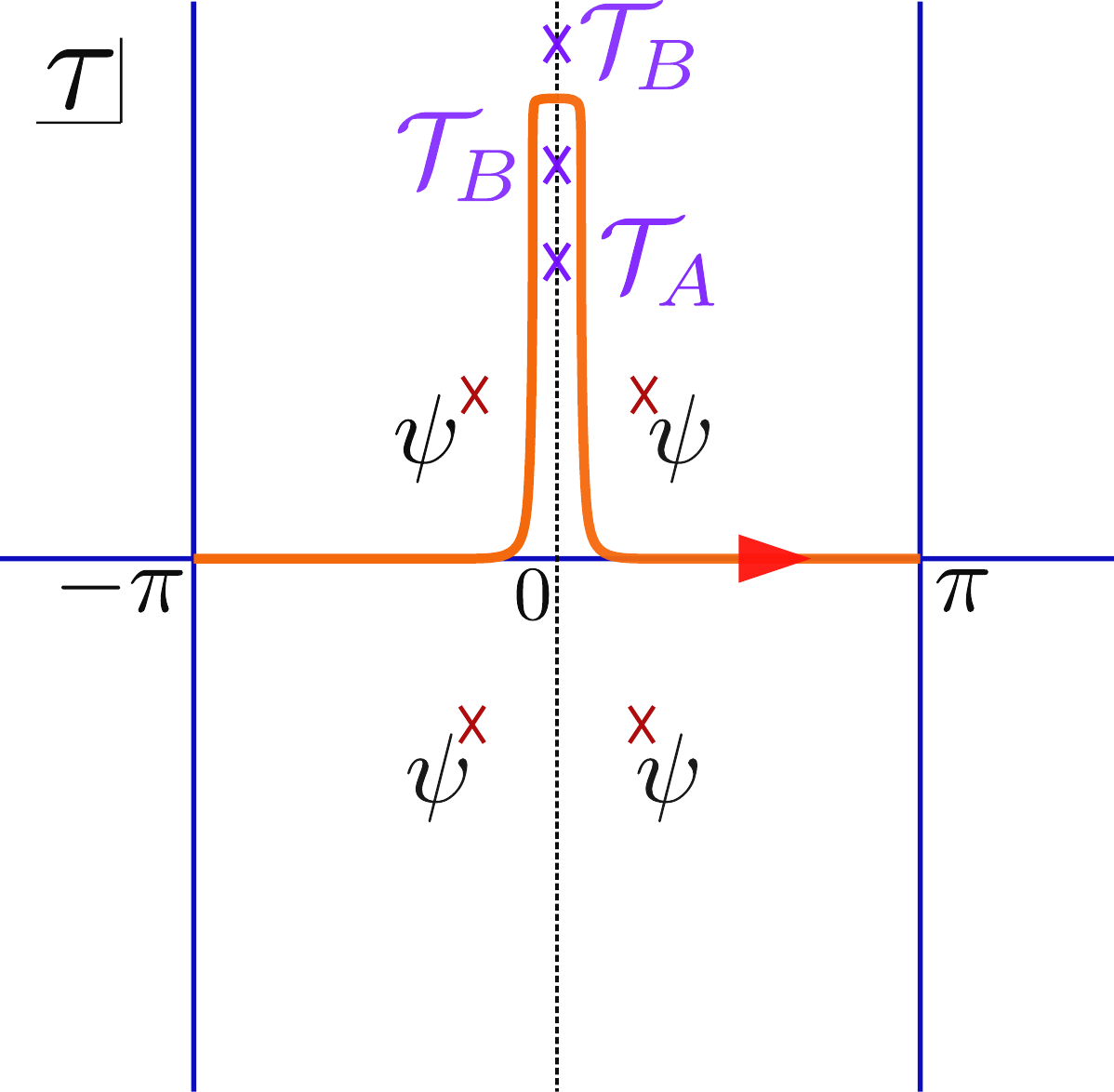}
\end{center}
\caption{Schwinger-Keldysh contour with singularities from the twist operators. Note that each operator produces two singularities in the complex $\tau$ plane corresponding to where the future and past lightcones of the insertion hit the interface. There is only one $\mathcal{T}_A$ because the future lightcone from point $A$ does not hit the interface, see figure \ref{fig:setup-twosided2}.\label{fig:SKtwist} 
}
\end{figure}

For the Lorentzian problem, we must allow $B,\bar{B}$ to be independent, and integrate the equation of motion on the Schwinger-Keldysh contour shown in figure \ref{fig:SKtwist}. This is the same contour as the background geometry in figure \ref{fig:SKbackground} but we have also shown the singular sources at the locations of the twist operators. These will lead to branch cuts in $w_{n}(\tau)$, so that at finite $n$, the gluing function $\theta_n(it)$ has both real and imaginary parts along the real-$t$ part of the contour.

\subsection{Equations for $n\to1$}

The equations simplify for $n\sim1$ so we can be more explicit. The twist field stress tensor and $R_n$ terms have explicit factor of $n-1$ as their coefficients. Hence, to the first order, they are evaluated on the background solution $w_{n=1} =e^{ i \theta(\tau)}$. 
Moreover, given the solution to the welding problem on any background $F_1, G_1$, there is a solution in closed form for the perturbative welding functions\cite{KIRILLOV1998735} $\delta F , \delta G$,
\begin{align}\label{pertweld}
&\delta F(v)= -\f{F_1(v)}{2\pi i} \oint ds \f{F_1'^2(s) \delta w(s)}{F_1(s) (F_1(s)- F_1(v)) w_1'(s) } \nonumber\\
&\delta G(w)= -\f{G_1(w)}{2\pi i} \oint ds \f{G_1'^2(w_1(s)) \delta w(s)}{ G_1(w(s)) \l(G_1(w_1(s))- G_1(w)\r) }
\end{align}
where $w_1 (s) = e^{i \theta(-i \log(s))}$, $\delta w(s) = \delta \l( e^{i \theta(-i \log(s))} \r)$, and  $v=e^{ i \tau}$.  It is easy to check that equation \eqref{pertweld} reproduces \eqref{sol wel forder} when perturbing around the trivial background gluing. 

Putting everything together, with $w_n (\tau) \approx w_1(\tau) + \delta w$ where $\delta w$ is $O(n-1)$, we find to first order 
\begin{align}\label{pertequation}
& \pa_\tau \delta \{ w_n,\tau\} + i \kappa (\delta \{F, v\} - \delta \{ \bar{F},\bar{v} \} )  = - (n-1)\pa_\tau R_1 + \f{24 \pi \kappa}{c} \delta \mathcal{F} , 
\end{align}
where   
\ba \delta {\cal F}=  i  \l. \l( (dz/dy)^2  T_{zz}^{\rm twist}   - (d\bz/d\by)^2 T_{\bz \bz}^{\rm twist}    \r) \r|_{y= i \tau} ,
\ea
and $z$ is the solution of the background welding problem. The right hand side in \eqref{pertequation} could be considered as the source term which depends only on the background solution whereas the the left hand side is only the function of $\delta w$. However, even near $n=1$, the equations for a general background are non-local and they are not reduced to differential equations.

\section{Derivation of QES from replica equations}\label{sec:qesreplica}

In this section we derive the QES condition from the replica equations of motion for $n \sim 1$. We will do this two different ways. First, we will derive the QES directly from the Schwarzian replica equations discussed in section \ref{sec:QESSch}. This was done for eternal black holes in \cite{Almheiri:2019qdq,Penington:2019kki}; here we will do it for a general background, then specialize to the shockwave. The other approach is to to solve the equations of motion locally near the defect, as in \cite{Lewkowycz:2013nqa, Dong:2017xht}. We review this derivation (specialized to JT gravity) in section \ref{sec:localQES}.

To derive the island rule, we also need to show that the entropy derived from the replica method is equal to the generalized entropy associated to the QES. This was demonstrated from the effective action for twist defects in \cite{Almheiri:2019qdq,Penington:2019kki}. In section \ref{sec:QESward} we give a simpler (but ultimately equivalent) argument based on the  Ward identity for a CFT coupled to gravity.

\subsection{QES from the Schwarzian equations} \label{sec:QESSch}
Our starting point is the replica equation of motion for $w_n(\tau)$ as $n \to 1$, given in \eqref{pertequation}. Working around a general background $w_1$, with $w_n = w_1 + \delta w$, the equation is 
\begin{align}\label{eqnearone}
 \p_\tau \{w_n, \tau \} & = - (n-1) \p_\tau R_1(\tau) +i  \f{8\pi G_N}{\phi_r}  (  T_{yy} -  T_{\by \by} ) |_{y = i \tau}  + O((n-1)^2),\\
R_1 (\tau) &=\frac{(1-|A|^2)^2 \l(w_1'(\tau)\r)^2}{(1- \bA w_1(\tau))^2(w_1(\tau)-A)^2} \ . \notag
\end{align}
The flux $i (  T_{yy} -  T_{\by \by} )$ has contributions from the background state, the twist operators, and the Schwarzian of the conformal welding map.

 It is well known that the Schwarzian theory has an $\text{SL}(2, \mathbb{R})$ symmetry. Using this symmetry, the left hand side of \eqref{eqnearone} vanishes if we integrate it against the following $\text{SL}(2)$ generators:
\ba\label{kerkillgeom}
\oint_{\cal C} dw_n \frac{(w_n(\tau))^\alpha}{(w_n'(\tau))^2} \pa_{\tau}\{ w_n ,\tau \}=0,
\ea 
where $\alpha \in \{0,1,2\}$. The equations \eqref{kerkillgeom} follow from the relation $ \f{ \pa_\tau\{ w_n(\tau), \tau \}}{w_n^\prime (\tau)}= \l( \f{1}{w_n^\prime} \l( \f{w_n^{\prime \prime}}{w_n^\prime} \r)^\prime \r)^\prime$ which can be used to show that integrand is a total derivative for $\alpha = 0,1,2$.
The strategy is to show that these identities, applied to the equation of motion, give the extremality condition. In Euclidean signature, the contour ${\cal C}$ of integration is the boundary of a unit disk, while in the Lorentzian setup, the integral is taken over the interface, i.e., over the Schwinger-Keldysh contour depicted in figure \ref{fig:SKtwist}. We will give the argument in Euclidean signature and generalize to Lorentzian at the end. 

Energy conservation relates the stress tensor across the interface as 
\ba\label{energyconservation}
\l. -\l( w_n'(\tau)^2 T_{ w w } -\bar{w}_n'(\tau)^2 T_{\bar{w} \bar{w}} \r)  \r|_{ w = w_n(\tau), \bar{w} = \bar{w}_n(\tau)} =  \l. {T}_{y y} - {T}_{\y \y}  \r|_{y= i\tau}.
\ea
Therefore, integrating  matter flux against $\text{SL}(2)$ generators, we have
\begin{align} \label{genonstrtensor}
&-\oint dw_n \frac{w_n^\alpha}{(w_n'(\tau))^2}  \l( {T}_{y y} - {T}_{\y \y}  \r) =  \l. \oint dw_n w_n^\alpha \l( T_{ww } - \f{\bar{w}_n'(\tau)^2}{w'_n(\tau)^2}T_{\bar{w}\bar{w}} \r)\r|_{ w = w_n(\tau), \bar{w} = \bar{w}_n(\tau)} \nonumber\\
& = \oint dw_n (w_n)^\alpha  T_{ww } - \oint d\bar{w}_n (\bar{w}_n)^{2-\alpha} T_{\bar{w} \bar{w}},
\end{align}
where in the second line we use the relation $\bar{w}_n = 1/w_{n}$ which holds along the interface.
 Since $\alpha \in \{0,1,2\}$, the equation \eqref{genonstrtensor} shows that any source contribution for the holomorphic and anti-holomorphic stress tensors that are analytic inside the disk vanishes when they are integrated against SL(2) kernels. In particular, the conformal welding stress tensors proportional to $\l. \{ G_n(w) , w\} \r|_{w=w_n}$, and $\l. \{ \bar{G}_n(\bar{w}) , \bar{w}\} \r|_{\bar{w}=\bar{w}_n}$, drop out from \eqref{genonstrtensor}. Similarly, matter sources such as operator insertions outside the unit disk do not contribute to \eqref{genonstrtensor}. This matches with the fact that in the local argument for finding the QES, reviewed below, the only important terms in the stress tensor are the residues near the dynamical twist defects.

The QES condition is obtained by integrating an appropriate linear combination of these kernels. One way to guess the correct kernel is as follows. We expect the $R$-term in \eqref{eqnearone} to integrate to give the dilaton term in the QES equation, $\p_A \phi$, because of the relation derived in appendix \ref{app:dilaton}:
\ba\label{dilatonintermR}
-\frac{\phi_r}{2\pi} \int_0^{2\pi} R_1 d\tau = \phi(A,\bA).
\ea
Therefore we look for a kernel that satisfies
\ba \label{kerdilaton}
\f{\phi_r}{2\pi}  \int_0^{2\pi} d\tau K_{A}(\tau) \pa_\tau R_1 = \pa_A \phi(A,\bA).
\ea
Starting with a general linear combination of the integrals for $\alpha= 0,1,2$ we find that this holds for the kernel
\ba \label{kernel}
K_A( \tau) = - \frac{1}{ w_1'(\tau)} \l( r (w_1-A)(1- w_1 \bA) + \frac{w_1 (1-w_1 \bA)}{A (1-|A|^2)}  \r) \ .
\ea
There is a similar kernel $\bar{K}_{\bar{A}}$ obtained by taking conjugates that integrates to $\p_{\bar{A}}\phi$. Here $r$ is an arbitrary complex number.

Another way to determine the correct kernel is starting from the Ward identity. We will try to design a kernel which, upon integrating against the flux $T_{yy} - T_{\by \by}$, produces the entropy term $\p_A S_{\rm CFT}$ in the extremality equation. The Ward identity in the CFT on the replica manifold relates the derivative of the entropy to the stress tensor \cite{Calabrese:2004eu, Calabrese:2009qy}. The relation is 
\newcommand{\btheta}{\bar{\theta}}
\begin{align}\label{Wardderivation}
\p_A S_{CFT} &=  -\p_n|_{n=1} \p_A \log Z \\
\label{eq:ward}&= i \p_n|_{n=1} \oint d\tilde{w} T_{\tilde w \tilde w}\p_A \tilde{w} - i \p_n|_{n=1} \oint d\bar{\tilde{w}} T_{\bar{\tilde{w}} \bar{\tilde{w}}} \p_A \bar{\tilde{w}}  \\
\label{eq:ward2}&= i \p_n|_{n=1} n \oint dw w' T_{ww} \frac{\p_A \tilde{w}}{\tilde{w}'} - i \p_n|_{n=1} n \oint d\bar{w} \bar{w}' T_{\bar{w} \bar{w}} \frac{\p_A \bar{\tilde{w}}}{\bar{\tilde{w}}'} \\
&= i \p_n|_{n=1} n \int_0^{2\pi} d\theta \theta' T_{\theta\theta} \frac{\p_A \tilde{w}}{\tilde{w}'}  - i \p_n|_{n=1} n \int_0^{2\pi} d\btheta \btheta' T_{\btheta\btheta} 
\frac{\p_A \bar{\tilde{w}}}{\bar{\tilde{w}}'}\\
&= - i \int_0^{2\pi}d\tau  \frac{\p_A \tilde{w}}{\tilde{w}'} \p_n|_{n=1}(T_{yy} - T_{\by \by} )  \ .
\end{align}
Here primes are $\tau$-derivatives. The first line is the replica calculation of the entanglement entropy; the second line is the usual Ward identity; in the third line we have gone from the global replica manifold $\tilde{w}$ to the quotient $w$ (there is no Schwarzian because we are not doing a Weyl transformation to remove the conformal factor in the $w$-metric); the fourth line is the coordinate change $w=e^{i\theta}$. In the last line we used the background equation of motion to drop the $(\p_n|_{n=1} n)$ term, and used the fact that $w = 1/\bar{w}$ along the contour of integration to combine the two integrals, and used the conservation of flux, discussed above, to rewrite the stress tensors in the $y$-coordinate.

 Using the explicit coordinate change \eqref{wvstildew}, assuming $|w|=1$, and setting $w = w_1(\tau)$, the kernel in the last line is
\be\label{derw}
\frac{\p_A \tilde{w}}{\tilde{w}'} = \frac{\p_A \bar{ \tilde{w}}}{\bar{\tilde{w}}'} = 
-\frac{1-\bar{A}w_1}{(1-|A|^2)w_1'} 
\ee
This is equal to $K_A$ if we assign $r = \frac{1}{A(|A|^2-1)}$. Therefore we have
\be\label{entropykernel}
\p_A S_{\rm CFT} =-i \int_0^{2\pi}d\tau  K_A \p_n|_{n=1}(T_{yy} - T_{\by \by} ) \ .
\ee
The ambiguous term proportional to $r$ in the kernel corresponds to the Ward identity for rescaling $\delta \tilde{w} \propto \tilde{w}$, and does not affect this integral. The identity \eqref{entropykernel} can also be checked by explicit evaluation of the integral.\footnote{In more detail: The stress tensor $T_{yy}$ has three contributions, discussed around \eqref{wardz}. The first is $T_{zz}^{\rm flat}$, the stress tensor in the metric $dz d\bar{z}$ on the welding plane. The expansion of this stress tensor near the twist point is known \cite{Calabrese:2004eu,Calabrese:2009qy}, and its residue is proportional to $\p_{A} S_{\rm CFT}^{\rm flat}$. The second contribution is from welding; this drops out of the integral \eqref{entropykernel} because it is analytic inside the disk. The third contribution comes from the Weyl factor and accounts for the extra term in $\p_A S_{\rm CFT} = \p_A\left( S_{\rm CFT}^{\rm flat} -\f{c}{6} \log( \Omega(A) ) \right)$.}

Combining everything, we found that integrating the equation \eqref{eqnearone} with kernels $K_{A}$ and $K_{\bA}$ yields the extremality conditions,
\ba
\pa_{A} \l( \f{\phi}{4 G_N} +S_{\rm CFT} \r) = \pa_{\bA} \l(  \f{\phi}{4 G_N}+S_{\rm CFT} \r)=0.
\ea
For the Lorentzian problem where we have independent $A, \bar{A}$ and $B,\bar{B}$, all of the integrals are done over the Schwinger-Keldysh contour shown in figure \ref{fig:SKtwist} rather than the unit disk, and the final equations are the same. For the entropy term, this follows by deforming equation \eqref{entropykernel} into Lorentzian signature by moving the point $(A,\bar{A})$ and simultaneously deforming the contour to prevent any singularities from crossing the contour as we smoothly move $(A,\bar{A})$. The same argument applies to the dilaton term; see appendix \ref{Sec:DerSK} for a more explicit calculation.

\subsection{Local Derivation of the QES}\label{sec:localQES}
We will now apply the methods of \cite{Lewkowycz:2013nqa, Dong:2017xht} to JT gravity to re-derive the QES condition by locally solving the equations of motion near the defect (see also the appendix of \cite{Hartman:2020swn}).  In this approach we directly use the dilaton equation of motion. In conformal gauge,
\ba
ds^2=e^{2\rho(w,\w)}dwd\bar{w}\, ,
\ea
the dilaton equations of motion in JT gravity are \cite{Maldacena:2016upp}
\bal\label{dileom}
&e^{2\rho}\pp_w(e^{-2\rho}\pp_w\phi)=\pp_w^2\phi-2\pp_w\rho\pp_w\phi=-2\pi T_{ww}\, ,\no
&e^{2\rho}\pp_{\w}(e^{-2\rho}\pp_{\w}\phi)=\pp_{\w}^2\phi-2\pp_{\w}\rho\pp_{\w}\phi=-2\pi T_{\w\w}\, ,
\eal
where we set $4G_N=1$. At $n=1$, we can take the metric of a standard hyperbolic disk,
\ba
e^{2\rho}=\f{4}{(1-|w|^2)^2}\, .
\ea
Expanding around $n\sim1$, we have dynamical twist defects inserted in the gravity region. We focus on a single dynamical branch point and determine its position by locally solving the equation of motion. We choose the coordinate $(w,\w)$ so that the branch point $A$ is placed at the origin $w=\w=0$, and solve the equation \eqref{dileom} and its barred version near the origin to  leading order in $n-1$.\footnote{$w$ in this subsection is a local coordinate near the defect. It is shifted and rescaled compared to the $w$ used elsewhere.} The quantities appearing in \eqref{dileom} are expanded around their background values as
\ba
\rho^{(n)}=\rho+(n-1)\d\rho\, , \q  \phi^{(n)}=\phi+(n-1)\d\phi\, ,\q  T_{ww}=T^{(1)}_{ww}+(n-1)\d T_{ww}\, ,
 \ea 
 and the first order equations are
\bal\label{dileom2}
\pp_w^2\d\phi-2\pp_w\d\rho\pp_w\phi-2\pp_w\rho\pp_w\d\phi&=-2\pi \d T_{ww}\no
\pp_{\w}^2\d\phi-2\pp_{\w}\d\rho\pp_{\w}\phi-2\pp_{\w}\rho\pp_{\w}\d\phi&=-2\pi \d T_{\w\w}\, .
\eal
The metric near the branch point on a single sheet of the replica manifold may be locally written as
\ba
ds^2\approx (w\w)^{\f{1-n}{n}}dwd\w\, ,
\ea
since the new coordinate $\widetilde{w}=w^{1/n}$ that covers a neighborhood of the defect in the full manifold gives a smooth metric $ds^2\propto d\widetilde{w}d\overline{\widetilde{w}}$.
 Accordingly, the conformal factor at first order is
\ba\label{drho}
\d\rho\approx -\f{1}{2}{\rm log}w\w\, .
\ea
The dilaton can be generally expanded near the branch point as 
 \bal
 \delta \phi\approx &\, a_{00}+a_{10} w+a_{01} \bar{w}+a_{11} w \bar{w}+a_{20} w^{2}+a_{02} \bar{w}^{2}+\cdots \\ & +\log (w \bar{w})\left(b_{00}+b_{10} w+b_{01} \bar{w}+b_{11} w \bar{w}+b_{20} w^{2}+b_{02} \bar{w}^{2}+\cdots\right)\, . \notag
 \eal
We can relate the residue of the  stress tensor $\d T_{ww}$ to the derivative of the entanglement entropy as
\bal\label{Ward2}
-2\pi \d T_{ww}&\approx-\f{\pp_w S_{\rm CFT}}{w},
\eal
which was derived in \eqref{wardz}.
Now plug these into the equation \eqref{dileom2} and solve the singular terms as $w,\w\ra 0$. The absence of the quadratic singularity fixes $b_{00}=0$ and the simple pole terms fixes $b_{10}=b_{01}=0$. Then the  the equation \eqref{dileom2} becomes
\bal\label{dTsing}
-2\pi \d T_{ww}\approx\f{\pp_w\phi}{w}\, ,\q
-2\pi \d T_{\w\w}\approx\f{\pp_{\w}\phi}{\w}\, .
\eal 
Comparing the residues of the simple poles in \eqref{Ward2} and \eqref{dTsing}, we obtain the conditions for the QES
 \ba\label{eqQES}
\pp_A\l(\frac{\phi}{4G_N}+S_{\rm CFT}\r)=\pp_{\bar{A}}\l(\frac{\phi}{4G_N}+S_{\rm CFT}\r)=0\, ,
\ea
where we have restored $G_N = 1/4$.

Let us make a brief comment on the relation with the arguments in the previous subsection using the contour integral.  The derivative of the entanglement entropy can be expressed by a contour integral derived from the Ward identity as \eqref{eq:ward}. By using \eqref{derw}, we can rewrite it as
\ba
\p_A S_{CFT} = i \oint dw \delta T_{ww}  +i  \oint d\bar{w} \bar{w}^2 \delta T_{\bar{w} \bar{w}},
\ea
in our gauge: $A=\bar{A}=0$. Now we plug the relation \eqref{dTsing} into this integral and pick the residue at $A$, then we obtain the condition for the QES \eqref{eqQES}.

The Ward identity used here is that for CFT in a fixed spacetime. The QES equation \eqref{eqQES} itself can be viewed as the full Ward identity in the CFT coupled to gravity; the vanishing of the derivatives is the statement of diffeomorphism invaraince. This is similar to the point of view in \cite{Dong:2017xht}. We will discuss the full gravitational Ward identity further in the next subsection.

\subsection{Island entropy from the gravitational Ward identity}\label{sec:QESward}
So far, we have derived the extremality condition, but we have not yet derived the island formula. The last step is to calculate the entropy from the on-shell action and verify \eqref{islandRT}. The entropy is obtained from the replica partition function by
\begin{align}\label{sfromz}
S &= -\left.\partial_n  \frac{Z_n}{(Z_1)^n} \right|_{n=1} =  -\left. \p_n \log \frac{Z_n}{(Z_1)^n } \right|_{n=1} \ .
\end{align}
This can be calculated directly from the action of the defect, and the result \cite{Almheiri:2019qdq,Penington:2019kki} is the island formula \eqref{islandRTdiv} . Here we will give a different, simpler derivation from the Ward identity. The point is that once we have derived the extremality condition from the equations of motion, the entropy is automatically correct as well.

Let us first recall how the Ward identity is used to calculate the entropy in a CFT without gravity \cite{Calabrese:2004eu,Calabrese:2009qy}. Consider the entropy of an interval with endpoints $(A,\bar{A})$ and $(B,\bar{B})$. This obeys the Ward identities\footnote{This version of the Ward identity looks different from \eqref{Wardderivation}, but actually they agree. To see this we note that the kernel $\frac{1-\bar{A}w}{1-|A|^2}$ vanishes at $w = 1/\bar{A}$ and is equal to one at $w = A$, so the effect of the kernel on the last line of \eqref{Wardderivation} is to remove the anti-holomorphic contribution to the integral. This argument also uses the fact that once the conformal factors are included, the stress tensor has no double pole near the defect, as in \eqref{wardz}.}
\begin{align}
\p_A S_{\rm CFT}(A,B) &= -\p_n \p_A \log Z_n |_{n=1} 
= -i\p_n \oint_A dw T_{ww}(w) |_{n=1} \\
\p_{\bar{A}} S_{\rm CFT}(A,B) &= -\p_n \p_{\bar{A}}  \log Z_n |_{n=1} 
= - i\p_n \oint_{\bar{A}} d\bar{w}  T_{\w\w}(\bar{w}) |_{n=1}\ , 
\end{align}
and similarly for $\p_B$ and $\p_{\bar{B}}$. We used these relations above in writing the stress tensor near the defects in equations \eqref{eq:ward} The Ward identities can be integrated to find $S$. For example, for a CFT (not coupled to gravity) in vacuum on the flat $dw d\bar{w}$ plane, $T_{ww} = -\frac{c}{24\pi}\{ \left( \frac{w-A}{B-w} \right)^{1/n}, w\}=-\f{c}{48\pi}\l(1-\f{1}{n^2}\r)\f{(A-B)^2}{(w-A)^2(w-B)^2}$, which gives $\p_A S = \frac{c}{6(A-B)}$. This integrates to the well known formula $S = \frac{c}{6}\log|A-B|^2 + \mbox{const}$ \cite{Calabrese:2004eu, Calabrese:2009qy}.

In a CFT coupled to gravity, $S$ is only a function of the point $(B,\bar{B})$ in the asymptotic region, since $(A,\bar{A})$ is determined dynamically by the equations of motion. So there is no Ward identity for $\p_A, \p_{\bar{A}}$, but the relations for $\p_B, \p_{\bar{B}}$ are unchanged. Thus
\begin{align}
\p_B S = -i \p_n \oint_B dv T_{vv}(v)|_{n=1}  
\end{align}
and similarly for $\p_{\bar{B}}$.
The matter stress tensor on the replica manifold is the same in gravity+CFT as it is in CFT, up to $O((n-1)^2)$. Therefore the right-hand side can be evaluated by the CFT Ward identity, and we have
\begin{align}\label{ssward}
\p_B S(B, \bar{B}) &= \left.\p_B S_{\rm CFT}(A,\bar{A}, B, \bar{B})\right|_{\rm QES} \\
\p_{\bar{B}} S(B, \bar{B}) &= \left.\p_{\bar{B}} S_{\rm CFT}(A,\bar{A}, B, \bar{B})\right|_{\rm QES} \notag
\end{align}
These equations are evaluated at the QES after taking the derivatives, $A = A(B,\bar{B}), \bar{A} = \bar{A}(B,\bar{B})$.
These are the Ward identities in gravity+CFT. The only difference from the CFT calculation is that now $A$ and $\bar{A}$ are not independent variables --- they are functions of $(B,\bar{B})$ determined by the QES condition. Note that for CFT, we had four Ward identities associated to $A, \bar{A}, B, \bar{B}$. For gravity we also have four Ward identities, two are given by \eqref{ssward} and the other two are the QES conditions.

The last step is to integrate this equation. It is easy to check that the solution is the generalized entropy at the extremum, as in the island rule:
\be
S = S_{\rm CFT}+\f{\phi}{4G_N}  \  .
\ee
To check this we simply take the derivative
\begin{align}
\p_B \left(\f{\phi}{4G_N}+ S_{\rm CFT} \right) &= 
\l(\frac{\p A}{\p B} \p_A  + \frac{\p\bar{A}}{\p B}\p_{\bar{A}}\r)\l(\f{\phi}{4G_N} + S_{\rm CFT}\r) + 
\p_B S_{\rm CFT}\\
&= \p_B S_{\rm CFT} \ , \notag
\end{align}
as needed.
The other terms dropped out by the extremality condition. (This argument actually fixes $S$ only up to an overall constant, which requires a direct evaluation of the action in one example).

This derivation can be illustrated by a simple example: the QES in empty AdS$_2$ at zero temperature \cite{Almheiri:2019yqk,Almheiri:2019qdq}. We refer to \cite[Section 4.1]{Almheiri:2019qdq} for details of the island in this example. For the interval $ y \in [-a,b]$, with $a,b>0$ so the left endpoint is in the gravity region and the right endpoint is outside, the generalized entropy is 
\be\label{sgenex}
S_{\rm gen}(a,b) =\f{\phi_r}{4G_N} \frac{1}{a} + \frac{c}{6} \log \frac{ (a+b)^2}{a}  + \mbox{const.} 
\ee
We are working at $t=0$, so $a=\bar{a}$, $b = \bar{b}$. The extremality condition $\p_a S_{\rm gen} = 0$ gives the QES as
\be
a(b) =   \frac{1}{4\kappa}(1 +2 b\kappa + \sqrt{1+  12b\kappa +4 b^2 \kappa^2} ) \ .
\ee
Now there are two ways to calculate the entropy. The first is to apply the island formula, i.e., plug this value of $a = a(b)$ into \eqref{sgenex}, 
\be\label{sisland1}
S(b) = S_{\rm gen}(a(b), b) \ .
\ee
The second is to integrate the Ward identity. We want to check that this gives the same answer. The Ward identity for gravity+CFT is \eqref{ssward}, which in this example states
\begin{align}
\p_b S(b) &=\left. \p_b S_{\rm CFT}\right|_{a=a(b)}\\
&= \frac{c}{3(b + a(b))} \notag
\end{align}
with $S_{\rm CFT} = \frac{c}{6}  \log \frac{ (a+b)^2}{a}$. 
Integrating this equation gives \eqref{sisland1} up to an integration constant, which can be fixed from the $b\to 0$ limit.

 \section{Factorization of the two-interval solution} \label{sec:fac}

So far we have discussed the setting of evaporating black holes. There is also an information paradox for eternal black holes \cite{Almheiri:2019yqk}, and replica wormholes can be applied in this context \cite{Almheiri:2019qdq}. This setup is simpler than an evaporating black hole because it has no operator insertions in the definition of the state, but it has an extra complication: to see the information paradox, one must consider the entropy of two disjoint intervals, with one on each side of the black hole \cite{Almheiri:2019yqk}. Otherwise the entropy is time-independent and there is no paradox.

It is reasonable to expect that at late times, the replica wormhole for this setup factorizes into two independent wormholes for the left and right regions, since this is a regime where the twist operators are in an OPE limit. This factorization property was advocated on physical grounds in \cite[Section 5]{Almheiri:2019qdq} in order to make contact with the information paradox. In this section we will confirm factorization of the wormhole geometry explicitly for $n \sim 1$ by explaining how to patch together two separate solutions of the Schwarzian equation, and recover the expected QES's.
 
When the replica manifold is branched along a single interval, with only one branch point in the gravity region, the topology of the gravity region in the replica manifold is unchanged --- it is a disk for all $n$. By contrast when there are two branch points in the gravity region, the replica manifold has wormholes with higher topology. This introduces moduli, and therefore new equations of motion, beyond the Schwarzian equation. We will see however that in the factorized limit these extra equations are not necessary. We will start with the one-sided problem, then describe how at late times two copies of the solution can be patched together.
 
 \subsection{  Single interval geometry in the eternal black hole}

All equations discussed in sections \ref{sec:repworm} - \ref{sec:qesreplica} for the one-interval case also apply to the situation with no shockwave operator insertions. This is the eternal black hole. The geometry is given by $w_1(t) =e^{ -t}$ (we set $\beta= 2\pi$), $F=G= \text{id}$. 

\begin{figure}
\centering
\includegraphics[scale=0.74]{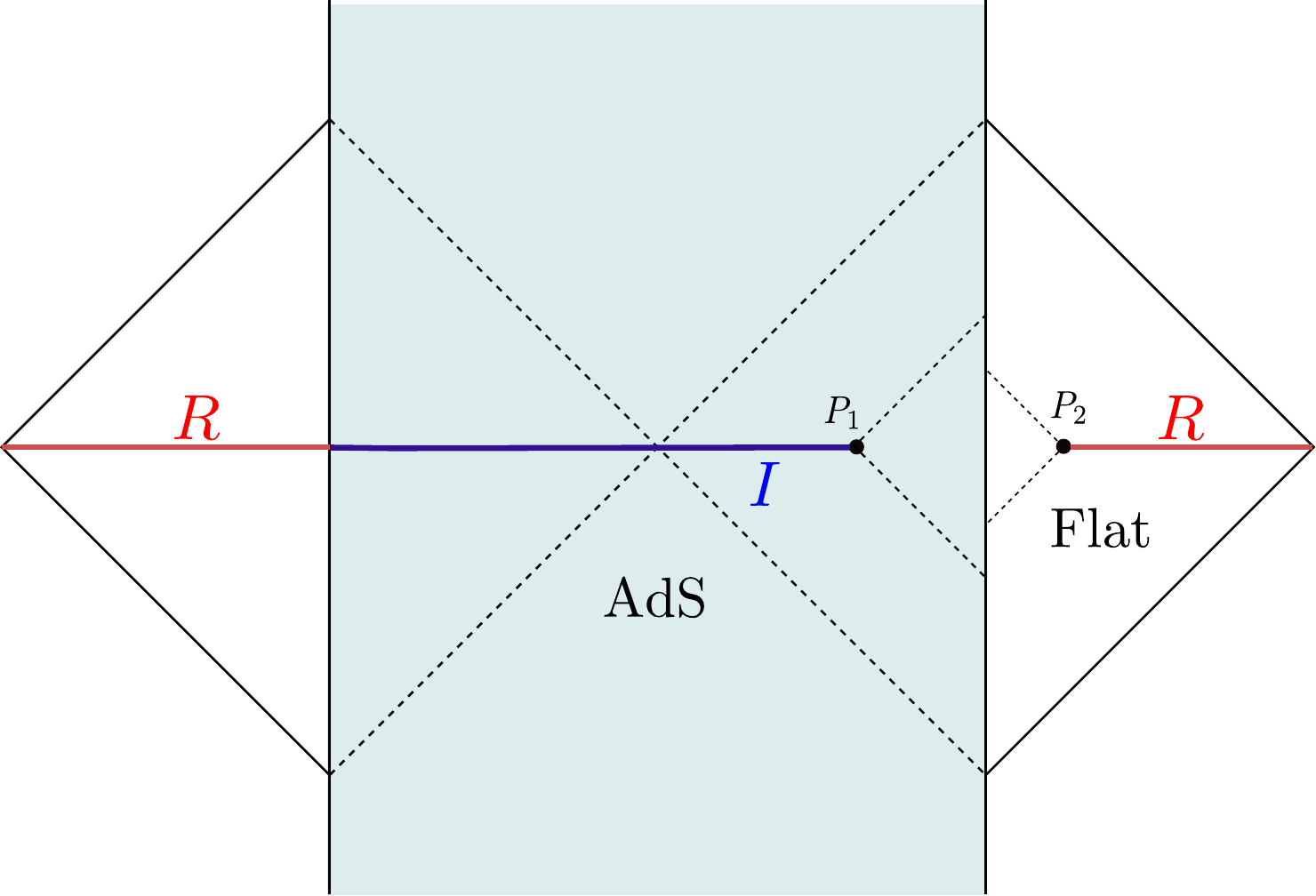}
\caption{Single-interval setup for the eternal black hole} \label{singleintervalpenrose}
\end{figure}

\begin{figure}
\centering
\begin{subfigure}{.5\textwidth}
  \centering
  \includegraphics[width=.9\linewidth]{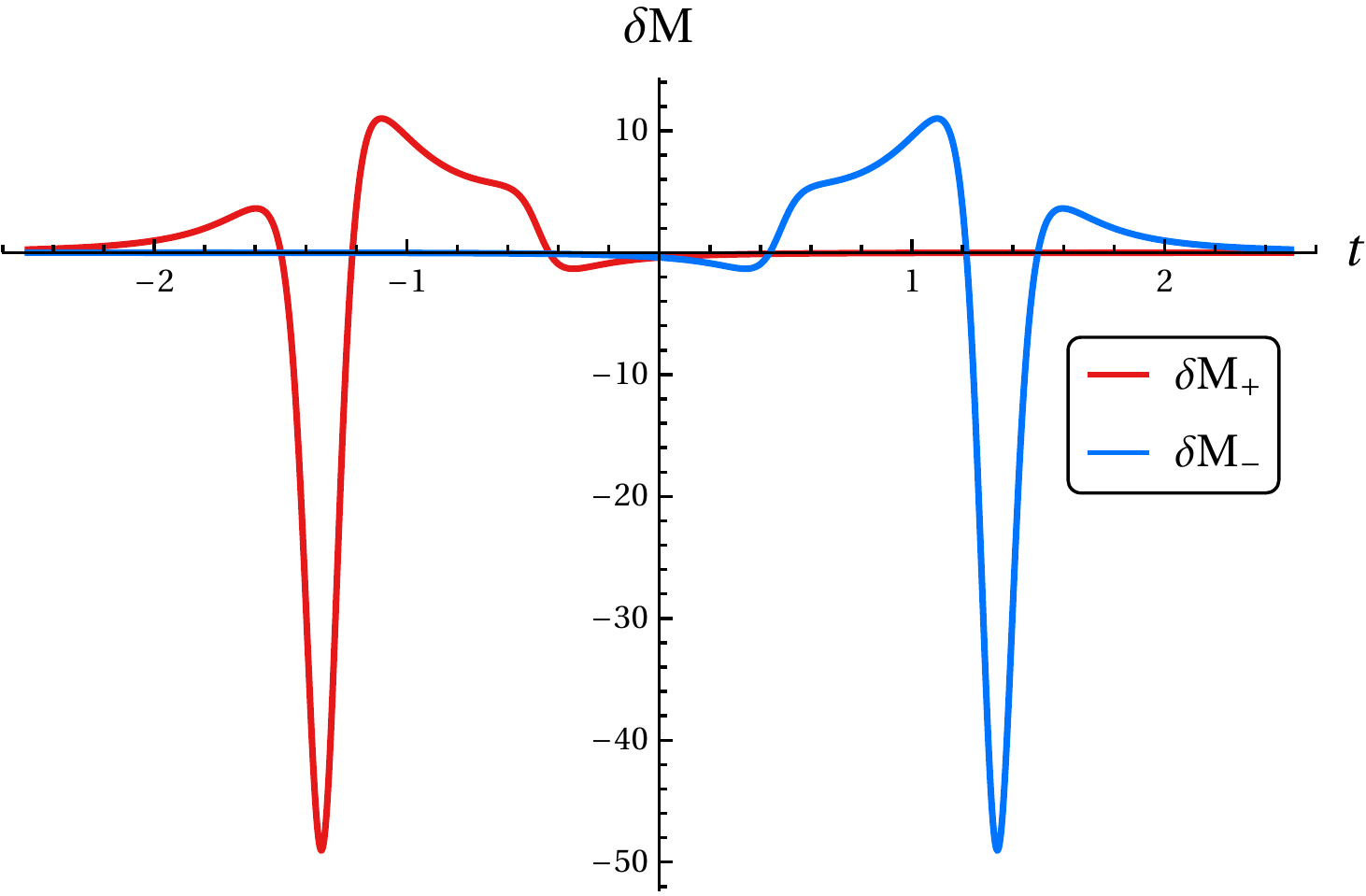}
  \caption{$\epsilon= 1/5$}
  \label{fig:sub1}
\end{subfigure}%
\begin{subfigure}{.5\textwidth}
  \centering
  \includegraphics[width=.8\linewidth]{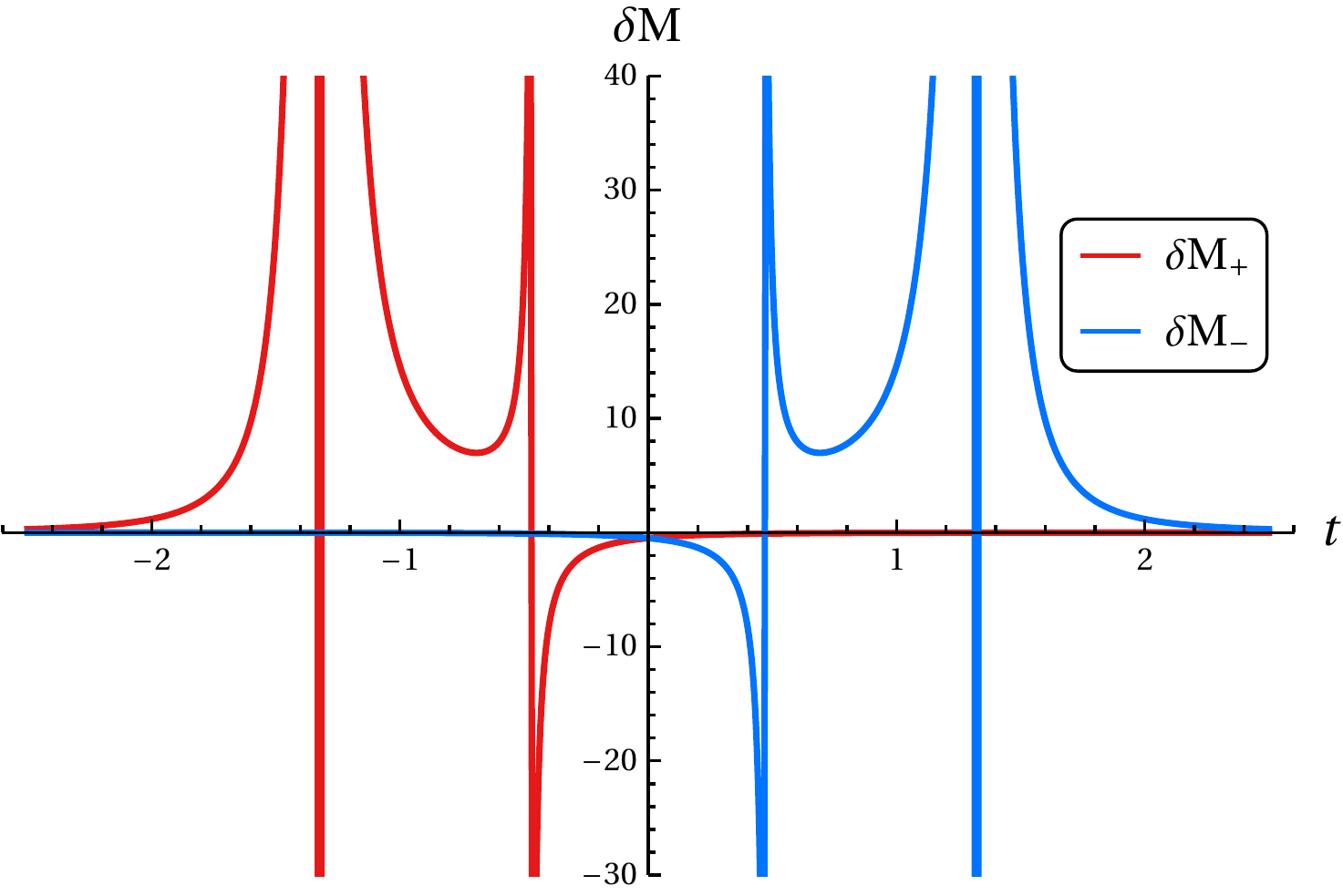}
  \caption{$\epsilon=1/100$}
  \label{fig:sub2}
\end{subfigure}
\caption{Solutions for the perturbative positive and negative Schwarzian $\delta M_+, \delta M_-$. The parameters are $ \kappa = 2/3, \sigma_b= \log(1.6+ i \epsilon)$. The first plot shows that with the appropriate $i\epsilon$ the solutions are decaying in both past and future, and the second plot shows the behavior as $\epsilon \to 0$.
}
\label{singleintervalplot}
\end{figure}

The  endpoints of the interval in $(\sigma, \tau)$ coordinates are given by 
\ba
P_1 = (\sigma_a, \tau_a), \qquad P_2 = (\sigma_b, \tau_b),
\ea
where $-\sigma_a,\sigma_b>0$ and $\tau_a$ is a complex number. The problem has a boost symmetry and therefore we can always set $\tau_b$=0.  The Lorentzian time and Euclidean time are related by $\tau= it$.  The equation of motion for the replica manifold is the same as equation \eqref{pertequation}.  Defining $\delta M$ by the expansion
\be
\{ w(\tau), \tau\} = \{ w_1(\tau), \tau\} - (n-1) \delta M \ ,
\ee
the equation for the perturbation around $n=1$ is
\ba\label{singleternal}
\pa_\tau \delta M - i \kappa \; \text{H}[\delta M] =   \kappa \mathcal{F} -\pa_\tau R_1 ,
\ea
where 
\begin{align}
\mathcal{F} &= - i \frac{e^{2 i\tau} (A-B)^2  }{(e^{i\tau} - A)^2 (e^{i \tau} -B)^2} + i \frac{e^{-2 i\tau} (\bar{A}-B)^2  }{(e^{-i\tau} - \bar{A})^2 (e^{-i \tau} -B)^2} \\
R_1 &= - \f{e^{ 2 i \tau} (1- |A|^2)^2}{(e^{ i \tau} - A)^2 (1- \bar{A} e^{i \tau})^2}, \notag
\end{align}
and $A=e^{i\tau_a+\sigma_a}, \bar{A}= e^{- i \tau_a +\sigma_a}, B= e^{\sigma_b}, \kappa= \frac{c}{24 \pi} \f{8\pi G}{\phi_r}$.   H is the  Hilbert transform which acts on Fourier modes as $H [ e^{ i m \tau}] = - \text{sgn}(m) e^{ i m \tau}$. This is the equation governing the geometry of the replica wormhole solution for a single interval near $n=1$.  Solving \eqref{singleternal} by matching the three Fourier modes $e^{ i \tau}, 1, e^{- i \tau}$, one finds the position of point $\sigma_a$ according to the island rule which gives $\tau_a=0$ and the condition \cite{Almheiri:2019yqk, Almheiri:2019qdq}
\ba\label{QESsingle}
 2 \kappa \f{\sinh(\f{\sigma_a+ \sigma_b}{2})}{\sinh (\f{-\sigma_a+\sigma_b}{2})} = \f{1}{\sinh(\sigma_a)}.
\ea
After imposing these conditions, the general solution to \eqref{singleternal} can be found in closed form. This is done in appendix \ref{app:eternal}, with the final answer for the positive and negative frequency components $\delta M_{\pm}$ appearing in \eqref{posM}-\eqref{negM}. The full answer is $\delta M = \delta M_+ + \delta M_-$. An example is plotted in figure \ref{singleintervalplot} in Lorentzian signature.

There are several features of this solution in Lorentzian signature that we want to highlight:

\begin{itemize}

\item   The series expansion as a sum over Euclidean Fourier modes $e^{i m \tau}$ diverges at the lightcones of the twist operators. This leads to singularities in Lorentzian signature, which have been regulated by an $i\epsilon$ shift in the plot. The $i\epsilon$ prescription corresponds to the contour in figure \ref{fig:SKtwist} (without any $\psi$ insertions).

\item Since the series diverges, it must be summed before continuing past the lightcone. This changes the naive behavior in an essential way. For example, positive Fourier modes $e^{i m \tau}$ with $m>0$ decay as $e^{-mt}$ as $t \to +\infty$. After doing the sum and analytic continuation around the lightcone singularity, we find that the sum over positive modes \textit{also} decays as $t \to -\infty$, $\delta M_+ \sim e^{\kappa t}$. This is the crucial fact we will use below to glue two one-interval solutions into a two-interval solution.

\item To obtain this solution we first set $a$ to the QES. In other words, we first solve for the $m=0,\pm 1$ Fourier modes. If we do not impose this condition, keeping $a$ general, then we can still solve the equation of motion for all of the other Fourier modes, but we find $\delta M$ diverges as $t \to \pm \infty$. Therefore imposing regularity at early/late Lorentzian times is equivalent to the extremality condition.

\end{itemize}

 \subsection{Two interval geometry in the eternal black hole}
 
\begin{figure}[h!]
\centering
\includegraphics[scale=0.74]{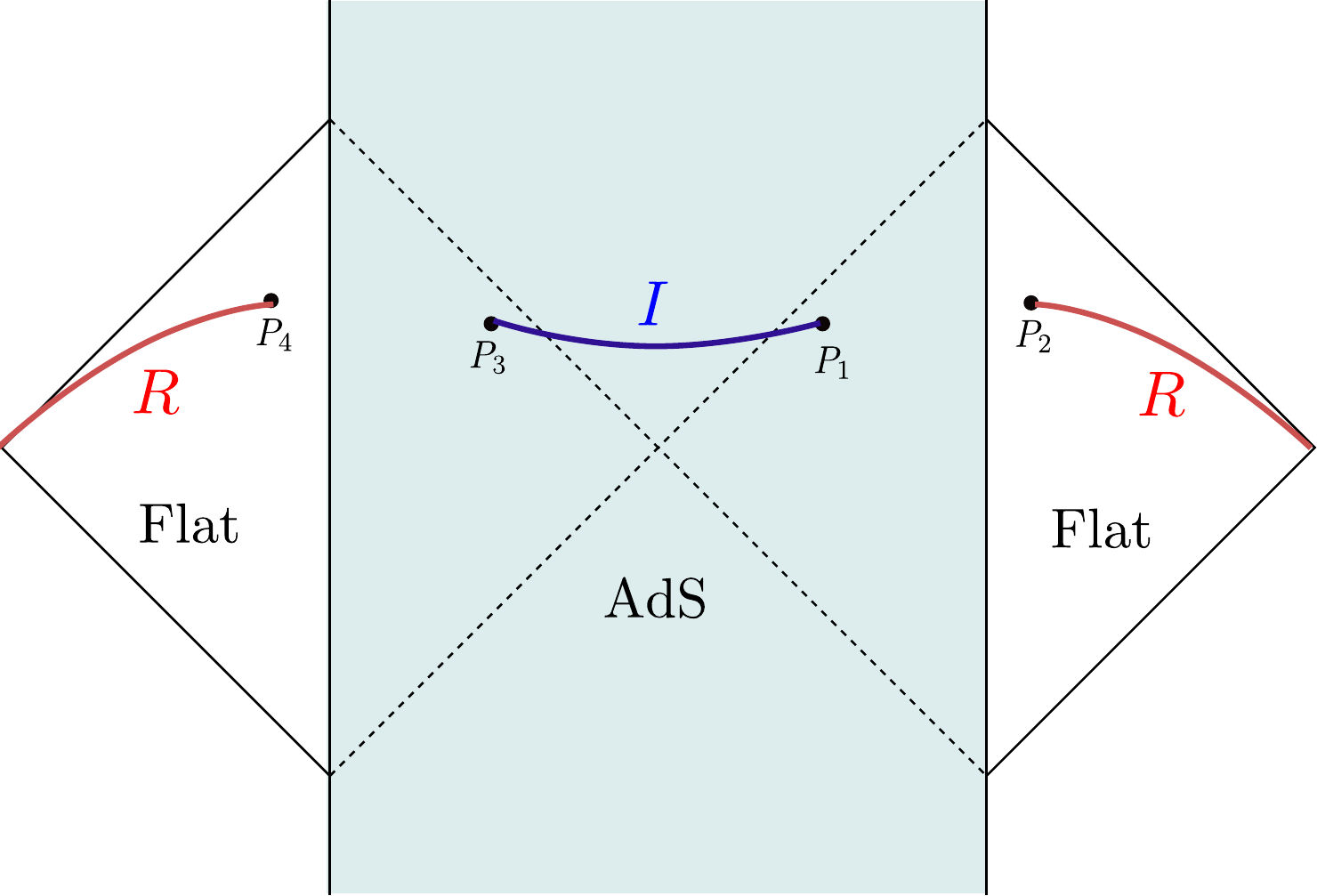}
\caption{Island for two intervals in the eternal black hole.} \label{fig:twointerval}
\end{figure}

 Now we consider the geometry near $n=1$ for two intervals in the eternal black hole. The coordinates in terms of $(\sigma, \tau)$ are given by
 \ba
P_1 = (\sigma_a, i t_a), \qquad P_2 = (\sigma_b, i t_b), \qquad P_3 = (\sigma_a, -i t_a - \pi), \qquad P_4= (\sigma_b, -i t_b - \pi),
\ea 
 and the setup is shown in figure \ref{fig:twointerval}. The  entropy of region $R$ without considering the island grows linearly with time $t_b$, $S_{\rm Hawking} \sim \f{\pi c}{6 \beta} t_b$. This follows from  the fact that the wormhole grows linearly with time\cite{Hartman:2013qma,Susskind:2014moa}. This indefinite entropy growth of a region is the Hawking paradox for the eternal black hole.  However, an island dominates after the Page time, and the island prescription
\ba 
S(R) = \text{min} \{ S_{\rm Hawking} (R), S_{\rm island} (R)\},
\ea
gives the unitary Page curve. 
At late times, the twist operators are in an OPE limit, so the matter entropy factorizes as
\ba 
S_{\rm QFT} (R \cup I) \approx 2 S_{\rm QFT} ([P_1, P_2]).
\ea 
In addition the  island prescripion sets $t_a= t_b$. The dilaton contribute to the entropy obviously just adds the two endpoints, so the upshot is that the generalized entropy with the island at late times is given by twice the single-interval answer. 

\begin{figure}[h!]
\centering
\includegraphics[scale=0.17]{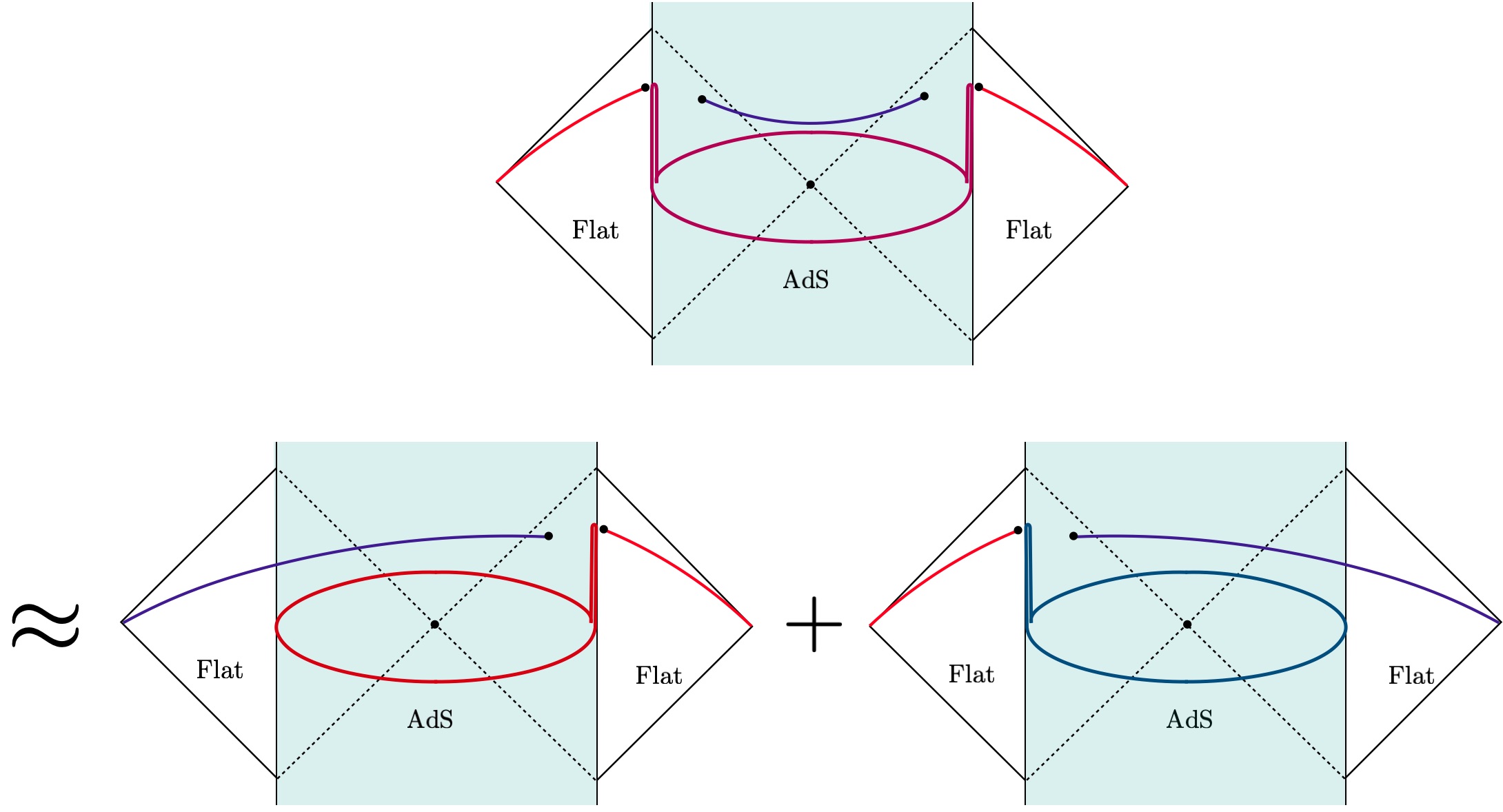}
\caption{At late times, the replica solution for the perturbation of the Schwarzian mode factorizes into the sum two single-interval solutions.} \label{cartoon_factorization}
\end{figure}

We want to argue that factorization at the level of geometries also holds.  Intuitively this is expected because the points $P_1, P_2$ are far from points $P_3, P_4$. Therefore, the matter stress tensor and the effects of conical defects factorizes to two single interval answers. 

To be more concrete, let us denote the perturbative Schwarzian of the single interval case due to points $[P_1, P_2]$ and $[P_3, P_4]$ by $\delta M_{[P_1, P_2]}$ and $\delta M_{[P_3, P_4]}$. We are now solving the Schwarzian equation along a contour similar to figure \ref{fig:SKtwist}, but with two Lorentzian pieces --- one on the right side of the black hole, and the other on the left. We must show that along this contour the perturbation is additive,
\be
\delta M \approx \delta M_{[P_1, P_2]} + \delta M_{[P_3, P_4]}  \ .
\ee
This is indeed a solution to the equation of motion. It was shown in the single-interval case that along the Lorentzian part of the contour, $\delta M_{[P_1, P_2]}$ is exponentially small for $ t \ll t_a$.  Therefore for $t_a > t_{\rm Page}$,
\be
\delta M_{[P_1, P_2]} \lesssim e^{-\kappa t_{\rm Page}} \ll 1 \ .
\ee
The same is true for $\delta M_{[P_3, P_4]} $.  Therefore we can add these two solutions, gluing them together in the Euclidean regime where both perturbations are negligible.
Practically, this factorization implies that for evaluating observables like the gravitational action or the matter effective action on a Schwinger-Keldysh contour, one could alternatively approximate the exact two intervals answer by the sum of two Schwinger-Keldysh contours for single interval geometries as shown in figure \ref{cartoon_factorization}. 

Let us make a comment about  number of equations needed to find all QES conditions. In an exact analysis of two intervals geometries, only a subset of QES conditions at points $P_1, P_3$ are expected to be given by integrating the boundary curve using three SL(2) kernels constructed in section \ref{sec:qesreplica}. The other set of equations are supposed to follow from variation with respect to moduli of the replica manifold. However, when we consider the limit corresponding to the factorized two interval geometry to single interval solutions, we find all QES conditions from integrating the boundary curve. The reason for finding extra equations is that for each single interval solution, we impose the boundary conditions at $t \to \pm \infty$ separately from the other single-interval solution. Therefore the extra equations, which should in principle come from the equations of motion for the moduli, came for free from regularity, i.e. from requiring the solution to factorize. 
 A better understanding the two interval replica manifold is an interesting question that we leave for future work.

To summarize, we have seen that two copies of the single-interval replica wormhole can be patched together to write the replica wormhole for the late-time, two-interval problem. It follows that we have also derived the QES's for two intervals in the eternal black hole, and therefore provided further justification for the island rule in this version of the information paradox. Let us note that the welding terms in the Schwarzian equation were essential for this to work, since these terms were responsible for the early-time decay of the Schwarzian perturbation.

\ \\

\bigskip

\textbf{Acknowledgments}
We thank Ahmed Almheiri, Juan Maldacena, Mukund Rangamani, and Edgar Shaghoulian for useful discussions. We also thank Tadashi Takayanagi for a careful reading of the draft and for valuable comments. KG thanks the Yukawa Institute for Theoretical Physics at Kyoto University for the hospitality in the final stages of this work. The work of TH is supported by  DOE grant DE-SC0020397. AT is supported by the US Air Force Office of Scientific Research under award number FA9550-19-1-0360 and by the Simons Foundation. 
 
 \ \\
 \bigskip
 
\noindent {\huge\textbf{Appendices}}

\appendix

\section{Factorization of the matter entropy}\label{app:matterentropy}

In this appendix we will check the assumption that we can use the factorized matter entropy in determining the nontrivial QES of the evaporating black hole in section \ref{sec2}. 
The relevant regime is $t\sim  \CO(1/\kappa)$ ($t>0$) and $u_0e^{-\kappa(t-L)/2}\gg 1$ with $\kappa\ll 1$. In this section, we set $\beta=2\pi$ and $y^-_B=y^+_B=t$ for simplicity. 
 
We will first consider the general CFTs and show  the factorization $\pp_{x^\pm_{A}}S_{\rm QFT}(I\cup R)\approx\pp_{x^\pm_{A}}S_{\rm QFT}([A,B])$, i.e, for the variation with respect to the right QES $A$. In addition to the general arguments,  we will also demonstrate the factorization in the case of the free Dirac fermion by explicit computations of $S_{\rm QFT}(I\cup R)$ not only for the variation with respect to $A$ but also for the left QES $A'$.

Let us first consider general CFTs. The entanglement entropy may be computed from the four-point function of the twist operators with dimension $h=\bar{h}=\f{c}{24}(n-1/n)$ inserted at the endpoints of the interval $I$ and $R$ using the replica trick
 \ba
 S_{\rm QFT}(I\cup R)=\lim_{n\ra 1}\frac{1}{1-n}\log \left[\frac{\langle\sg(z_{A'})\bar{\sg}(z_{A})\sg(z_B)\bar{\sg}(z_{B'})\rangle}{(\Omega(z_{A'}) \Omega(z_{A})\Omega(z_{B}) \Omega(z_{B'}))^{-2h}}\right]\, .
 \ea
Here the complex coordinates $z,\z$ are related to the  light-cone coordinates in Lorentzian regime by $\z=z^+$ and $z=1/z^-$. They are defined for $A$ and $B$ as (\ref{zzbar}) i.e., $z^- = x(y^-), z^+ =e^{y^+}$. For $A'$ and $B'$, they are defined as $z^- =e^{y^-}, z^+ =e^{y^+}$.  $\Omega$ is the Weyl factor in $ds^2=\Omega^{-2}dzd\z$.  The four-point  function may be expressed in terms of the cross-ratios,
\ba
\eta=\f{z_{BA}z_{B'A'}}{z_{BA'}z_{B'A}},  \quad
\bar{\eta}=\f{\bz_{BA}\bz_{B'A'}}{\bz_{BA'}\bz_{B'A}}
\ea
where $z_{ij}=z_j-z_i$, as
\ba
\langle\sg(z_{A'})\bar{\sg}(z_{A})\sg(z_B)\bar{\sg}(z_{B'})\rangle=|z_{B'A}|^{-4h}|z_{BA'}|^{-4h}\langle\sg_{A'}(1)\bar{\sg}_{A}(\eta,\bar{\eta})\sg_{B}(0)\bar{\sg}_{B'}(\infty)\rangle\, .
\ea
Now let us evaluate the cross-ratio $\eta$, which is related to the $z^-$ coordinate, at the positions of QESs obtained by the analyses in section \ref{sec2}. As we send $B'\ra\infty$, i.e, $v^-_{B'}=0$, the left QES $A'$ is placed at the bifurcation point $v^-_{A'}=\infty$. Moreover, by using $u_0 \gg 1$ we obtain
\ba
\eta\approx \f{x^-_{BA}}{x^-_A}\approx \f{4}{3u_0}e^{-\f{2u_0}{\kappa}(1-e^{-\f{\kappa}{2} (t-L)})}\approx 0\, ,
\ea
where for the first approximation we used the condition for the left QES $A'$ and $B'\ra \infty$, and for the second approximation we used  the condition for the right QES (\ref{locQES2}) and the asymptotic expansion of the map $x(t)$ in (\ref{Asymptx}).   Therefore we need to study the entanglement entropy in the regime $\eta \approx 0$.

We expand $\langle\sg_{A'}(1)\bar{\sg}_{A}(\eta,\bar{\eta})\sg_{B}(0)\bar{\sg}_{B'}(\infty)\rangle$ in Virasoro conformal blocks in the $\eta\ra 0$ OPE channel as  
\bal
\langle\sg_{A'}(1)\bar{\sg}_{A}(\eta,\bar{\eta})\sg_{B}(0)\bar{\sg}_{B'}(\infty)\rangle
=(\eta\bar{\eta})^{-2h}\sum_{h_p,\bar{h}_p}c_p{\cal F}(h_p,\eta)\overline{{\cal F}}(\bar{h}_p,\bar{\eta})\, ,
\eal 
where the block admits a series expansion around $\eta=0$ as
\ba\label{block}
{\cal F}(h_p,\eta)=\eta^{h_p}\sum_{q=0}^{\infty}{\cal F}_q(h_p) \eta^q\, .
\ea
Therefore the entanglement entropy can be decomposed as
\ba
S_{\rm QFT}(I\cup R)=S^{\rm fact.}_{\rm QFT}+S^{\rm non\mathchar`-fact.}_{\rm QFT}\, ,
\ea
where $S^{\rm fact.}_{\rm QFT}=S_{\rm QFT}([A,B])+S_{\rm QFT}([A',B'])$ is the factorized part of the entropy we used to find the QES in the previous section, and $S^{\rm non\mathchar`- fact.}_{\rm QFT}$ is the remaining ``non-factorized'' part on the entropy $S_{\rm QFT}(I\cup R)$ defined respectively as 
\bal\label{fact_nonfact}
 S^{\rm fact.}_{\rm QFT}&\equiv \f{c}{6}\log \f{|z_{BA}|^2|z_{B'A'}|^2}{\Omega_{A'} \Omega_{A} \Omega_{B} \Omega_{B'}}\no
S^{\rm non\mathchar`-fact.}_{\rm QFT}&\equiv \lim_{n\ra 1}\f{1}{1-n}\log \sum_{h_p,\bar{h}_p} c_p {\cal F}(h_p,\eta)\overline{{\cal F}}(\bar{h}_p,\bar{\eta}) \notag\\
&\approx \lim_{n\ra 1}\f{1}{1-n}\log \overline{{\cal F}}(0,\bar{\eta})  \ ,
\eal
where in the last line we have used $\eta \to 0$, which projects onto the vacuum block.
As we can see, in general the entanglement entropy itself doesn't factorize completely due to $S^{\rm non\mathchar`-fact.}_{\rm QFT}$ even in the limit $\eta\ra 0$ due to the Virasoro descendants of the vacuum. Nevertheless we can show the factorization of its derivative (\ref{factorization}) in the regime of interest by using the QES condition obtained from $S^{\rm fact.}_{\rm QFT}$.

First  we consider the derivative with respect to $x^-_A$  in the limit $\eta\ra 0$. As we can see from the expression (\ref{fact_nonfact}), the factorized part $S^{\rm fact.}_{\rm QFT}$ gives the leading singularity in the limit $\eta\ra0$ as 
\ba
\pp_{x^-_{A}}S_{\rm QFT}(I \cup R)\approx \f{c}{6}\f{1}{x^-_A-x^-_B}
\ea
while the non-factorized part gives terms subleading in $\eta$. 
Therefore in the regime of interest $u_0\gg 1\LR \eta\approx 0$ we can confirm the approximation
\bal
\pp_{x^-_{A}}S_{\rm QFT}(I\cup R)\approx\pp_{x^-_{A}}S_{\rm QFT}([A,B])\, .
\eal
Next let us consider the variation with respect to $x^+_A$. 
Let us remind ourselves that the leading contribution to the derivative of the factorized part $\pp_{x^+_{A}}S_{\rm QFT}([A,B])$ comes from the derivative of the Weyl factor 
\ba
\pp_{x^+_{A}}S_{\rm QFT}([A,B])\approx -\f{c}{6}\f{\Omega'(x_A)}{\Omega(x_A)}\approx \f{c}{12}\f{1}{x_\infty-x^+_A}+\CO(x_\infty-x^-_A)\, .
\ea
The non-factorized part $S^{\rm non\mathchar`-fact.}_{\rm QFT}(\bar{\eta})$ (after taking $\eta\ra 0$ limit) depends only on $\z=z^+=e^{y^+}$, so it is independent of $x(t)$ and $u_0$. Thus the dependence on $x(t)$ of $\pp_{x^+_A} S^{\rm non\mathchar`-fact.}_{\rm QFT}$ comes only from the derivative $\pp_{x^+_A}=\f{1}{x'(y^+_A)}\pp_{y^+_A}$. By using the asymptotic expression of $x'(t)$ (\ref{Asymptx2}), we obtain
\ba
\pp_{x^+_A} S^{\rm non\mathchar`-fact.}_{\rm QFT}\sim \f{1}{u_0e^{-\kappa (t-L)/2}} \f{1}{x_\infty-x^+_A}\times \pp_{y^+_A}S^{\rm non\mathchar`-fact.}_{\rm QFT}(\bar{\eta})
\ea
where $S^{\rm non\mathchar`-fact.}_{\rm QFT}(\bar{\eta})$ is non-singular and independent of $x(t)$ and $u_0$. Therefore using $u_0e^{-\f{\kappa}{2} (t-L)}\gg 1$, we find
\ba
\pp_{x^+_{A}}S^{\rm fact.}_{\rm QFT}\gg\pp_{x^+_{A}}S^{\rm non\mathchar`-fact.}_{\rm QFT}\, ,
\ea
and confirm the relation
\ba
\pp_{x^+_{A}}S_{\rm QFT}(I\cup R)\approx\pp_{x^+_{A}}S_{\rm QFT}([A,B])\, .
\ea
{\bf Example: Free Dirac fermion}\\
We explicitly demonstrate the factorization of the entanglement entropy using the free Dirac fermion. The entanglement entropy of the free Dirac fermion  is given by \cite{Casini:2005rm}
\bal
S_{\text {fermions }}(I\cup R)&=\frac{c}{6} \log \left[\frac{\left|z_{AA'} z_{BA} z_{B'B} z_{B'A'}\right|^{2}}{\left|z_{BA'} z_{B'A}\right|^{2} \Omega_{A'} \Omega_{A} \Omega_{B} \Omega_{B'}}\right]\no
&=\frac{c}{6} \log \frac{\left|z_{AB}|^2| z_{A'B'}\right|^{2}}{\Omega_{A'} \Omega_{A} \Omega_{B} \Omega_{B'}}+\frac{c}{6} \log |1-\eta|^2\, .
\eal
As we did above, we divide the entropy into the factorized part end non-factorized part as
\ba
S_{\text {fermions }}(I\cup R)=S^{\rm fact.}_{\text {fermions }}+S^{\rm non\mathchar`-fact.}_{\text {fermions }}\, ,
\ea
where
\ba
S^{\rm fact.}_{\text {fermions }}(I\cup R)\equiv\frac{c}{6} \log \frac{\left|z_{AB}|^2| z_{A'B'}\right|^{2}}{\Omega_{A'} \Omega_{A} \Omega_{B} \Omega_{B'}}\, ,\q 
S^{\rm non\mathchar`-fact.}_{\text {fermions }}\equiv \frac{c}{6} \log |1-\eta|^2\, .
\ea
First we compute the derivatives  with respect to  $x^-_A$  as
\ba
\pp_{x^-_A} S^{\rm fact.}_{\text {fermions }}\sim\f{c}{6}\f{1}{x^-_A-x^-_B}\, ,\q
\pp_{x^-_A} S^{\rm non\mathchar`-fact.}_{\text {fermions }}\sim\f{c}{6}\l(\f{1}{x^-_A-v^-_{A'}}-\f{1}{x^-_A-v^-_{B'}}\r)\, .
\ea
The leading contribution comes from the singularity $\eta\approx 0$ in the factorized part $S^{\rm fact.}_{\text {fermions }}$. Thus to compute the QES, we can ignore the non-factorized part of the entanglement entropy. 

The derivatives  with respect to  $x^+_A$ are
\bal
\pp_{x^+_A} S^{\rm fact.}_{\text {fermions }}&\sim\f{c}{12}\f{1}{x_\infty-x^+_A}\, ,
\eal 
and
\ba
\pp_{x^+_A} S^{\rm non\mathchar`-fact.}_{\text {fermions }}=\f{c}{6x'(y^+_A)}\l(\f{1}{v^+_A-v^+_{A'}}-\f{1}{v^+_A-v^+_{B'}}\r)\approx  -\f{c}{6}\f{1}{u_0e^{-\kappa(t-L)/2}}\f{1}{x_\infty-x^+_A}e^{-y^+_A}\, .\ea
We have $\pp_{x^+_A} S^{\rm fact.}_{\text {fermions }}\gg\pp_{x^+_A} S^{\rm no\mathchar`-fact.}_{\text {fermions }}$ for  $u_0e^{-\kappa(t-L)/2}\gg 1$, thus the factorized part gives the leading contribution. 

Finally we will check $\pp_{v^\pm_{A'}}S_{\rm fermion}(I\cup R)\approx\pp_{v^\pm_{A'}}S_{\text {fermions }}([A',B'])$ (which we did not check in general CFTs). Since the derivatives of the factorized entropy $S_{\text {fermions }}([A',B'])$  evaluated at the QES $v^+_{A'}=0, v^-_{A'}=\infty$ are given by
\ba
\pp_{v^+_{A'}} S_{\text {fermions }}([A',B'])|_{\rm QES}\sim \f{c}{6}\f{1}{v^-_{A'}-v^+_{A'}}\biggl|_{\rm QES}=0\, ,
\ea
and
\ba
\pp_{v^-_{A'}} S_{\text {fermions }}([A',B'])|_{\rm QES}\sim\f{c}{6}\l.\l(\f{1}{v^-_{A'}-v^-_{B'}}+\f{1}{v^+_{A'}-v^-_{A'}}\r)\r|_{\rm QES}=0\, ,\ea
 we need to check $\pp_{v^\pm_{A'}} S^{\rm non\mathchar`-fact.}_{\text {fermions }}\approx 0$ at the QES in the $\kappa \ra 0$ limit. The variation of $S^{\rm non\mathchar`-fact.}_{\text {fermions }}$ with respect to $v^-_{A'}$ may be evaluated at the QES $A'$ as
\ba
\pp_{v^-_{A'}} S^{\rm non\mathchar`-fact.}_{\text {fermions }}|_{\rm QES}\sim\f{c}{6}\l.\l(\f{1}{v^-_{A'}-x^-_A}-\f{1}{v^-_{A'}-x^-_B}\r)\r|_{\rm QES}=0\, .
\ea
Thus we confirmed that $S^{\rm non\mathchar`-fact.}_{\text {fermions }}$ doesn't change the position of the right QES $A'$: $v^-_{A'}=0$.
The variation with respect to $v^+_{A'}$ is evaluated as
\ba
\pp_{v^+_{A'}} S^{\rm non\mathchar`-fact.}_{\text {fermions }}|_{\rm QES}=\f{c}{6}\l.\l(\f{1}{v^+_{A'}-v^+_{A}}-\f{1}{v^+_{A'}-v^+_{B}}\r)\r|_{\rm QES}=-\f{c}{6}e^{-y^+_A}(1-e^{y^+_A-y^+_B})|_{\rm QES}\, .
\ea
Since we have $y^+_A-y^+_B<0$ and $e^{-y^+_{A'}}\sim\CO( e^{-1/\kappa})$ at the QES, this gives zero up to a tiny correction $\CO( e^{-1/\kappa})$ we can neglect in the $\kappa\ra 0$ limit. In this way we can check the factorization (\ref{factorization}).

\section{Shockwave solution at finite $\delta$, small $E_{\psi}$}\label{app:secorderdelta}

In this appendix, we describe the shockwave solution to leading order in $E_{\psi} = h_{\psi}/\delta$ at finite $\delta$. In this limit, the form of welding is known and the mixing between positive and negative frequency modes is tractable. For simplicity, we consider the zero temperature black holes for this discussion.

\begin{figure}[h!]
    \centering
    \begin{subfigure}[b]{0.4\textwidth}
        \includegraphics[width=\textwidth]{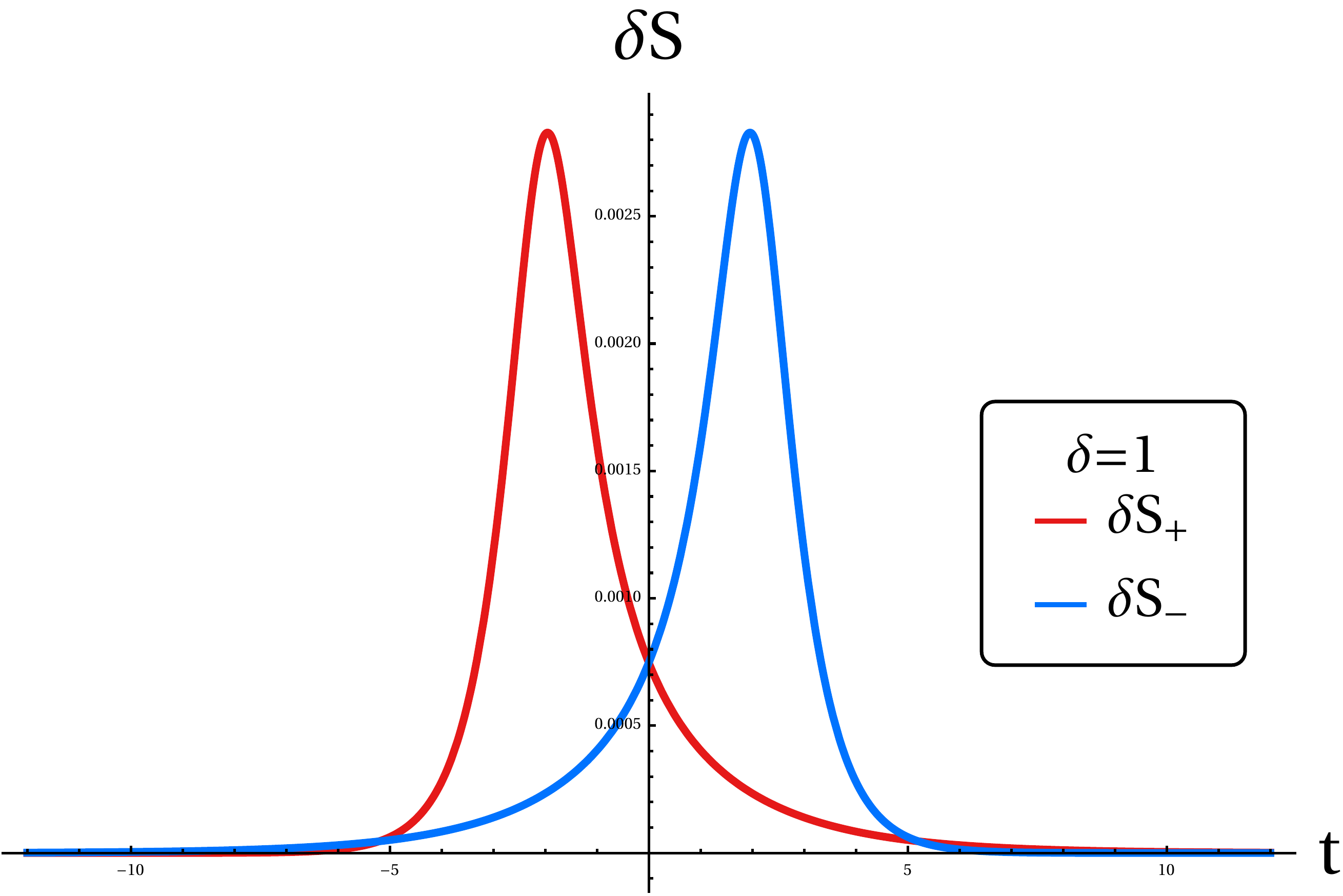}
        \caption{}
        \label{fig:}
    \end{subfigure}
~
    \begin{subfigure}[b]{0.4\textwidth}
        \includegraphics[width=\textwidth]{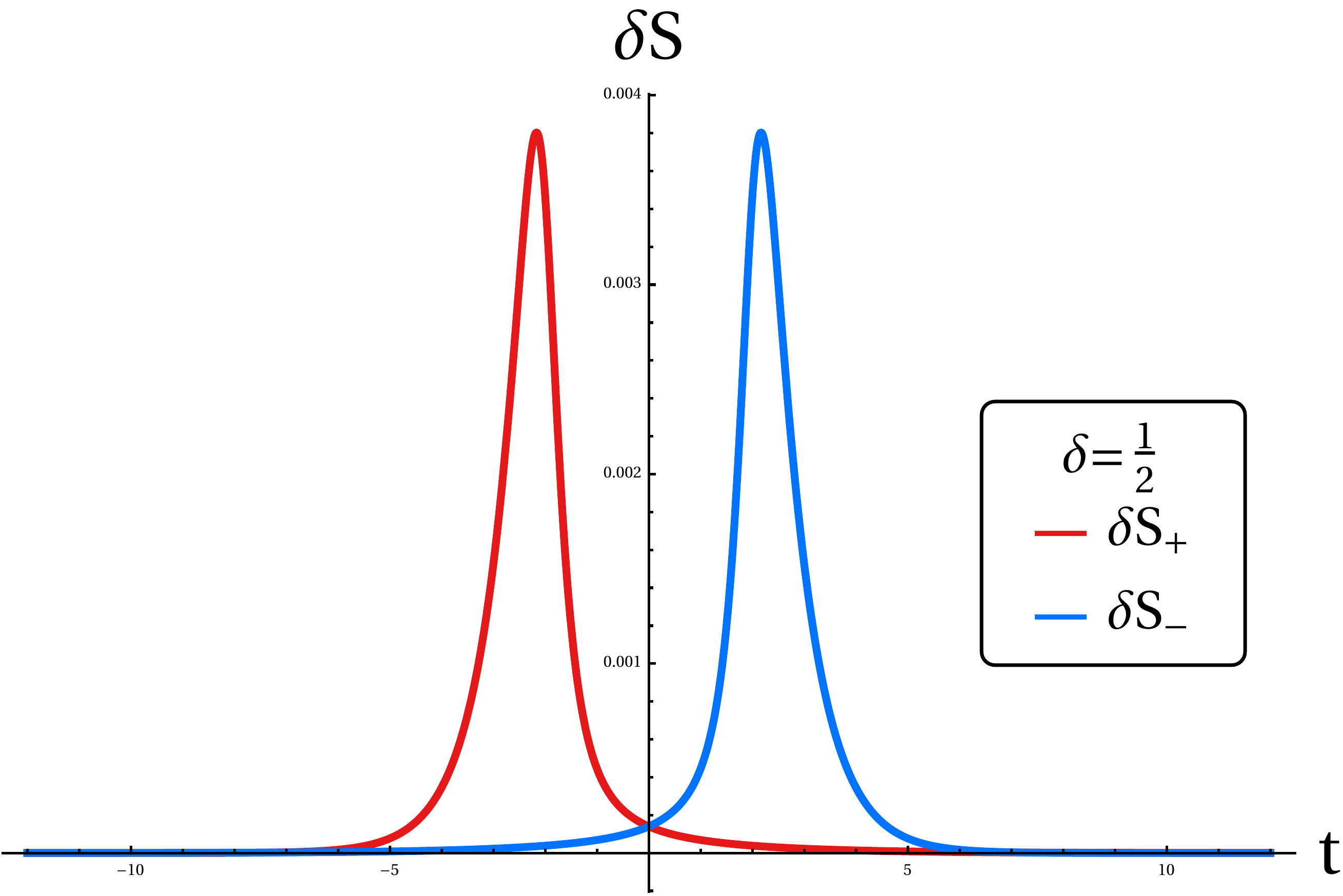}
        \caption{}
        \label{fig:}
    \end{subfigure}
  ~
    \begin{subfigure}[b]{0.4\textwidth}
        \includegraphics[width=\textwidth]{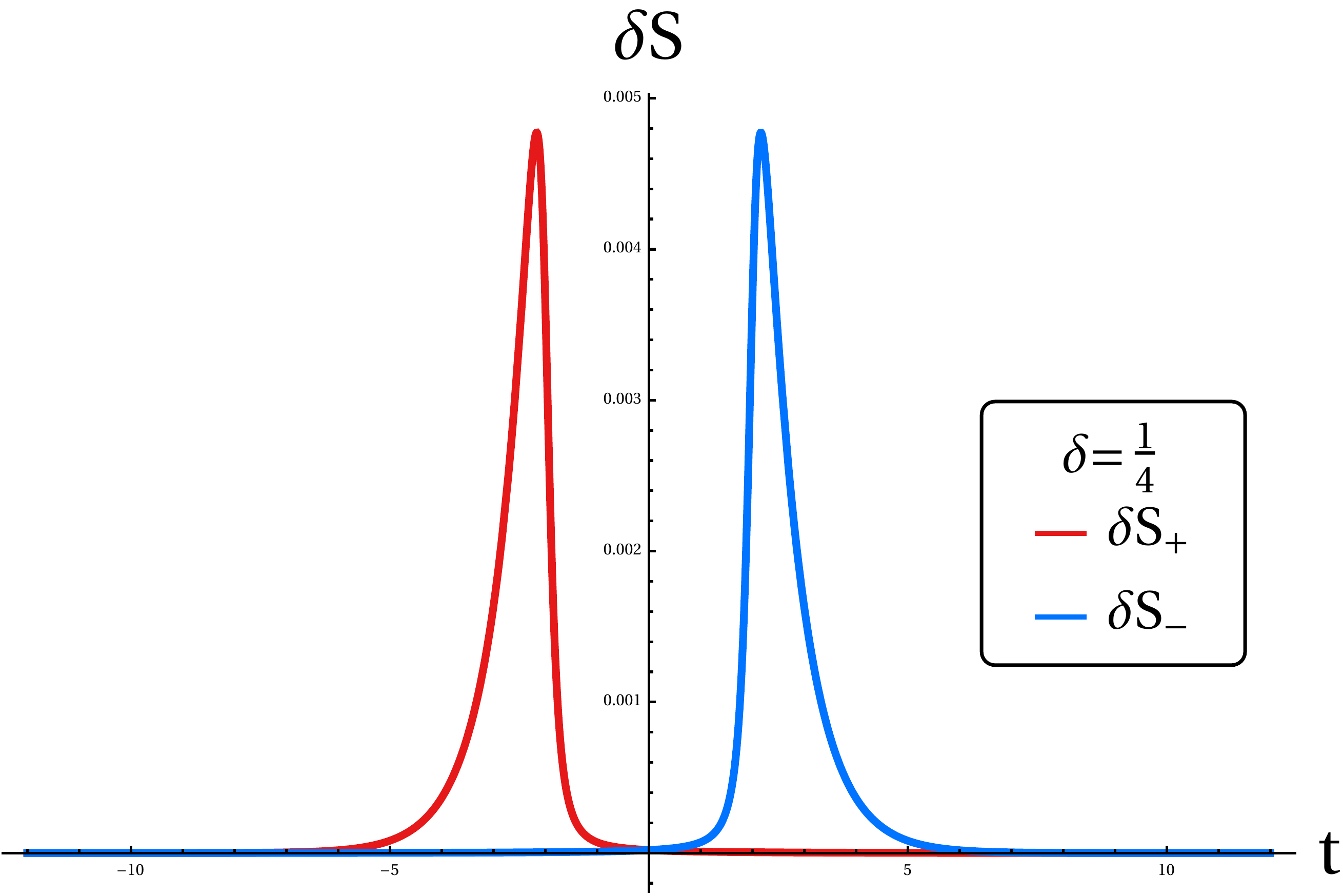}
        \caption{}
        \label{fig:}
    \end{subfigure}
    \caption{Plots for different values of $\delta$ for some values of parameters $\kappa=3/2,L=2, E_\psi= 1/10, \f{c}{24\pi}=1$}\label{fig:different deltas}
\end{figure}

After the analytic continuation $\tau \to  i t$,  the equation \eqref{small delta corr} in the limit $\beta \to \infty$ becomes
\bal\label{small dela Lor}
\pa_t \delta S_{\pm}  \mp  \kappa \delta S_{\pm} = - \f{24\pi \kappa}{c} \mathcal{F}^L_{\pm},
\eal
where 
\bal\label{flux Lor}
\mathcal{F}^L_+ = \frac{h_\psi}{2\pi} \frac{\sin^2\delta}{(\cos(\delta) - \cosh(L+t) )^2}, \qquad \mathcal{F}^L_- = -\frac{h_\psi}{2\pi} \frac{\sin^2\delta}{(\cos(\delta) - \cosh(L- t) )^2}.
\eal
These equations were also derived in \cite[Appendix C]{Almheiri:2019qdq}.
The numerical solutions are plotted in figure \ref{fig:different deltas} for different values of $\delta$. The full $\delta S $ is determined by $\delta S= \delta S_+ + \delta S_-$. The region where positive and negative modes overlap gets smaller as $\delta$ decreases.

As we increase the shock energy $h_\psi/\delta$, there are corrections to the welding term in the differential equation \eqref{small dela Lor}. However, as argued in section \ref{sec:3}, these corrections are controlled by the mixing between positive and negative modes and if take $\delta \to 0$ first, the leading welding term is given by \eqref{small dela Lor}, and for the finite shock energy we find \eqref{flux-Lor1} and \eqref{flux-Lor2}. 

\section{Dilaton from boundary curve}\label{app:dilaton}
In this appendix, we demonstrate that on any background $w_1(\tau)$ created by an arbitrary number of operator insertions outside the gravity region, the dilaton $\phi$ has a simple form
\begin{align}\label{dildisk}
\phi(A,\bar{A}) &= -\f{\phi_r}{2\pi} \oint R_1 d\tau\, ,
\end{align}
where
\begin{align}
R_1 &= \frac{(1-|A|^2)^2 \l(w_1'(\tau)\r)^2}{(1- \bA w_1(\tau))^2(w_1(\tau)-A)^2}\, .
\end{align}
The dilaton has a well-known closed form solution \eqref{dilaton} in terms of the stress tensor. The goal here is to write the dilaton only in terms of the boundary curve. In subsection \ref{app:dil-euc} this is derived in Euclidean signature. The derivation in \ref{app:dil-euc} is general and relies only on the analytic properties of the stress tensor inside the gravity region. For solutions like sharp shockwaves which satisfy the no-mixing condition in real time, we compute the dilaton directly in terms of the boundary curve $x_1(t)$ in \ref{app:dil-lor}. Finally, using the Schwinger-Keldysh contour, we show in \ref{Sec:DerSK} that the Euclidean method is equivalent to the Lorentzian calculation.

\subsection{Euclidean}\label{app:dil-euc}

 Here we evaluate the dilaton at an arbitrary point $(A,\bar{A})$ in the $w$-coordinate inside the unit disk. The argument is easier if we apply a $\text{SL}(2)$ transformation to the disk and work instead on a plane. Let us define a new coordinate $Z= X+i Y$ from the relation $ - \frac{Z(\tau)-Z_0}{Z(\tau)+\bar{Z}_0} = \frac{w_1(\tau)-A}{1- w_1(\tau)\bar{A}}$, where $Z_0$ is an arbitrary constant that we assumed for simplicity is real and $Z_0<0$. The metric is the hyperbolic metric in the Poincare coordinates
\ba
 ds^2 = \frac{4 dZ d\bar{Z}}{(Z+\bar{Z})^2}.
 \ea
  The boundary curve is  $(X(\tau), Y(\tau))$. The intrinsic metric condition on the boundary curve gives
\ba
\frac{1}{\epsilon^2} = g_{\tau \tau} = \frac{Y'(\tau)^2+ X'(\tau)^2}{X(\tau)^2} \qquad  \Rightarrow\qquad  X(\tau) = \epsilon Y'(\tau) +\mathcal{O}(\epsilon^2)\, .
\ea
The equations for the dilaton in the $Z$-plane are given by Euclidean version of \eqref{dilaton} as written in \cite{Maldacena:2016upp}. The closed form solution for these equations is
\ba\label{dilinstress}
\phi(Z, \bar{Z}) = \Phi_0 + 8\pi G_N \int^{Z} d\tilde{Z} \f{ (\tilde{Z}+ \bar{Z})(Z- \tilde{Z})}{(Z+\bar{Z})} T_{ZZ}(\tilde{Z}) + (Z \leftrightarrow \bar{Z}, T_{ZZ} \leftrightarrow T_{\bar{Z}\bar{Z}} ),
\ea
where 
\ba
\Phi_0 = \frac{a_1 + a_2 Z+ a_3 \bar{Z} + a_4 Z \bar{Z} }{Z+\bar{Z}},
\ea
is the homogenous solution. Changing the lower limit of the integral in \eqref{dilinstress} corresponds to a change in the homogeous part of the dilaton which would be fixed by the value of the dilaton  at the boundary. The dilaton at point $(Z_0, Z_0)$ is written compactly as
\bal\label{dilatz0}
\phi(Z_0,Z_0) = \Phi_0(Z_0, Z_0) + 8 \pi G_N \int^{Z_0} dx \f{Z_0^2- x^2}{Z_0} {\rm Re} T (x, 0).
\eal

If the operators creating the geometry are all inserted outside the gravity region, each components of the stress tensor $T_{ZZ}, T_{\bar{Z} \bar{Z}}$ are analytic functions in the gravity region. In the Schwarzian limit, the boundary lies at $X=0$ and therefore the  real part of the stress tensor can be written in terms of the imaginary part,
\ba\label{poissonkernel}
{\rm Re} T( X,Y) = \frac{1}{\pi} \int_{-\infty}^{+\infty} dY_1 \f{(Y_1 - Y)}{X^2 + (Y-Y_1)^2} \text{Im}T (0, Y_1).
\ea
 The kernel appearing in \eqref{poissonkernel} is nothing but the Poisson kernel for harmonic functions in two dimensions. The imaginary part of the stress tensor is the energy-momentum flux and it is related to the boundary curve $Z(\tau)=  i Y(\tau)+ \mathcal{O}(\epsilon)$ as
\ba\label{imstrstens}
-2\text{Im} T = \f{\phi_r}{8 \pi G}\f{\pa_\tau \{ Z(\tau), \tau\}}{ Z'(\tau)^2} = \f{\phi_r}{8 \pi G} \f{1}{Z'(\tau)}\left( \f{1}{Z'(\tau)} \left( \f{Z''(\tau)}{Z'(\tau)} \right)' \right)'.
\ea 
 Combining \eqref{dilatz0}, \eqref{imstrstens} and \eqref{poissonkernel}, the dilaton can be written in terms of the derivatives of the boundary curve. Doing the integration by parts and dropping the boundary terms, we find 
\ba\label{finaldil}
\phi(Z_0,Z_0) = - \frac{2\phi_r}{\pi} \int_{0}^{2\pi} d\tau \f{Z_0^2 Z'(\tau)^2}{(Z(\tau)^2- Z_0^2)^2}.
\ea
Note that $Z(\tau)$ is purely imaginary. By a $\text{SL}(2)$ transformation, the dilaton in a general point $(Z_0, \bar{Z}_0)$ is written as
\ba\label{dilarbitrary}
\phi(Z_0, \bar{Z}_0) = - \f{\phi_r}{2\pi } \int_{0}^{2\pi} d\tau \f{(Z_0+ \bar{Z}_0)^2 Z'(\tau)^2}{(Z(\tau) - Z_0)^2 (Z(\tau)+ \bar{Z}_0)^2}.
\ea
To make sure there is no homogenous term in \eqref{dilarbitrary}, let us take $Z_0 \to X(u) + i Y(u)$, $\bar{Z}_0 \to X(u) - i Y(u)$. The equation \eqref{dilarbitrary} then becomes
\ba
\phi(X(u), Y(u)) = - \f{\phi_r}{2\pi} \int_{0}^{2\pi} d\tau \f{4 \epsilon^2 Y'(u)^2 Z'(\tau)^2}{ ((Y(\tau) - Y(u))^2 + \epsilon^2 Y'(u)^2)^2} = \phi_r/\epsilon,
\ea
where we used $\lim_{\alpha \to 0} \f{\alpha^2}{(r^2+ \alpha^2)^2} = \f{\pi}{2 \alpha} \delta (r)$. This shows that \eqref{dilarbitrary} has the right boundary condition and in particular, there is no extra homogeous term in the final expression. Once the transformation $ - \frac{Z-Z_0}{Z+\bar{Z}_0} = \frac{w_1-A}{1- w_1\bar{A}}$ is applied back to \eqref{dilarbitrary}, we find \eqref{dildisk} which is the dilaton written in the disk coordinates. Note that computing \eqref{dildisk} by residues would require additional assumptions on analytic properties of $w_1(\tau)$ in the complex $\tau$-plane. Here it is only assumed that $w_1(\tau)$ is a smooth real function and therefore \eqref{dildisk} cannot be simplified further.

\subsection{Lorentzian -- direct calculation in shockwave}\label{app:dil-lor}

In the Lorentzian discussion for computing the generalized entropy, we used a particular form for the dilaton expressed in terms of the boundary curve \eqref{dilaton2}.
In this subsection, we derive this equation from the expression \eqref{dilaton}.

\begin{figure}
\centering
\includegraphics[scale=0.35]{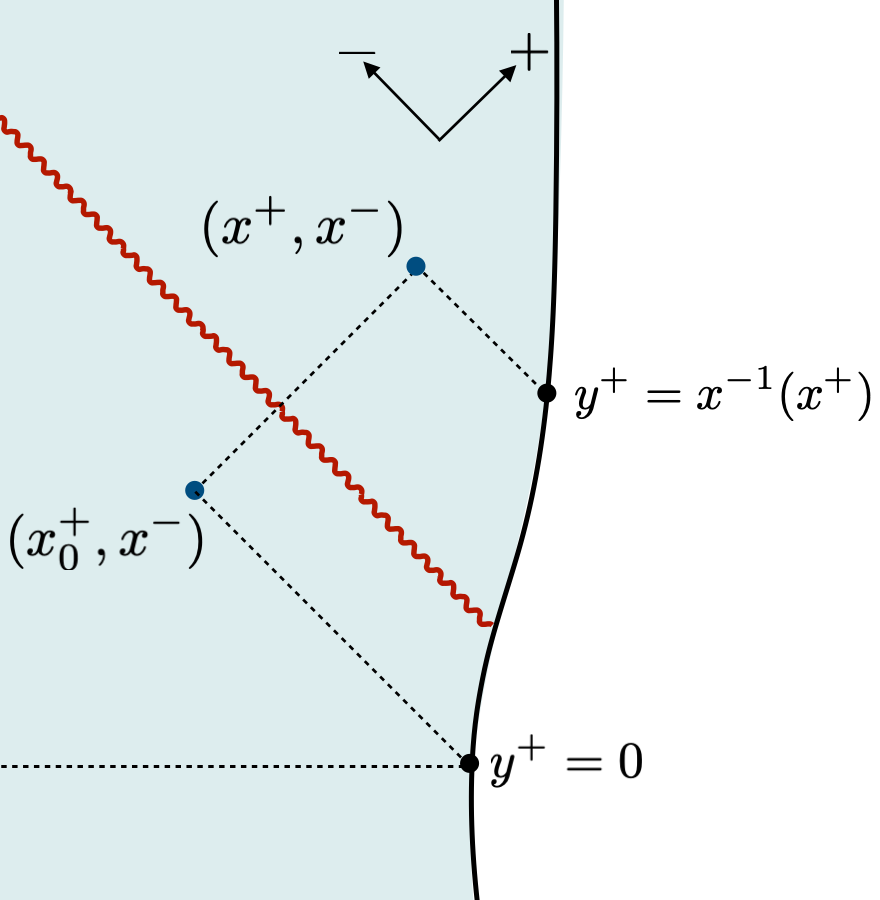}
\caption{}\label{Dilatonderive}
\end{figure}

As we saw in section \ref{sec:3}, the full stress tensor for $t>0$, $\delta \to 0$ has $T_{x^+x^+}\neq 0, T_{x^-x^-}=0$. We will focus on such cases, for which the dilaton is given by
\bal \label{dilatonD1}
\phi(x^+,x^-)=-\f{2\pi\phi_r}{\beta}\f{x^++x^-}{x^+-x^-}&+\f{8\pi G_N}{x^+-x^-}\int^{x^+}_{x^+_0}dx (x^+-x)(x^--x)T_{x^+x^+}(x)\, .
\eal
We take $x^+_0=x(0)$ by using the gluing map $x(t)$, which is placed before the shockwave, and we have $T_{x^+x^+}(x^+_0)=0$.
We evaluate the integral by changing the integration variable from $x$ to $t\equiv x^{-1}(x)$ (see figure \ref{Dilatonderive}) and  by using the Schwarzian equation of motion at time $t$:
\ba
-\pp_{t}\l(\f{\phi_r}{8\pi G_N}\{x(t),t\}\r) =x'(t)^2T_{x^+x^+}(x).\,
\ea
Using the following identity $\frac{\pp_t \{ x(t), t \}}{x'(t)} = ( \frac{1}{x'} ( \frac{x''}{x'})' )' $, we can compute this integral as
\bal \label{dil-lor}
\phi(x^+)&=-\f{2\pi \phi_r}{\beta}\f{x^++x^-}{x^+-x^-}-\f{\phi_r}{x^+-x^-}\int^{y^+}_{0}dt (x^+-x(t))(x^--x(t))\frac{\pp_t \{ x(t), t \}}{x'(t)}\no &=-\f{2\pi\phi_r}{\beta}\f{x^++x^-}{x^+-x^-}-\f{\phi_r}{x^+-x^-}\int^{y^+}_{0}dt (x^+-x(t))(x^--x(t))\l( \frac{1}{x'(t)} \l( \frac{x''(t)}{x'(t)}\r)' \r)' \no
&=\phi_r\l(-2\f{x'(y^+)}{x^+-x^-}+\f{x''(y^+)}{x'(y^+)}\r),\,  \qquad y^+\equiv x^{-1}(x^+)\, ,
\eal
where in the last equality we used the initial conditions $x(0)=1, x'(0) = 2\pi/\beta, x''(0) = \l(2\pi/\beta\r)^2, x'''(0) = \l(2\pi/\beta\r)^3$. One can check that for an eternal black hole with a boundary curve $x(y^+) =e^{\f{2\pi}{\beta}y^+}$, the dilaton is given by the expected answer:
\ba
\phi =-\f{2\pi\phi_r}{\beta}\f{x^++x^-}{x^+-x^-}\, .
\ea

\subsection{Schwinger-Keldysh contour}\label{Sec:DerSK}

Let us show that the dilaton evaluated by \eqref{dilarbitrary} in the shockwave background is equal to the direct computation done in \ref{app:dil-lor}. Since the expression for the dilaton was derived for an arbitrary background in Euclidean signature, it is expected that the Lorentzian version of the formula is also given by Schwinger-Keldysh contour \eqref{dil-lor}. Here the boundary curve parametrized with $\tau$. The Euclidean variables in \ref{app:dil-euc} are related to Lorentzian ones in \ref{app:dil-lor} by the analytic continuation $ it= \tau,   Z(i t) = i x(t)$. For the shockwave background, the only singularity in the integrand $R_1$ comes from residue of \eqref{dilarbitrary} at $\tau= i x^{-1}(x^+)$. By computing the residue we find
\begin{figure}
\centering
\includegraphics[scale=0.5]{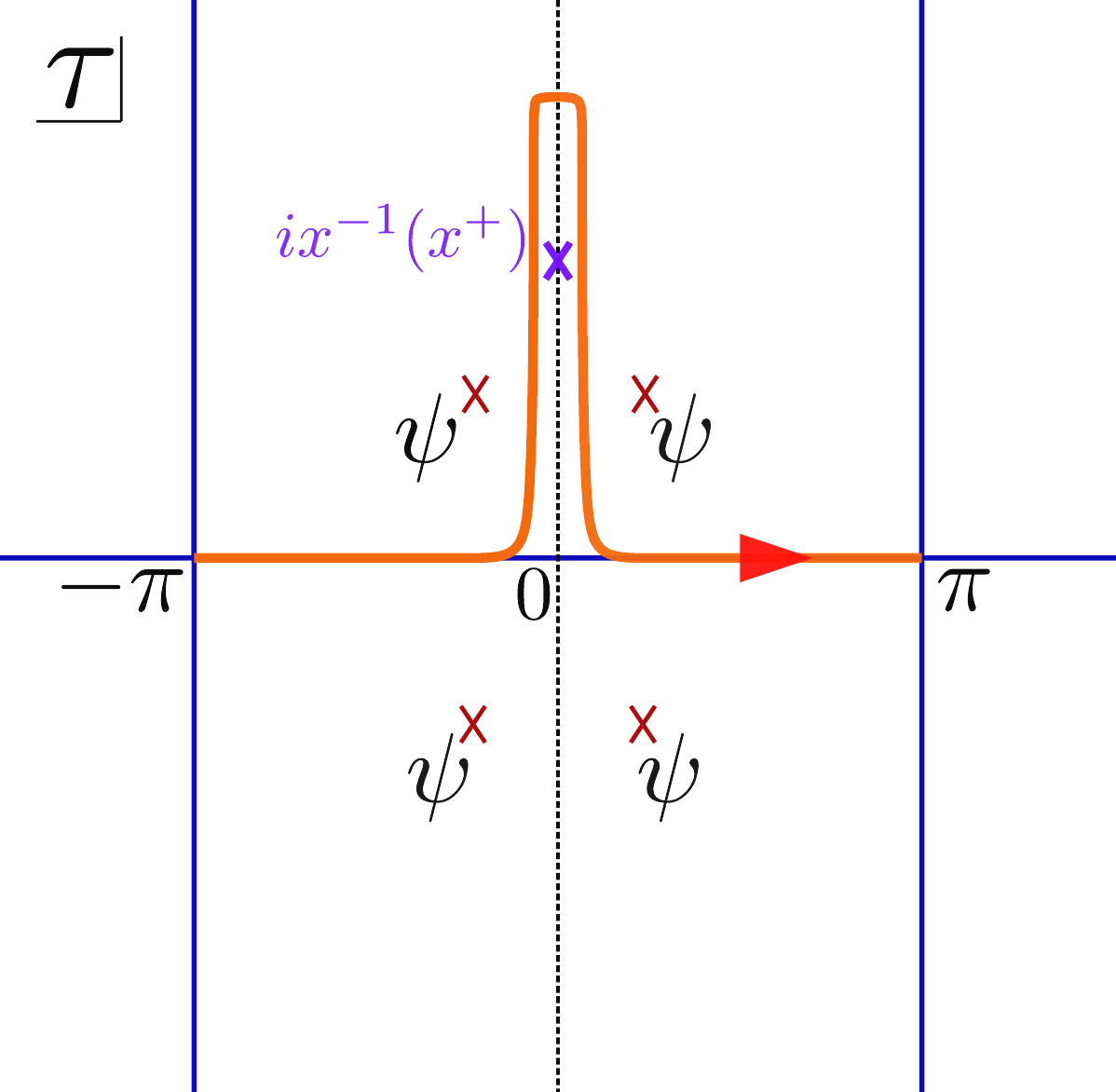}
\caption{Schwinger-Keldysh contour for the dilaton. Note that \eqref{dildisk} is independent of the point $B$. Therefore the only singularity on the Schwinger-Keldysh contour comes when $x(-i \tau) = x^+$.  }\label{sk-contour}
\end{figure}
\bal
 \phi_r &\text{Res} \left. \frac{x'(t)^2 (x^+ -x^-)^2 }{ (x^+ - x(t))^2 (x^- - x(t))^2} \right|_{t = x^{-1}(x^+)} = \nonumber\\
 &  =\phi_r \left(-2 \frac{x'(y^+)}{x^+ - x^- } + \frac{x''(y^+)}{x'(y^+)}\right), \qquad y^+\equiv x^{-1}(x^+).
\eal
This is exactly \eqref{dilaton}, so the two expressions match. 
When we integrate $R_1$ against the $\text{SL}(2)$ kernels to find the QES, all boundary terms from integration by parts drop out since the Schwinger-Keldysh contour is closed. By repeating the argument in section \ref{sec:QESSch}, we  find \eqref{dilatonintermR} is valid when the island appears in Lorentzian signature. 

\section{Details of replica geometry for one interval in eternal black hole}\label{app:singleinterval}\label{app:eternal}

Starting from equation \eqref{singleternal}, we expand the right hand side in terms of Fourier modes
\begin{align}
\pa_\tau R_1 =& - i \sum _{m=2}^{\infty } (m-1) m \bar{A}^m e^{i m \tau} - i \f{2}{1-|A|^2}  \sum _{m=2}^{\infty } m \bar{A}^m e^{i m\tau} \nonumber \\
& + i \f{2 |A|^2}{1- |A|^2}  \sum _{m=-1}^{\infty } m A^m e^{ - i m \tau} + i \sum _{m=0}^{\infty } m (m+1) A^m e^{ - i m \tau},
\end{align}
and 
\begin{align}
\mathcal{F} =& - i \sum_{m=0}^{ \infty} (m+1)  A^m  e^{-i m\tau}- i \frac{2 A}{B- A} \sum_{m=-1}^{\infty} A^m e^{- i m\tau}    \nonumber \\
&- i \sum_{m=2}^{\infty} (m-1) B^{-m} e^{ i m \tau}  - i\f{2 B}{ B-A}\sum_{m=2}^{\infty} e^{ i m \tau} B^{-m}   \nonumber\\
& + i \sum_{m=0}^{\infty}(m+1) \bar{A}^m e^{i m \tau} + i \f{2 \bar{A}}{\bar{B}- \bar{A}} \sum_{m=-1}^{\infty} \bar{A}^m e^{i m\tau}  \nonumber\\
&+ i \sum_{m=2}^{\infty} (m-1) \bar{B}^{-m} e^{- i m \tau} + i \f{2 \bar{B}}{ \bar{B} - \bar{A}} \sum_{m=2}^{\infty}  \bar{B}^{-m}e^{ - i m \tau}.
\end{align}
The left hand side of equation  \eqref{singleternal} does not have modes $m=0, \pm 1$. Solving the equation for those modes sets the QES condition  $\bar{A}/\bar{B}= A/B$ and \eqref{QESsingle}. Due to the boost symmetry we set $\bar{B} = B$. By solving \eqref{singleternal} for other modes, one finds
\bal\label{posM}
&\delta M_+=\no &\frac{1 }{\kappa+1} Ae^{i\tau}\l(-\frac{A+B}{A-B}\kappa \, _2F_1(1,\kappa+1;\kappa+2;A e^{i\tau})+ \f{1+A^2}{1-A^2} \, _2F_1(2,\kappa+1;\kappa+2;A e^{i\tau})\r)
\no&+\frac{\kappa}{\kappa+1}\f{e^{i\tau}}{B} \l(\frac{A+B}{A-B}\, _2F_1\left(1,\kappa+1;\kappa+2;\frac{e^{i\tau}}{B}\right)-\, _2F_1\left(2,\kappa+1;\kappa+2;\frac{e^{i\tau}}{B}\right)\r)
\no &
+\frac{Ae^{i\tau}}{(1-A e^{i\tau})^2}-\frac{2e^{i\tau}}{\kappa+1} \left(\kappa\frac{1-A B}{A-B}-\frac{A}{A^2-1}\right),
\eal
and
\bal\label{negM}
&\delta M_-=\no &\frac{1 }{\kappa+1} Ae^{-i\tau}\l(-\frac{A+B}{A-B}\kappa \, _2F_1(1,\kappa+1;\kappa+2;A e^{-i\tau})+ \f{1+A^2}{1-A^2} \, _2F_1(2,\kappa+1;\kappa+2;A e^{-i\tau})\r)
\no&+\frac{\kappa}{\kappa+1}\f{e^{-i\tau}}{B} \l(\frac{A+B}{A-B}\, _2F_1\left(1,\kappa+1;\kappa+2;\frac{e^{-i\tau}}{B}\right)-\, _2F_1\left(2,\kappa+1;\kappa+2;\frac{e^{-i\tau}}{B}\right)\r)
\no &
+\frac{Ae^{-i\tau}}{(1-A e^{-i\tau})^2}-\frac{2e^{-i\tau}}{\kappa+1} \left(\kappa\frac{1-A B}{A-B}-\frac{A}{A^2-1}\right).
\eal

\end{spacing}
\small
\bibliographystyle{ourbst}
\bibliography{jtshock.bib}

\providecommand{\href}[2]{#2}\begingroup\raggedright\begin{thebibliography}{10}

\bibitem{Almheiri:2019psf}
A.~Almheiri, N.~Engelhardt, D.~Marolf and H.~Maxfield, {{The entropy of bulk
  quantum fields and the entanglement wedge of an evaporating black hole}},
  2019, [\href{http://arxiv.org/abs/arXiv:1905.08762}{{arXiv:1905.08762
  [hep-th]}}].

\bibitem{Penington:2019npb}
G.~Penington, {{Entanglement Wedge Reconstruction and the Information
  Paradox}},  2019,
  [\href{http://arxiv.org/abs/arXiv:1905.08255}{{arXiv:1905.08255 [hep-th]}}].

\bibitem{Ryu:2006bv}
S.~Ryu and T.~Takayanagi, {{Holographic derivation of entanglement entropy from
  AdS/CFT}}, \href{http://dx.doi.org/10.1103/PhysRevLett.96.181602}{Phys. Rev.
  Lett. {\bf 96}, 181602, 2006},
  [\href{http://arxiv.org/abs/arXiv:hep-th/0603001}{{arXiv:hep-th/0603001
  [hep-th]}}].

\bibitem{Hubeny:2007xt}
V.~E. Hubeny, M.~Rangamani and T.~Takayanagi, {{A Covariant holographic
  entanglement entropy proposal}},
  \href{http://dx.doi.org/10.1088/1126-6708/2007/07/062}{JHEP {\bf 07}, 062,
  2007}, [\href{http://arxiv.org/abs/arXiv:0705.0016}{{arXiv:0705.0016
  [hep-th]}}].

\bibitem{Lewkowycz:2013nqa}
A.~Lewkowycz and J.~Maldacena, {{Generalized gravitational entropy}},
  \href{http://dx.doi.org/10.1007/JHEP08(2013)090}{JHEP {\bf 08}, 090, 2013},
  [\href{http://arxiv.org/abs/arXiv:1304.4926}{{arXiv:1304.4926 [hep-th]}}].

\bibitem{Barrella:2013wja}
T.~Barrella, X.~Dong, S.~A. Hartnoll and V.~L. Martin, {{Holographic
  entanglement beyond classical gravity}},
  \href{http://dx.doi.org/10.1007/JHEP09(2013)109}{JHEP {\bf 09}, 109, 2013},
  [\href{http://arxiv.org/abs/arXiv:1306.4682}{{arXiv:1306.4682 [hep-th]}}].

\bibitem{Faulkner:2013ana}
T.~Faulkner, A.~Lewkowycz and J.~Maldacena, {{Quantum corrections to
  holographic entanglement entropy}},
  \href{http://dx.doi.org/10.1007/JHEP11(2013)074}{JHEP {\bf 11}, 074, 2013},
  [\href{http://arxiv.org/abs/arXiv:1307.2892}{{arXiv:1307.2892 [hep-th]}}].

\bibitem{Engelhardt:2014gca}
N.~Engelhardt and A.~C. Wall, {{Quantum Extremal Surfaces: Holographic
  Entanglement Entropy beyond the Classical Regime}},
  \href{http://dx.doi.org/10.1007/JHEP01(2015)073}{JHEP {\bf 01}, 073, 2015},
  [\href{http://arxiv.org/abs/arXiv:1408.3203}{{arXiv:1408.3203 [hep-th]}}].

\bibitem{Almheiri:2019hni}
A.~Almheiri, R.~Mahajan, J.~Maldacena and Y.~Zhao, {{The Page curve of Hawking
  radiation from semiclassical geometry}},
  \href{http://dx.doi.org/10.1007/JHEP03(2020)149}{JHEP {\bf 03}, 149, 2020},
  [\href{http://arxiv.org/abs/arXiv:1908.10996}{{arXiv:1908.10996 [hep-th]}}].

\bibitem{JACKIW1985343}
R.~Jackiw, {Lower dimensional gravity},
  \href{http://dx.doi.org/https://doi.org/10.1016/0550-3213(85)90448-1}{Nuclear
  Physics B {\bf 252}, 343 -- 356, 1985}.

\bibitem{TEITELBOIM198341}
C.~Teitelboim, {Gravitation and hamiltonian structure in two spacetime
  dimensions},
  \href{http://dx.doi.org/https://doi.org/10.1016/0370-2693(83)90012-6}{Physics
  Letters B {\bf 126}, 41 -- 45, 1983}.

\bibitem{Almheiri:2019yqk}
A.~Almheiri, R.~Mahajan and J.~Maldacena, {{Islands outside the horizon}},
  2019, [\href{http://arxiv.org/abs/arXiv:1910.11077}{{arXiv:1910.11077
  [hep-th]}}].

\bibitem{Almheiri:2019qdq}
A.~Almheiri, T.~Hartman, J.~Maldacena, E.~Shaghoulian and A.~Tajdini, {{Replica
  Wormholes and the Entropy of Hawking Radiation}},
  \href{http://dx.doi.org/10.1007/JHEP05(2020)013}{JHEP {\bf 05}, 013, 2020},
  [\href{http://arxiv.org/abs/arXiv:1911.12333}{{arXiv:1911.12333 [hep-th]}}].

\bibitem{Penington:2019kki}
G.~Penington, S.~H. Shenker, D.~Stanford and Z.~Yang, {{Replica wormholes and
  the black hole interior}},  2019,
  [\href{http://arxiv.org/abs/arXiv:1911.11977}{{arXiv:1911.11977 [hep-th]}}].

\bibitem{Hartman:2020swn}
T.~Hartman, E.~Shaghoulian and A.~Strominger, {{Islands in Asymptotically Flat
  2D Gravity}}, \href{http://dx.doi.org/10.1007/JHEP07(2020)022}{JHEP {\bf 07},
  022, 2020}, [\href{http://arxiv.org/abs/arXiv:2004.13857}{{arXiv:2004.13857
  [hep-th]}}].

\bibitem{Hartman:2020khs}
T.~Hartman, Y.~Jiang and E.~Shaghoulian, {{Islands in cosmology}},  2020,
  [\href{http://arxiv.org/abs/arXiv:2008.01022}{{arXiv:2008.01022 [hep-th]}}].

\bibitem{Chen:2020tes}
Y.~Chen, V.~Gorbenko and J.~Maldacena, {{Bra-ket wormholes in gravitationally
  prepared states}},  2020,
  [\href{http://arxiv.org/abs/arXiv:2007.16091}{{arXiv:2007.16091 [hep-th]}}].

\bibitem{Chen:2020uac}
H.~Z. Chen, R.~C. Myers, D.~Neuenfeld, I.~A. Reyes and J.~Sandor, {{Quantum
  Extremal Islands Made Easy, Part I: Entanglement on the Brane}},  2020,
  [\href{http://arxiv.org/abs/arXiv:2006.04851}{{arXiv:2006.04851 [hep-th]}}].

\bibitem{Chen:2020jvn}
H.~Z. Chen, Z.~Fisher, J.~Hernandez, R.~C. Myers and S.-M. Ruan, {{Evaporating
  Black Holes Coupled to a Thermal Bath}},  2020,
  [\href{http://arxiv.org/abs/arXiv:2007.11658}{{arXiv:2007.11658 [hep-th]}}].

\bibitem{Chen:2020hmv}
H.~Z. Chen, R.~C. Myers, D.~Neuenfeld, I.~A. Reyes and J.~Sandor, {{Quantum
  Extremal Islands Made Easy, Part II: Black Holes on the Brane}},  2020,
  [\href{http://arxiv.org/abs/arXiv:2010.00018}{{arXiv:2010.00018 [hep-th]}}].

\bibitem{Balasubramanian:2020xqf}
V.~Balasubramanian, A.~Kar and T.~Ugajin, {{Islands in de Sitter space}},
  2020, [\href{http://arxiv.org/abs/arXiv:2008.05275}{{arXiv:2008.05275
  [hep-th]}}].

\bibitem{Hollowood:2020cou}
T.~J. Hollowood and S.~P. Kumar, {{Islands and Page Curves for Evaporating
  Black Holes in JT Gravity}},
  \href{http://dx.doi.org/10.1007/JHEP08(2020)094}{JHEP {\bf 08}, 094, 2020},
  [\href{http://arxiv.org/abs/arXiv:2004.14944}{{arXiv:2004.14944 [hep-th]}}].

\bibitem{Hollowood:2020kvk}
T.~J. Hollowood, S.~Prem~Kumar and A.~Legramandi, {{Hawking Radiation
  Correlations of Evaporating Black Holes in JT Gravity}},  2020,
  [\href{http://arxiv.org/abs/arXiv:2007.04877}{{arXiv:2007.04877 [hep-th]}}].

\bibitem{Geng:2020qvw}
H.~Geng and A.~Karch, {{Massive islands}},
  \href{http://dx.doi.org/10.1007/JHEP09(2020)121}{JHEP {\bf 09}, 121, 2020},
  [\href{http://arxiv.org/abs/arXiv:2006.02438}{{arXiv:2006.02438 [hep-th]}}].

\bibitem{Alishahiha:2020qza}
M.~Alishahiha, A.~Faraji~Astaneh and A.~Naseh, {{Island in the Presence of
  Higher Derivative Terms}},  2020,
  [\href{http://arxiv.org/abs/arXiv:2005.08715}{{arXiv:2005.08715 [hep-th]}}].

\bibitem{Hashimoto:2020cas}
K.~Hashimoto, N.~Iizuka and Y.~Matsuo, {{Islands in Schwarzschild black
  holes}}, \href{http://dx.doi.org/10.1007/JHEP06(2020)085}{JHEP {\bf 06}, 085,
  2020}, [\href{http://arxiv.org/abs/arXiv:2004.05863}{{arXiv:2004.05863
  [hep-th]}}].

\bibitem{Anegawa:2020ezn}
T.~Anegawa and N.~Iizuka, {{Notes on islands in asymptotically flat 2d dilaton
  black holes}}, \href{http://dx.doi.org/10.1007/JHEP07(2020)036}{JHEP {\bf
  07}, 036, 2020},
  [\href{http://arxiv.org/abs/arXiv:2004.01601}{{arXiv:2004.01601 [hep-th]}}].

\bibitem{Gautason:2020tmk}
F.~F. Gautason, L.~Schneiderbauer, W.~Sybesma and L.~Thorlacius, {{Page Curve
  for an Evaporating Black Hole}},
  \href{http://dx.doi.org/10.1007/JHEP05(2020)091}{JHEP {\bf 05}, 091, 2020},
  [\href{http://arxiv.org/abs/arXiv:2004.00598}{{arXiv:2004.00598 [hep-th]}}].

\bibitem{Almheiri:2020cfm}
A.~Almheiri, T.~Hartman, J.~Maldacena, E.~Shaghoulian and A.~Tajdini, {{The
  entropy of Hawking radiation}},  2020,
  [\href{http://arxiv.org/abs/arXiv:2006.06872}{{arXiv:2006.06872 [hep-th]}}].

\bibitem{Dong:2016hjy}
X.~Dong, A.~Lewkowycz and M.~Rangamani, {{Deriving covariant holographic
  entanglement}}, \href{http://dx.doi.org/10.1007/JHEP11(2016)028}{JHEP {\bf
  11}, 028, 2016},
  [\href{http://arxiv.org/abs/arXiv:1607.07506}{{arXiv:1607.07506 [hep-th]}}].

\bibitem{Dong:2020uxp}
X.~Dong, X.-L. Qi, Z.~Shangnan and Z.~Yang, {{Effective entropy of quantum
  fields coupled with gravity}},
  \href{http://dx.doi.org/10.1007/JHEP10(2020)052}{JHEP {\bf 10}, 052, 2020},
  [\href{http://arxiv.org/abs/arXiv:2007.02987}{{arXiv:2007.02987 [hep-th]}}].

\bibitem{Krishnan:2020fer}
C.~Krishnan, {{Critical Islands}},  2020,
  [\href{http://arxiv.org/abs/arXiv:2007.06551}{{arXiv:2007.06551 [hep-th]}}].

\bibitem{VanRaamsdonk:2020tlr}
M.~Van~Raamsdonk, {{Comments on wormholes, ensembles, and cosmology}},  2020,
  [\href{http://arxiv.org/abs/arXiv:2008.02259}{{arXiv:2008.02259 [hep-th]}}].

\bibitem{Balasubramanian:2020coy}
V.~Balasubramanian, A.~Kar and T.~Ugajin, {{Entanglement between two disjoint
  universes}},  2020,
  [\href{http://arxiv.org/abs/arXiv:2008.05274}{{arXiv:2008.05274 [hep-th]}}].

\bibitem{Marolf:2020xie}
D.~Marolf and H.~Maxfield, {{Transcending the ensemble: baby universes,
  spacetime wormholes, and the order and disorder of black hole information}},
  \href{http://dx.doi.org/10.1007/JHEP08(2020)044}{JHEP {\bf 08}, 044, 2020},
  [\href{http://arxiv.org/abs/arXiv:2002.08950}{{arXiv:2002.08950 [hep-th]}}].

\bibitem{Engelhardt:2020qpv}
N.~Engelhardt, S.~Fischetti and A.~Maloney, {{Free Energy from Replica
  Wormholes}},  2020,
  [\href{http://arxiv.org/abs/arXiv:2007.07444}{{arXiv:2007.07444 [hep-th]}}].

\bibitem{Stanford:2020wkf}
D.~Stanford, {{More quantum noise from wormholes}},  2020,
  [\href{http://arxiv.org/abs/arXiv:2008.08570}{{arXiv:2008.08570 [hep-th]}}].

\bibitem{Akers:2020pmf}
C.~Akers and G.~Penington, {{Leading order corrections to the quantum extremal
  surface prescription}},  2020,
  [\href{http://arxiv.org/abs/arXiv:2008.03319}{{arXiv:2008.03319 [hep-th]}}].

\bibitem{Marolf:2020rpm}
D.~Marolf and H.~Maxfield, {{Observations of Hawking radiation: the Page curve
  and baby universes}},  2020,
  [\href{http://arxiv.org/abs/arXiv:2010.06602}{{arXiv:2010.06602 [hep-th]}}].

\bibitem{Goel:2020yxl}
A.~Goel, L.~V. Iliesiu, J.~Kruthoff and Z.~Yang, {{Classifying boundary
  conditions in JT gravity: from energy-branes to $\alpha$-branes}},  2020,
  [\href{http://arxiv.org/abs/arXiv:2010.12592}{{arXiv:2010.12592 [hep-th]}}].

\bibitem{Maldacena:2016upp}
J.~Maldacena, D.~Stanford and Z.~Yang, {{Conformal symmetry and its breaking in
  two dimensional Nearly Anti-de-Sitter space}},
  \href{http://dx.doi.org/10.1093/ptep/ptw124}{PTEP {\bf 2016}, 12C104, 2016},
  [\href{http://arxiv.org/abs/arXiv:1606.01857}{{arXiv:1606.01857 [hep-th]}}].

\bibitem{Engelsoy:2016xyb}
J.~Engelsoy, T.~G. Mertens and H.~Verlinde, {{An investigation of AdS$_{2}$
  backreaction and holography}},
  \href{http://dx.doi.org/10.1007/JHEP07(2016)139}{JHEP {\bf 07}, 139, 2016},
  [\href{http://arxiv.org/abs/arXiv:1606.03438}{{arXiv:1606.03438 [hep-th]}}].

\bibitem{Jensen:2016pah}
K.~Jensen, {{Chaos and hydrodynamics near AdS$_2$}},  2016,
  [\href{http://arxiv.org/abs/arXiv:1605.06098}{{arXiv:1605.06098 [hep-th]}}].

\bibitem{Nozaki:2014hna}
M.~Nozaki, T.~Numasawa and T.~Takayanagi, {{Quantum Entanglement of Local
  Operators in Conformal Field Theories}},
  \href{http://dx.doi.org/10.1103/PhysRevLett.112.111602}{Phys. Rev. Lett. {\bf
  112}, 111602, 2014},
  [\href{http://arxiv.org/abs/arXiv:1401.0539}{{arXiv:1401.0539 [hep-th]}}].

\bibitem{Asplund:2014coa}
C.~T. Asplund, A.~Bernamonti, F.~Galli and T.~Hartman, {{Holographic
  Entanglement Entropy from 2d CFT: Heavy States and Local Quenches}},
  \href{http://dx.doi.org/10.1007/JHEP02(2015)171}{JHEP {\bf 02}, 171, 2015},
  [\href{http://arxiv.org/abs/arXiv:1410.1392}{{arXiv:1410.1392 [hep-th]}}].

\bibitem{Roberts:2014ifa}
D.~A. Roberts and D.~Stanford, {{Two-dimensional conformal field theory and the
  butterfly effect}},
  \href{http://dx.doi.org/10.1103/PhysRevLett.115.131603}{Phys. Rev. Lett. {\bf
  115}, 131603, 2015},
  [\href{http://arxiv.org/abs/arXiv:1412.5123}{{arXiv:1412.5123 [hep-th]}}].

\bibitem{Hartman:2015lfa}
T.~Hartman, S.~Jain and S.~Kundu, {{Causality Constraints in Conformal Field
  Theory}},  2015,
  [\href{http://arxiv.org/abs/arXiv:1509.00014}{{arXiv:1509.00014 [hep-th]}}].

\bibitem{Caputa:2015waa}
P.~Caputa, J.~Sim\'on, A.~\v{S}tikonas, T.~Takayanagi and K.~Watanabe,
  {{Scrambling time from local perturbations of the eternal BTZ black hole}},
  \href{http://dx.doi.org/10.1007/JHEP08(2015)011}{JHEP {\bf 08}, 011, 2015},
  [\href{http://arxiv.org/abs/arXiv:1503.08161}{{arXiv:1503.08161 [hep-th]}}].

\bibitem{Anous:2016kss}
T.~Anous, T.~Hartman, A.~Rovai and J.~Sonner, {{Black Hole Collapse in the 1/c
  Expansion}}, \href{http://dx.doi.org/10.1007/JHEP07(2016)123}{JHEP {\bf 07},
  123, 2016}, [\href{http://arxiv.org/abs/arXiv:1603.04856}{{arXiv:1603.04856
  [hep-th]}}].

\bibitem{Afkhami-Jeddi:2017rmx}
N.~Afkhami-Jeddi, T.~Hartman, S.~Kundu and A.~Tajdini, {{Shockwaves from the
  Operator Product Expansion}},
  \href{http://dx.doi.org/10.1007/JHEP03(2019)201}{JHEP {\bf 03}, 201, 2019},
  [\href{http://arxiv.org/abs/arXiv:1709.03597}{{arXiv:1709.03597 [hep-th]}}].

\bibitem{Dong:2017xht}
X.~Dong and A.~Lewkowycz, {{Entropy, Extremality, Euclidean Variations, and the
  Equations of Motion}}, \href{http://dx.doi.org/10.1007/JHEP01(2018)081}{JHEP
  {\bf 01}, 081, 2018},
  [\href{http://arxiv.org/abs/arXiv:1705.08453}{{arXiv:1705.08453 [hep-th]}}].

\bibitem{Calabrese:2004eu}
P.~Calabrese and J.~L. Cardy, {{Entanglement entropy and quantum field
  theory}}, \href{http://dx.doi.org/10.1088/1742-5468/2004/06/P06002}{J. Stat.
  Mech. {\bf 0406}, P06002, 2004},
  [\href{http://arxiv.org/abs/arXiv:hep-th/0405152}{{arXiv:hep-th/0405152
  [hep-th]}}].

\bibitem{Almheiri:2014cka}
A.~Almheiri and J.~Polchinski, {{Models of AdS$_{2}$ backreaction and
  holography}}, \href{http://dx.doi.org/10.1007/JHEP11(2015)014}{JHEP {\bf 11},
  014, 2015}, [\href{http://arxiv.org/abs/arXiv:1402.6334}{{arXiv:1402.6334
  [hep-th]}}].

\bibitem{Fiola:1994ir}
T.~M. Fiola, J.~Preskill, A.~Strominger and S.~P. Trivedi, {{Black hole
  thermodynamics and information loss in two-dimensions}},
  \href{http://dx.doi.org/10.1103/PhysRevD.50.3987}{Phys. Rev. D {\bf 50},
  3987--4014, 1994},
  [\href{http://arxiv.org/abs/arXiv:hep-th/9403137}{{arXiv:hep-th/9403137}}].

\bibitem{Hayden:2007cs}
P.~Hayden and J.~Preskill, {{Black holes as mirrors: Quantum information in
  random subsystems}},
  \href{http://dx.doi.org/10.1088/1126-6708/2007/09/120}{JHEP {\bf 09}, 120,
  2007}, [\href{http://arxiv.org/abs/arXiv:0708.4025}{{arXiv:0708.4025
  [hep-th]}}].

\bibitem{Caputa:2014eta}
P.~Caputa, J.~Simon, A.~Stikonas and T.~Takayanagi, {{Quantum Entanglement of
  Localized Excited States at Finite Temperature}},
  \href{http://dx.doi.org/10.1007/JHEP01(2015)102}{JHEP {\bf 01}, 102, 2015},
  [\href{http://arxiv.org/abs/arXiv:1410.2287}{{arXiv:1410.2287 [hep-th]}}].

\bibitem{Mumford}
E.~Sharon and D.~Mumford, {{2D-Shape Analysis Using Conformal Mapping}},
  \href{http://dx.doi.org/10.1007/s11263-006-6121-z}{International Journal of
  Computer Vision {\bf 70}, 55, 2006}.

\bibitem{Jana:2020vyx}
C.~Jana, R.~Loganayagam and M.~Rangamani, {{Open quantum systems and
  Schwinger-Keldysh holograms}},
  \href{http://dx.doi.org/10.1007/JHEP07(2020)242}{JHEP {\bf 07}, 242, 2020},
  [\href{http://arxiv.org/abs/arXiv:2004.02888}{{arXiv:2004.02888 [hep-th]}}].

\bibitem{Hartman:2013mia}
T.~Hartman, {{Entanglement Entropy at Large Central Charge}},  2013,
  [\href{http://arxiv.org/abs/arXiv:1303.6955}{{arXiv:1303.6955 [hep-th]}}].

\bibitem{Fitzpatrick:2014vua}
A.~Fitzpatrick, J.~Kaplan and M.~T. Walters, {{Universality of Long-Distance
  AdS Physics from the CFT Bootstrap}},
  \href{http://dx.doi.org/10.1007/JHEP08(2014)145}{JHEP {\bf 08}, 145, 2014},
  [\href{http://arxiv.org/abs/arXiv:1403.6829}{{arXiv:1403.6829 [hep-th]}}].

\bibitem{He:2014mwa}
S.~He, T.~Numasawa, T.~Takayanagi and K.~Watanabe, {{Quantum dimension as
  entanglement entropy in two dimensional conformal field theories}},
  \href{http://dx.doi.org/10.1103/PhysRevD.90.041701}{Phys. Rev. D {\bf 90},
  041701, 2014}, [\href{http://arxiv.org/abs/arXiv:1403.0702}{{arXiv:1403.0702
  [hep-th]}}].

\bibitem{Kusuki:2018wpa}
Y.~Kusuki, {{Light Cone Bootstrap in General 2D CFTs and Entanglement from
  Light Cone Singularity}},
  \href{http://dx.doi.org/10.1007/JHEP01(2019)025}{JHEP {\bf 01}, 025, 2019},
  [\href{http://arxiv.org/abs/arXiv:1810.01335}{{arXiv:1810.01335 [hep-th]}}].

\bibitem{Mirbabayi:2020fyk}
M.~Mirbabayi, {{A 2-Replica Wormhole}},  2020,
  [\href{http://arxiv.org/abs/arXiv:2008.09626}{{arXiv:2008.09626 [hep-th]}}].

\bibitem{Dong:2020iod}
X.~Dong and H.~Wang, {{Enhanced corrections near holographic entanglement
  transitions: a chaotic case study}},  2020,
  [\href{http://arxiv.org/abs/arXiv:2006.10051}{{arXiv:2006.10051 [hep-th]}}].

\bibitem{Marolf:2020vsi}
D.~Marolf, S.~Wang and Z.~Wang, {{Probing phase transitions of holographic
  entanglement entropy with fixed area states}},  2020,
  [\href{http://arxiv.org/abs/arXiv:2006.10089}{{arXiv:2006.10089 [hep-th]}}].

\bibitem{KIRILLOV1998735}
A.~Kirillov, {Geometric approach to discrete series of unirreps for vir.},
  \href{http://dx.doi.org/https://doi.org/10.1016/S0021-7824(98)80007-X}{Journal
  de Mathématiques Pures et Appliquées {\bf 77}, 735 -- 746, 1998}.

\bibitem{Calabrese:2009qy}
P.~Calabrese and J.~Cardy, {{Entanglement entropy and conformal field theory}},
  \href{http://dx.doi.org/10.1088/1751-8113/42/50/504005}{J. Phys. {\bf A42},
  504005, 2009}, [\href{http://arxiv.org/abs/arXiv:0905.4013}{{arXiv:0905.4013
  [cond-mat.stat-mech]}}].

\bibitem{Hartman:2013qma}
T.~Hartman and J.~Maldacena, {{Time Evolution of Entanglement Entropy from
  Black Hole Interiors}}, \href{http://dx.doi.org/10.1007/JHEP05(2013)014}{JHEP
  {\bf 05}, 014, 2013},
  [\href{http://arxiv.org/abs/arXiv:1303.1080}{{arXiv:1303.1080 [hep-th]}}].

\bibitem{Susskind:2014moa}
L.~Susskind, {{Entanglement is not enough}},
  \href{http://dx.doi.org/10.1002/prop.201500095}{Fortsch. Phys. {\bf 64},
  49--71, 2016}, [\href{http://arxiv.org/abs/arXiv:1411.0690}{{arXiv:1411.0690
  [hep-th]}}].

\bibitem{Casini:2005rm}
H.~Casini, C.~Fosco and M.~Huerta, {{Entanglement and alpha entropies for a
  massive Dirac field in two dimensions}},
  \href{http://dx.doi.org/10.1088/1742-5468/2005/07/P07007}{J. Stat. Mech. {\bf
  0507}, P07007, 2005},
  [\href{http://arxiv.org/abs/arXiv:cond-mat/0505563}{{arXiv:cond-mat/0505563}}].

\end{thebibliography}\endgroup

\end{document}